\documentclass[]{aa}
\usepackage[utf8]{inputenc}
\usepackage[]{color}
\usepackage[font=small]{caption}
\usepackage[labelformat = empty,position=top]{subcaption}
\usepackage[export]{adjustbox}

\newcommand{\tom}{\tilde\omega}
\newcommand{\tos}{\tilde s}
\begin{document}

\title{Forward modelling of brightness variations in Sun-like stars}
\subtitle{I. Emergence and surface transport of magnetic flux}
  \author{E. I\c{s}\i k\inst{1,2}
     \and S. K. Solanki\inst{1,3} 
     \and N. A. Krivova\inst{1} 
     \and A. I. Shapiro\inst{1}     
          }
  \institute{Max-Planck-Institut f\"ur Sonnensystemforschung, 
              Justus-von-Liebig-Weg 3, 37077 G\"ottingen, Germany \\
              \email{[isik,solanki]@mps.mpg.de}
         \and
             Feza G\"ursey Center for Physics and Mathematics, 
             Bo\u{g}azi\c{c}i University, Kuleli 34684, Istanbul, Turkey
        \and
              {School of Space Research, Kyung Hee University, Yongin, Gyeonggi-Do, 
              446-701, Republic of Korea}
             }
\date{\today}

\abstract{
    The latitudinal distribution of starspots deviates from the solar pattern 
    with increasing rotation rate. Numerical simulations of 
    magnetic flux emergence and transport can 
    help model the observed stellar activity patterns and 
    {the associated brightness} variations. 
}{
    We set up {a composite} model for the processes of flux emergence and 
    transport on Sun-like stars to simulate stellar brightness variations for various 
    levels of magnetic activity {and rotation rates}. 
}{
    {Assuming that the distribution of magnetic flux at the base of the 
     convection zone {follows solar scaling relations},}
    we {calculate the emergence latitudes and tilt angles of bipolar regions 
    at the surface for} various rotation rates, {using thin-flux-tube 
    simulations}. Taking {these two quantities} as input to a surface flux 
    transport (SFT) model, we simulate the diffusive-advective evolution of the radial 
    field at the stellar surface, {including effects of active region nesting}. 
}{
    As the rotation rate increases, (1) magnetic flux emerges at higher latitudes and an 
    inactive gap opens around the equator, reaching a half-width of $20^\circ$ {for }
    $8\Omega_\sun$; and (2) the tilt angles of freshly emerged bipolar 
    regions show stronger variations {with} latitude. 
     Polar spots {can} form  
    {at} $8\Omega_\sun$ by accumulation of follower-polarity flux 
   {from decaying bipolar regions}. 
    {From $4\Omega_\sun$ to $8\Omega_\sun$}, 
    the {maximum} spot coverage {changes from} 
   {3 to 20}\%, respectively, compared to {0.4}\% {in the solar model}.
   {Nesting of activity can lead to strongly non-axisymmetric spot distributions.}
}{
    On Sun-like stars {rotating at} $8\Omega_\sun$ ($P_{\rm rot}\simeq 3$ days), 
    polar spots can {form, owing to} higher {levels of} flux emergence 
    rate {and tilt angles}. Defining spots by a 
    threshold field strength yields global spot coverages that are roughly 
    consistent with stellar observations.
    }

\keywords{stars: activity -- stars: magnetic field --
                stars: solar-type -- starspots -- 
                methods: numerical -- magnetohydrodynamics (MHD)}

\maketitle

\section{Introduction}
The advent of space-borne photometry, which is designed 
primarily to detect extrasolar planets, has made short-term stellar 
activity signals measurable \citep{koch10}. 
Brightness variations in Sun-like stars on timescales ranging from hours to 
decades are driven by convective flows and magnetic fields at the surface, 
while the observed patterns of variability also depend on the stellar rotation 
and related quantities \citep{aigrain04,sasha14,sasha17}. 

Dark spots and bright faculae observed on the Sun are formed by 
relatively strong magnetic flux concentrations threading the photosphere. 
To investigate radiative variations driven by magnetic 
activity and rotation on Sun-like stars, it is important to numerically 
simulate the distribution 
and the evolution of surface magnetic fields at large scales. 

The latitudinal distribution of starspots is one of the  
extensively studied features of magnetic activity on rapidly 
rotating cool stars, via Doppler and Zeeman-Doppler imaging 
\citep{strassmeier09,dl09}.
For Sun-like stars in particular, the mean latitude of cool spots and 
magnetic fields has been observed to increase with the rotation rate. 
Spots near or at the rotational poles are observed  
on rapidly rotating Sun-like stars with rotation periods below about 3 days 
\citep{jeffers02,marsden06,jarvinen07,waite15}. 

As a possible explanation for polar spots, 
\citet{ss92} suggested a mechanism in which 
the action of the axially directed Coriolis force in the 
rotating frame of a rising flux tube becomes comparable to or 
even dominates the outward buoyancy force for sufficiently 
fast rotation. \citet{msch96} 
demonstrated how faster rotation deflects rising tubes  
towards high latitudes on Sun-like stars, and \citet{granzer00} 
obtained latitudinal distributions for a range of stellar masses. 

A surface mechanism which can also contribute to the formation 
of polar spots was suggested by \citet{st01} using a 
random-walk model of flux dispersal and transport. 
In this model, a strong high-latitude concentration of 
magnetic fields of a single polarity was formed by the trailing 
polarities of bipolar regions when the flux emergence rate was 
30 times the solar value. 
A ring-like structure was formed around 
a polar spot of opposite polarity, which had a much longer decay time 
than what a diffusion model would infer. 

Another attempt was made by \citet{isik07}, 
who used a {diffusive} surface flux transport (SFT) 
model subject to observed stellar surface differential 
rotation estimates and stellar radii. They 
suggested that the long lifetimes of polar spots can be explained 
by the mid-latitude emergence of 
a large bipolar region with a large tilt angle, presumably owing 
to strong action of the Coriolis effect. 
In both \citet{st01} and \citet{isik07}, a largely unipolar polar spot 
was produced by the diffusion of flux from the follower-polarities 
of bipolar magnetic regions (BMRs) at low or medium latitudes. 

\citet{voho06} also used an SFT model to show that a sufficiently 
fast poleward meridional flow at the surface 
can also lead to polar spots, but with intermingled polarities. 
In an attempt to reproduce the flip-flop 
activity variations that have been reported for some active stars \citep{berdyugina05}, 
\citet{ek05} applied a non-axisymmetric mean-field dynamo model, which 
produced an asymmetric distribution of magnetic field 
concentrations near the rotational poles. 

\citet{msch96} modelled the latitudinal distribution of emerging flux on solar-mass 
stars, and \citet{isik07b, isik11} studied {the combined} effects of 
the dynamo, emergence, and SFT processes on stellar cycles. 
{{It is known that the average inclination of bipolar magnetic 
regions with respect to the east-west direction (the tilt angle) is crucial in 
maintaining the solar activity cycle \citep{baumann04, jiang10}.}}
However, a systematic {modelling} of Sun-like stars with 
{physically consistent computation} of tilt angles {and the} subsequent 
surface {flux} transport has not yet been undertaken. 

To better understand the implications of the rotation-activity  
relation not too far from the solar regime and to provide a 
testing platform of forward modelling 
for {photometric} reconstructions, here we construct a modelling 
framework. It incorporates 
the observed properties of the solar cycle, 
the effects of rotation on rising flux 
tubes in the convection zone, and surface flux transport. 
In this paper, we present 
the method and discuss the impact of increasing rotation rate and 
activity on the surface patterns of magnetic flux. A following 
paper will be devoted to syntheses of 
light curves based on the magnetic flux emergence 
(Sect.~\ref{ssec:rise}) and transport models (Sect.~\ref{ssec:sft})
presented here. 
{The purpose of the model presented here is restricted to 
reproducing brightness variations of Sun-like stars in high-precision 
data such as from \textit{Kepler} \citep{koch10} and TESS \citep{ricker15}, 
particularly from very short timescales up to several stellar rotations. 
We do not, however, rule out a later extension to also model spectroscopic 
(e.g. radial velocity) and spectropolarimetric changes {by including 
horizontal fields}. }
In Sect.~\ref{sec:res} we present results from 
our scaled stellar models, namely temporal and latitudinal 
patterns of activity. The limitations of the model and the 
relevance of our results with 
respect to the observations of active Sun-like stars is 
discussed in Sect.~\ref{sec:dis}. 

\section{Model setup}

To model the emergence and the evolution of the surface magnetic field, 
we first generated {an 11-year} semi-synthetic sunspot group 
record based on the statistics of solar cycle 22 (Sect.~\ref{sssec:record1}). 
{With increasing stellar rotation rate, we linearly scaled the 
flux emergence rate, in agreement with the observed rotation-activity relation. 
Stronger solar cycles are known to show a tendency for higher mean emergence latitudes \citep{solanki08, jcss11a}. We therefore used an empirical relation to extrapolate the mean latitude of 
emergence to higher activity levels (Sect.~\ref{sssec:record2}). 
We assumed that the resulting distribution represents the butterfly diagram of flux-tube 
eruptions at the base of the convection zone of a G2V-type star with a given activity level. Using flux-tube simulations, we then computed the 
emergence latitudes and tilt angles of rising loops 
for a {given} stellar rotation rate (Sect.~\ref{ssec:rise}). 
{A schematic presentation of the algorithm 
is given in detail in Appendix~\ref{sec:app0}.}
}

{The latitudinal distribution of emerging 
flux is therefore determined by two independent effects in our approach: 
$(i)$ The base latitudinal 
distribution changes with the \emph{flux emergence rate} 
(extrapolation of the empirical solar relation between the activity level and the mean 
latitude); and $(ii)$ \emph{the rotation rate} affects the action of the Coriolis force, 
which tends to deflect rising flux tubes poleward. }

The {resulting} starspot {emergence} records are then used as input 
to the SFT model 
to obtain the evolution of surface magnetic flux (Sect.~\ref{ssec:sft}). 
In the following sub{-}sections we describe the various steps in 
this chain in greater detail. 

\subsection{Synthetic starspot group records}
\label{ssec:record}

\subsubsection{{The} input solar cycle}
\label{sssec:record1}

Using the RGO/SOON records between 1700 and 2010, 
\citet[][hereafter JCSS11]{jcss11a} found statistical relationships 
between various characteristics of sunspot groups.  
They generated a semi-synthetic Spot Group Record 
(SGR) 
based on these relationships. We adopted the solar SGR 
generation procedures described in JCSS11 and 
\citet[][hereafter CJS16; see their Sect.~2.2]{cjs16} 
to simulate a relatively strong, artificial cycle lasting 
about 11 years. We describe the details of this procedure in 
{Appendix~\ref{sec:app}}. 

{In summary}, the synthetic cycle module provides 
the following quantities {related to sunspot group emergence} 
as a function of time (cycle phase):
the total number of  sunspot groups (Eq.~\ref{eq:rg});
the emergence times and latitudes, which are determined by
the mean latitude of sunspot groups (Eqs.~\ref{eq:lat}-\ref{eq:latscale});
the width of the latitudinal spread (Eq.~\ref{eq:sigma});
random longitudes (nesting is introduced in Sect.~\ref{sssec:record3});
sunspot group areas drawn from a composite empirical distribution
(Eq.~\ref{eq:areadist}).

Instead of using the tilt angles synthesised within the JCSS11 model 
from empirical relationships, we use the results of the {flux-tube} simulations 
to be described in Sect.~\ref{ssec:rise}. For the solar case, the results are  
similar to the tilt angle relationship assumed by JCSS11, 
while for more rapidly rotating stars {very} different tilt angle {distributions} 
are obtained. 

{Having defined the base model of the solar cycle}, we now describe {how we 
determine the} flux emergence rate and the mean latitude {of flux eruption}, 
for more active {Sun-like} stars.

\subsubsection{Scaling {the cycle} on more active stars}
\label{sssec:record2}

To scale the stellar {cycle amplitude}, $S_\star$,
we define $\tos\equiv S_\star/S_\sun$, {where $S$ is 
in units of the {maximum of the} annual running mean of the 
sunspot group number, $S_\sun$ 
is set to 156, which is the observed value for Cycle 22 (see Appendix~\ref{ssec:appa}), and 
$\tos$ is then a relative measure of the emergence frequency (or rate) of BMRs; 
{for simplicity}, this quantity will be called the emergence 
frequency or flux emergence rate (in solar units) throughout the rest of the paper, as they 
are proportional quantities.}
To evaluate $\tos$, we focused on two extreme cases: 
$(i)$ a constant $\tos=1$, to isolate the rotational 
effects on the (otherwise solar) emergence pattern; and 
$(ii)$ a linear scaling 
$\tos=\tom:=\Omega_\star/\Omega_\sun$, following 
the observed proportionality $Bf=50v_{\rm eq}$ between the field 
strength ($Bf$, where $f$ is the filling factor) and the equatorial 
rotational velocity $v_{\rm eq}$ of Sun-like stars, 
estimated by \citet{reiners12}. To first order, 
$Bf$ is proportional to the total magnetic flux on the stellar 
disc. The scaling therefore provides a rough representation 
of the rotation-activity relationship of G dwarfs. 

{The effect} that stronger solar cycles start at higher latitudes 
{has already been considered by JCSS11}. 
We extrapolated this effect to higher activity levels ($\tos>1$){
 and expressed the mean latitude of flux eruptions from the base of the stellar 
convection zone,  $\langle\lambda\rangle_\star$,} 
by modifying Eq.~(\ref{eq:latscale}), as 
\begin{equation}
    \langle\lambda\rangle_\star = 12.2 + k\tos S_\sun, 
    \label{eq:meanlat}
\end{equation}
where we assumed $k=0.022$ for $\tos<8$, as in 
Eq.~(\ref{eq:latscale}). We chose $k=0.014$ for $\tos=8$ 
to limit the maximum latitude of the input cycle to $73^\circ$. 
This roughly corresponds 
to the high-latitude edge of the main region of instability of flux tubes (see 
Sect.~\ref{sssec:stab}). 

\begin{figure}
    \centering
    \includegraphics[width=\columnwidth]{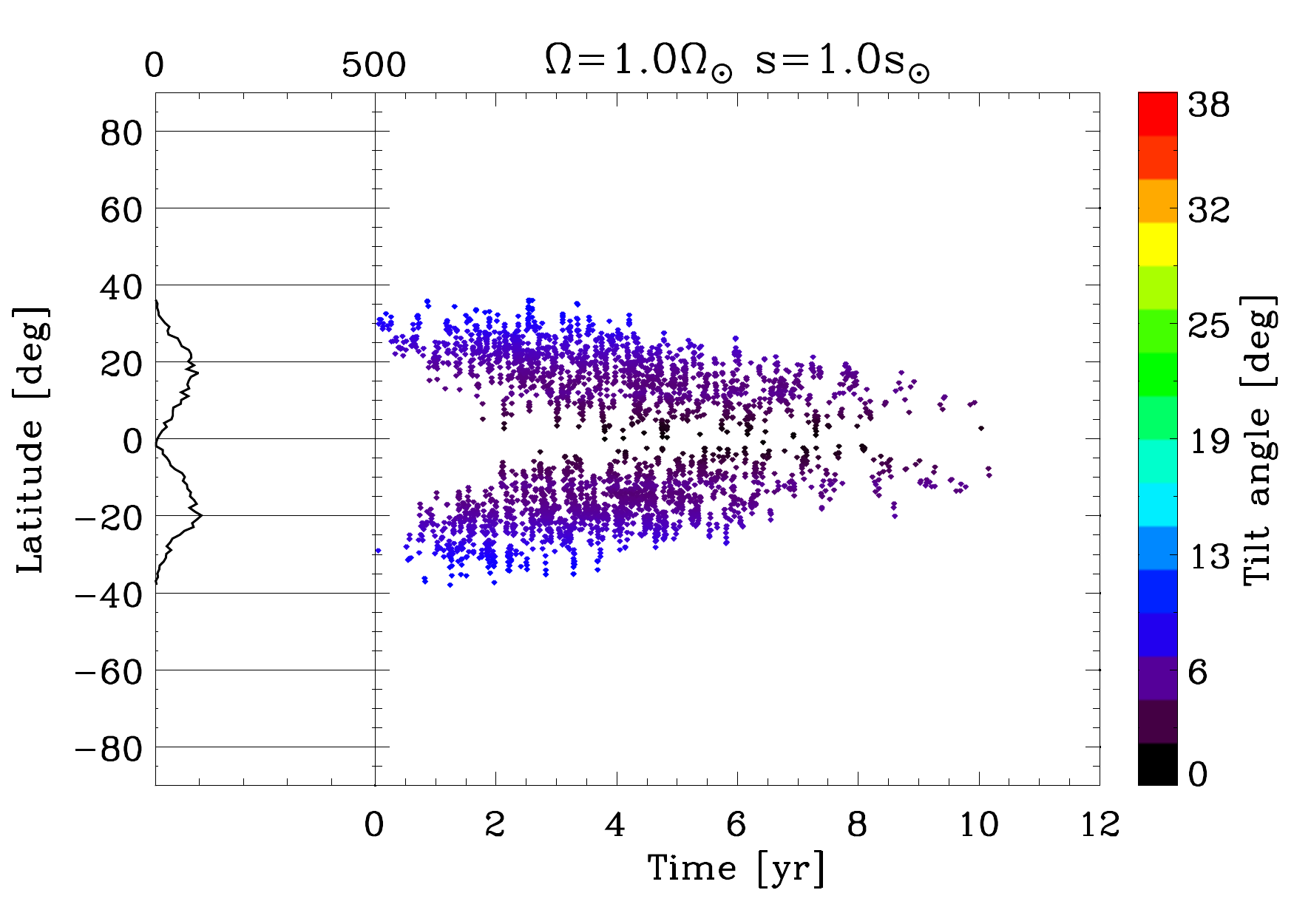}
    \caption{Semi-synthetic emergence record (SGR)
    for solar cycle 22, using emergence latitudes and tilt 
    angles resulting from the remapping procedure 
    for the solar rotation and flux emergence rate $(\tom,\tos)=(1.0,1.0)$. 
    The colour bar shows the tilt angles with a rather high saturation 
    level, {so} that the tilt angles can be compared 
    with faster-rotating cases. The plot on the left shows the histogram of emergence 
    latitudes.}
    \label{fig:bfly-ref}
\end{figure}

\begin{figure*}
\centering
\begin{subfigure}[t]{0.01\textwidth}
(a)
\end{subfigure}
\begin{subfigure}[t]{0.4\textwidth}
\includegraphics[width=\linewidth,valign=t]{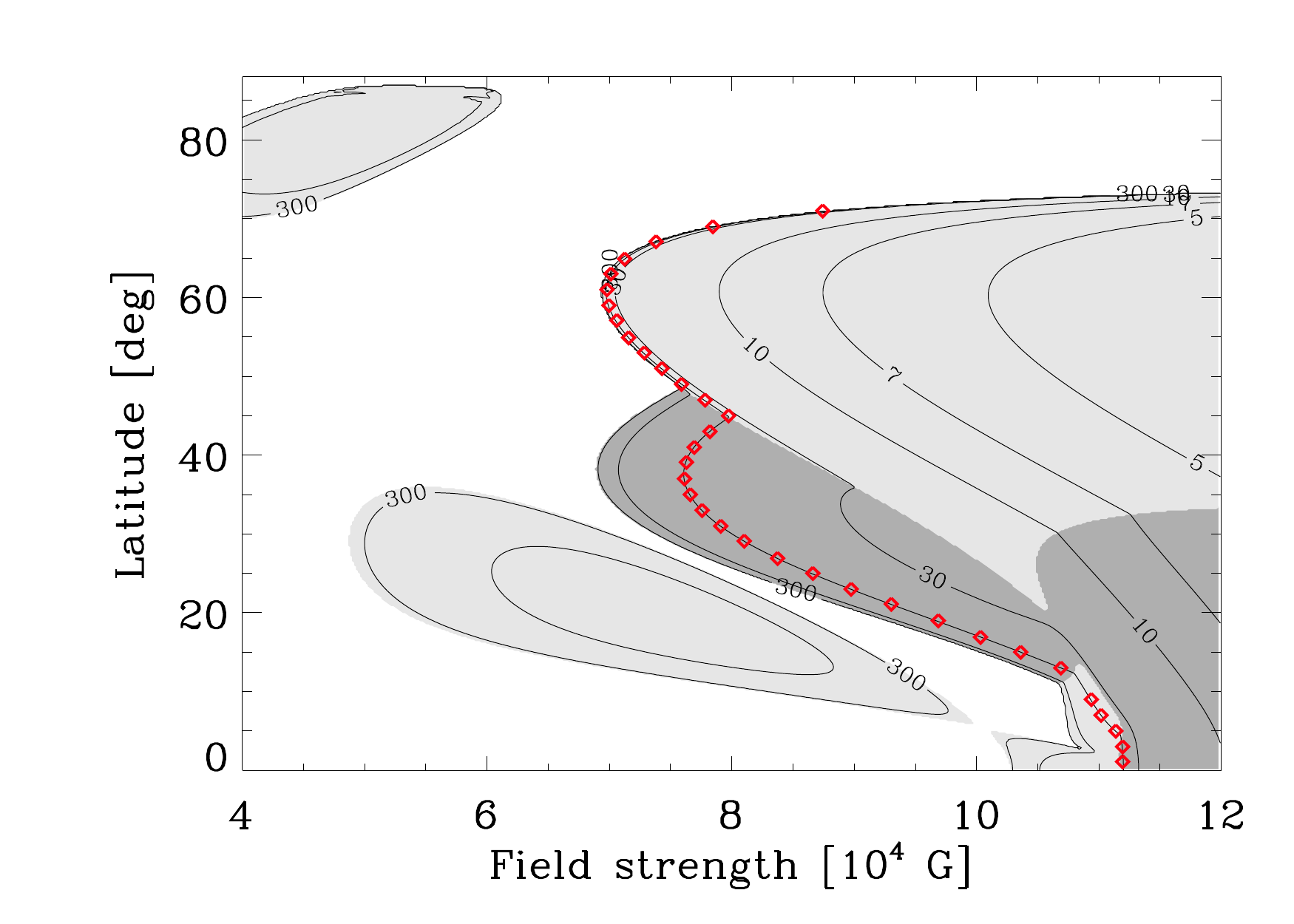}
\end{subfigure}
\begin{subfigure}[t]{0.01\textwidth}
(b)
\end{subfigure}
\begin{subfigure}[t]{0.4\textwidth}
\includegraphics[width=\linewidth,valign=t]{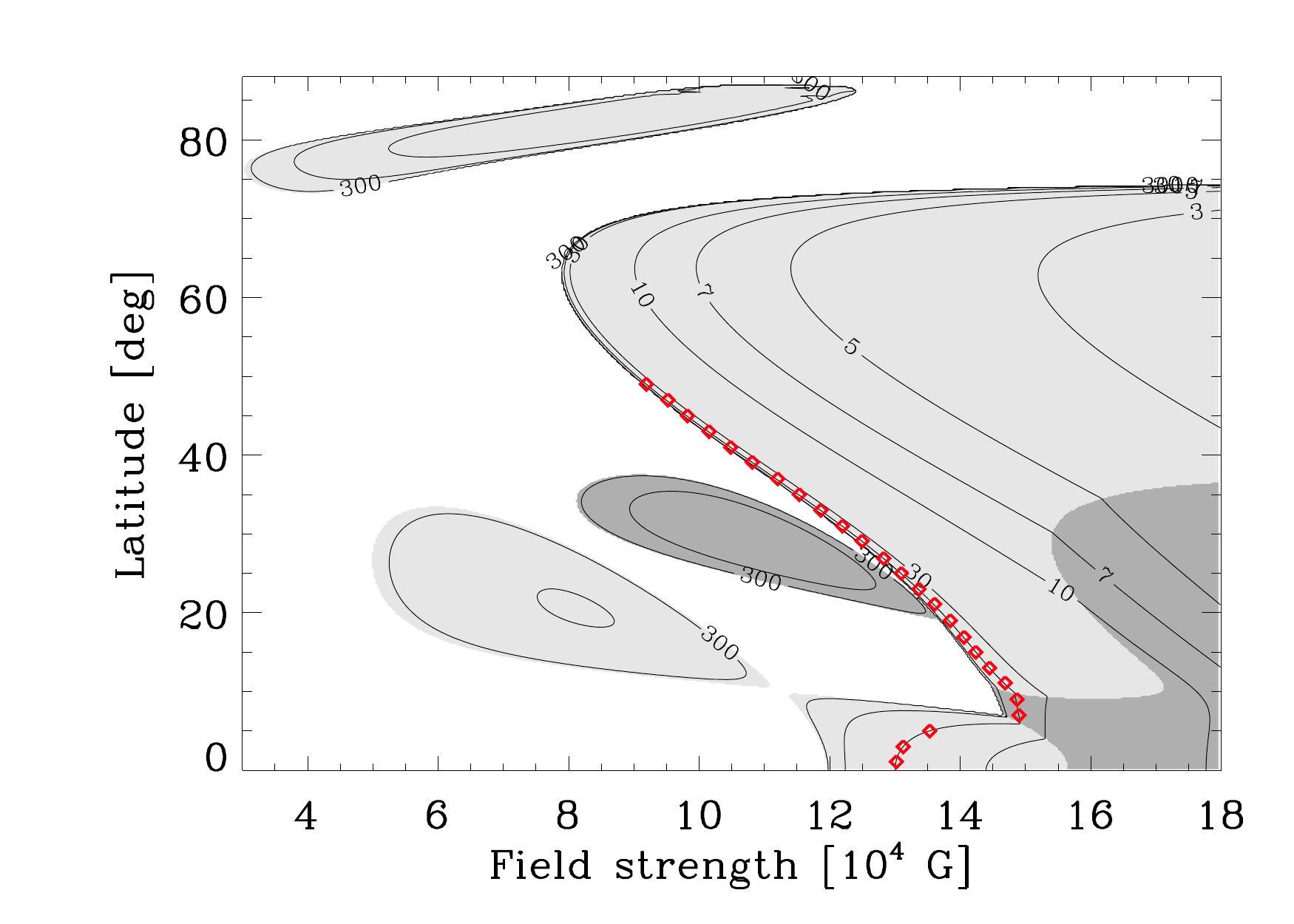}
\end{subfigure} \\
\begin{subfigure}[t]{0.01\textwidth}
(c)
\end{subfigure}
\begin{subfigure}[t]{0.4\textwidth}
\includegraphics[width=\linewidth,valign=t]{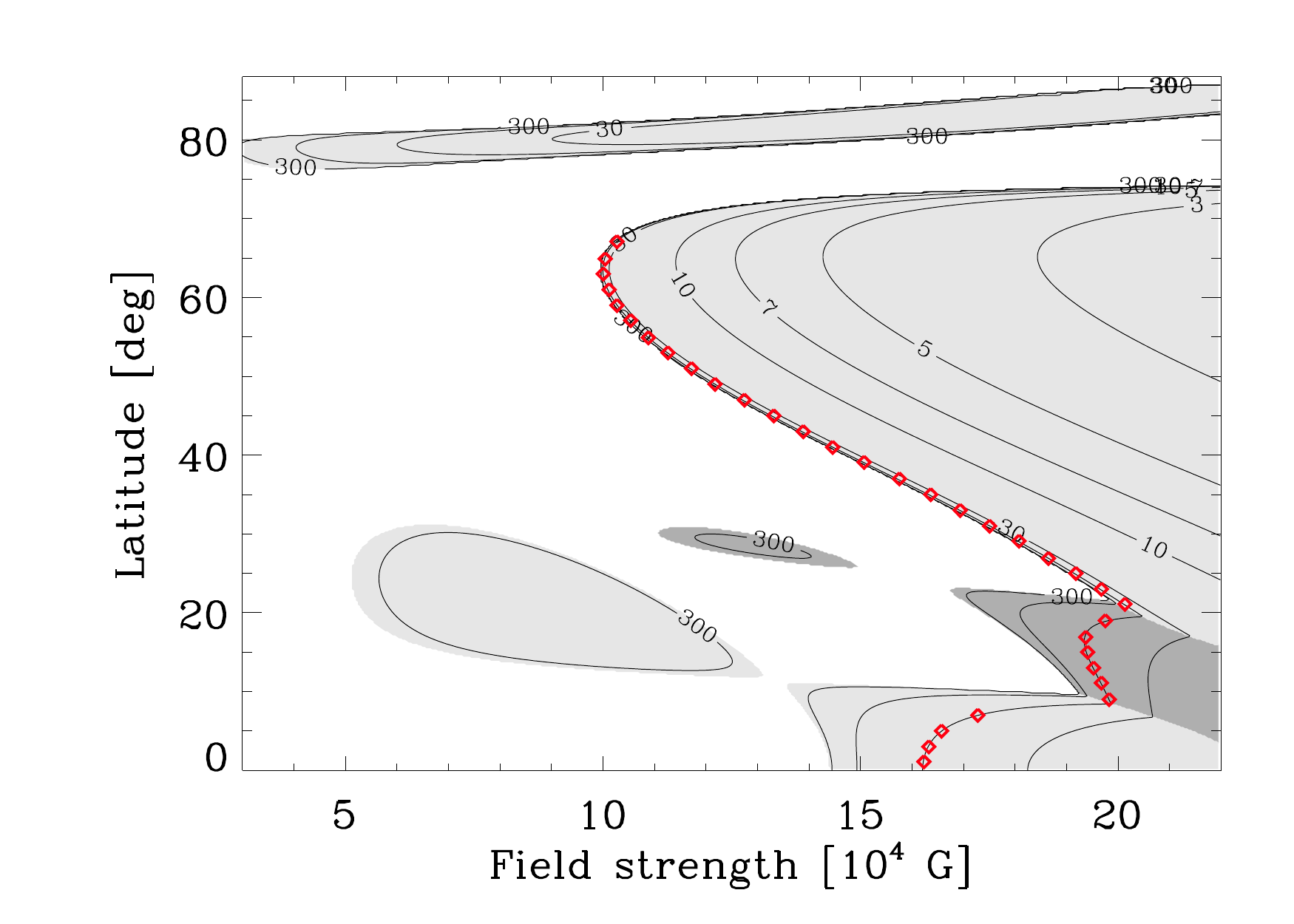}
\end{subfigure}
\begin{subfigure}[t]{0.01\textwidth}
(d)
\end{subfigure}
\begin{subfigure}[t]{0.4\textwidth}
\includegraphics[width=\linewidth,valign=t]{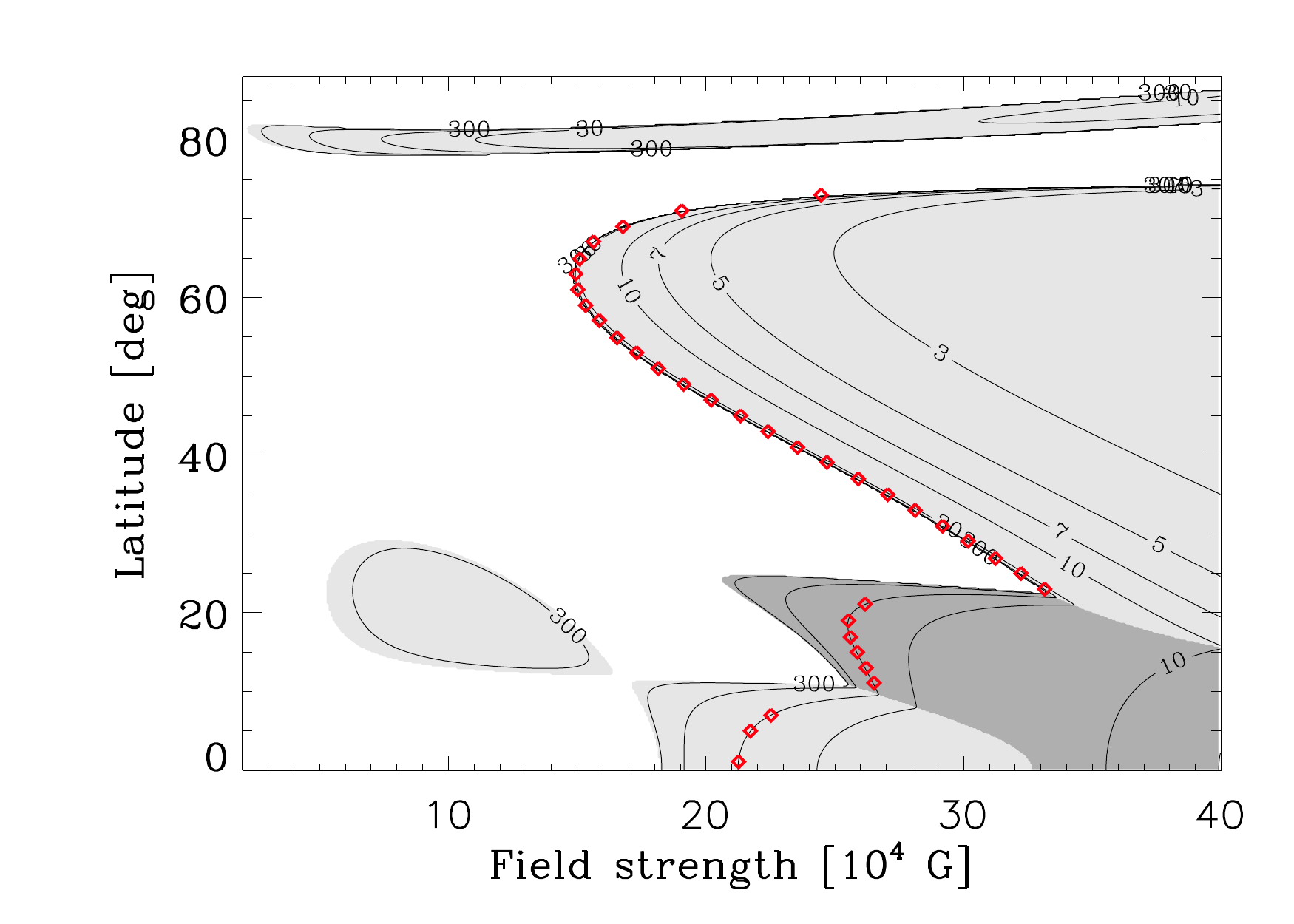}
\end{subfigure}
    \caption{The initial latitudes and field strengths of flux tubes (red diamonds) 
    chosen for flux-tube rise computations, 
    plotted over linear stability diagrams for flux tubes at the middle of the 
    overshoot region, for (a) $\tom=1$, (b) $\tom=2$, (c) $\tom=4$, and (d) $\tom=8$, 
    with differential rotation $\Delta\Omega_\star=\Delta\Omega_\sun$. 
    The contour lines denote the growth 
    time in days. The linearly stable regime is shown in white, and the unstable regime 
    is shaded with light grey where the fastest-growing mode is $m=1$, and with 
    dark-grey for $m=2$. 
    The red diamonds show the initial tube parameters chosen for the numerical 
    simulations, at a growth time of 50 days.}
\label{fig:stab}
\end{figure*}

{Rather than assuming the surface emergence record described above 
to represent the distribution at the base of the convection zone, 
we corrected the eruption latitudes at the base of the convection zone 
for the weak poleward deflection of rising flux tubes 
at the solar rotation rate (see Appendix~\ref{sec:app0}).}

\subsubsection{Nesting of active regions}
\label{sssec:record3}

{It is known that sunspot groups tend to emerge within `nests' of 
activity \citep[e.g.][]{castenmiller86}. To simulate the observed clumping of 
active-region 
emergence, we modified the longitudes and latitudes obtained in the previous steps 
{by setting a probability that a given BMR would be part of a nest, which we call nesting probability} (Appendix~\ref{sec:nests}). When this 
effect is included, the resulting cycle variation of low-order multipoles such as the equatorial 
dipole as well as the open flux better represent the observed variations.}

Figure~\ref{fig:bfly-ref} shows the resulting solar reference 
butterfly diagram with $(\tom,\tos)=(1.0,1.0)$, 
{with a nesting probability of {70\%}. The only 
significant difference from the synthetic cycle-22 butterfly diagram of the JCSS11 
model is the clustered emergence pattern.}

\subsection{The rise of flux tubes in the convection zone} 
\label{ssec:rise}

We model the latitudinal distribution of emerging magnetic flux 
on Sun-like stars using numerical simulations of buoyantly rising 
magnetic flux tubes. We adopt the thin flux tube approximation 
\citep{spruit81} and model flux tubes leading to sunspot groups as 
initially toroidal flux rings in mechanical equilibrium \citep{cale98} 
in the form given by \citet{fms95} and \citet{cale95}. At the equilibrium 
state, the flux tube is
assumed to lie in the stably stratified overshoot region 
at the base of a solar-like, non-local-mixing-length convection zone 
model, which is adopted from \citet{ss91}.

We take the angular rotation rate 
$\Omega$ as a function of radius $r$ and latitude $\lambda$, 
\begin{eqnarray}
    \Omega(r,\lambda)/\Omega_\star &=& 1+2\frac{c_3}{\tilde\omega}-\left[1+{\rm erf}\left(\frac{r-r_0}{d}\right)\right] \nonumber \\
        && \cdot\frac{1}{\tilde{\omega}}\left(c_1\sin^4\lambda+c_2\sin^2\lambda+c_3\right),
\label{eq:drc}
\end{eqnarray}
where $\Omega_\star$ is the equatorial rotation rate at the 
stellar surface, $c_1=0.0876$, $c_2=0.0535$, and $c_3=-0.0182$. 
The base of the convective overshoot 
region in the stratification model is at $r_0=0.724R_\sun$; it is taken as 
the centre of the tachocline here. Furthermore, $d=0.075R_\sun$ is 
the width of the error function defining the thickness of 
the tachocline and the constants $(c_1,c_2,c_3,d)$ are chosen {such} that 
Eq.~(\ref{eq:drc}) closely mimics helioseismic inversions of 
solar internal rotation \citep{schou98}. 
Observational studies indicate that surface differential 
rotation increases rather weakly with the rotation rate as 
$\Delta\Omega_\star\propto\Omega^n_\star$, where $n$ was estimated 
to be 0.15 by \citet{barnes05}, and 0.2 for G stars by \citet{ba16}.
For simplicity, we set $\Delta\Omega_\star=\Delta\Omega_\sun$ 
in both the radial and latitudinal directions. 
The factor $\tom^{-1}$ and the term $2c_3\tom^{-1}$ in 
Eq.~(\ref{eq:drc}) account for {keeping the same (solar) 
differential rotation rate in both the radius and latitude, as $\Omega_\star$ 
increases.}

\subsubsection{Initial properties of flux tubes}
\label{sssec:stab}

Following \citet{fms95}, we solve the sixth-order dispersion 
relation for linear perturbations of a toroidal flux ring in mechanical 
equilibrium, taking the thermodynamical quantities from the stratification 
model and the rotation profile from Eq.~(\ref{eq:drc}). 

Figure~\ref{fig:stab} shows the linear stability diagrams 
for toroidal flux tubes in the middle of the overshoot 
region ($r_{\rm mid}=0.728 R_\sun$) as a function of the initial 
latitude $\lambda_0$ and the field strength $B_0$, for different 
rotation rates and the same Sun-like stratification. 
In light- and dark-shaded regions, the fastest-growing 
wave mode has an azimuthal wavenumber of $m=1$ and $m=2$, respectively.
The radial location is about 5000~km beneath the base of the 
convection zone (the term `base' signifies the depth at 
which the convective heat flux changes its sign). 
The red dots in Fig.~\ref{fig:stab} show 
$\lambda_0$ and $B_0$ of flux tubes chosen for the non-linear simulations 
(Sect.~\ref{sssec:lookup}) with $2^\circ$ latitudinal steps, 
all corresponding to a characteristic linear growth time of 50 days.
This growth time ensures that 
the tubes are sufficiently close to the onset of instability 
and that the Joy's law resulting from the simulations 
for $\tom=1$ matches well with the solar observations. 
We did not consider the islands of instability 
because (1) the corresponding growth time is not reached in the lower-latitude 
island; and (2) the high-latitude island is not reached by the input butterfly 
diagram in any case, except for $\tos=8$. To be conservative, we preferred to limit 
the simulations to the main region of instability by decreasing the factor $k$ 
in Eq.~(\ref{eq:meanlat}) for $\tos=8$. 

The maximum latitude at step I is $73^\circ$ for $\tos=8$ 
(Sect.~\ref{sssec:record2}). 
To cover the entire latitude range in the input solar 
cycle for all flux emergence rates, we set up flux-tube 
simulations with initial latitudes at the base of the convection 
zone up to $\lambda_0=73^\circ$ for $\tom=1$, to move from step I to II. 
This is the same maximum latitude as for the input 
model at the surface 
because the flux tubes rise almost radially for $\tom=1$, 
especially at high latitudes (Fig.~\ref{fig:joys}).

The stability diagrams in Fig.~\ref{fig:stab} show that the onset 
of instability is shifted towards higher field strengths for higher 
rotation rates, by a factor of about 3 for $\tom=8$, compared to the 
solar value, owing to the enhanced Coriolis force component directed 
towards the rotation axis, which has a stabilising effect. 
The dynamics of the tube is governed predominantly by the 
buoyancy, curvature, and Coriolis forces in the rotating frame. 
The tube radius is relevant only for the drag force, which 
only weakly affects the resulting emergence properties.
For all the simulations, we set the initial cross-sectional radius 
to 2000~km, about 3.6 \%\ of the local pressure scale height. 
With this radius and the initial field strengths corresponding to 
a growth time of 50~days (Fig.~\ref{fig:stab}), 
the magnetic flux within a tube is typically about $10^{22}$~Mx 
for $\tom=1$ and $3\times 10^{22}$~Mx for $\tom=8$.\footnote{We note 
that the fluxes of individual flux tubes with the chosen $B_0$ and 
$\lambda_0$ (Sect.~\ref{ssec:rise}) are in the same range with the BMRs 
used in the SFT (Sect.~\ref{ssec:sft}). 
Whenever a surface source is introduced in the surface flux transport simulation, 
however, its flux is determined only by the empirically synthesised size 
distribution.} The flux-tube rise simulations serve only to obtain  
emergence latitudes and tilt angles, which are only slightly affected 
by the initial tube radius via the drag force.

\subsubsection{Simulations of flux tube emergence}
\label{sssec:lookup}

\begin{figure}
    \includegraphics[width=\columnwidth]{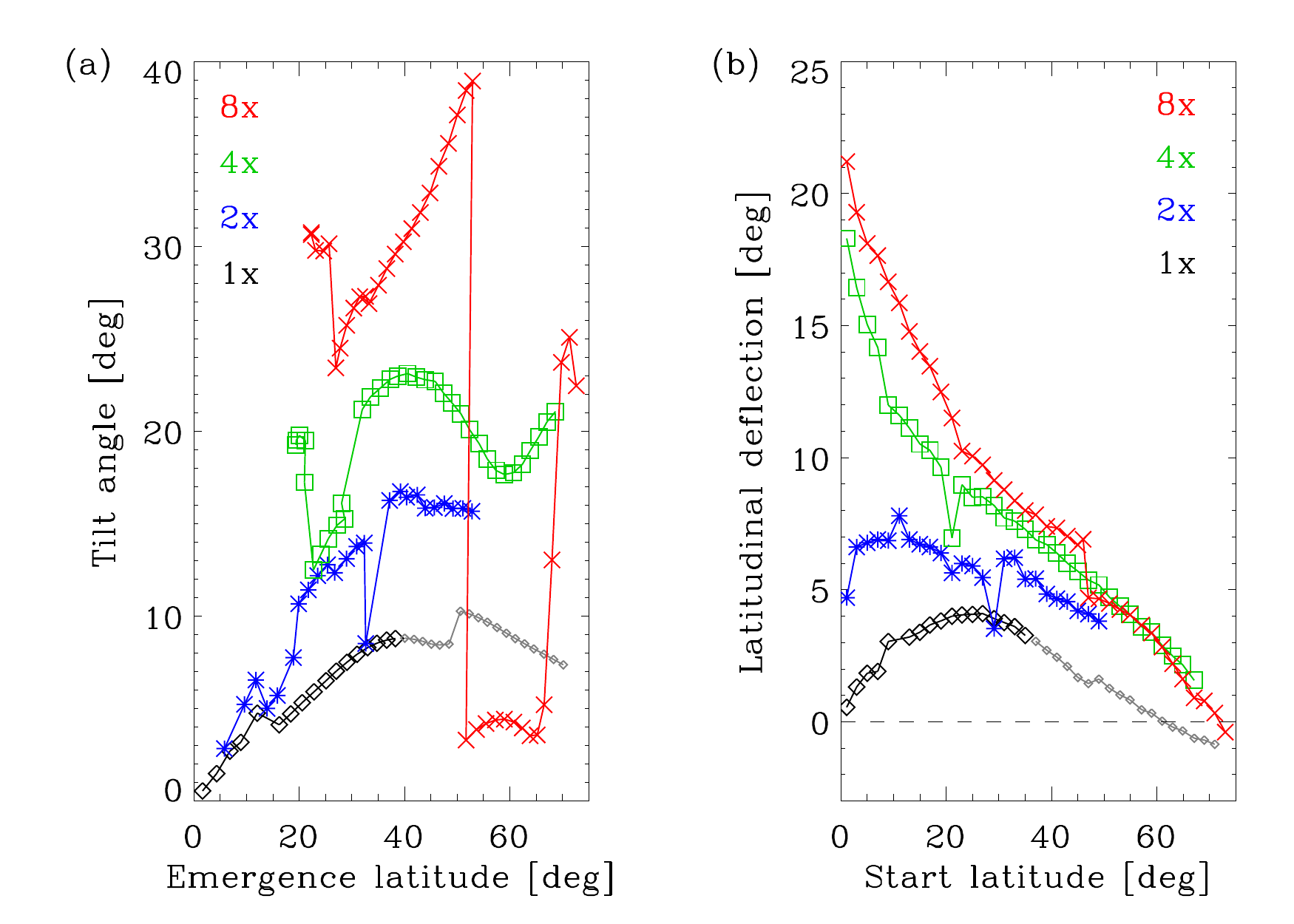}
    \caption{(a) The tilt angle of emerging flux 
    loops vs. their emergence latitude for $\tom=1$ (black diamonds), 
    $\tom=2$ (blue asterisks), $\tom=4$ (green squares), 
    and $\tom=8$ (red crosses).  
    (b) Latitudinal deflections (poleward is positive) as a function of the initial latitude at the base of the convection zone. The initial $\lambda_0$ and $B_0$ are as in Fig.~\ref{fig:stab}. Emergence 
    latitudes for which no active-region-sized BMR emerges for $(\tom,\tos)=(1,1)$ 
    are shown by small grey diamonds (see text). }
    \label{fig:joys}
\end{figure}

To model the rise of flux tubes through the convection zone, 
we carried out simulations starting from the initial parameters mentioned above 
(Sect.~\ref{sssec:stab}, see Fig.~\ref{fig:stab}). We used the code developed 
by \citet{moreno86} 
and extended to three-dimensional (3D) geometry by \citet{cale95}. It solves the 
dynamical evolution of a one-dimensional (1D), initially 
toroidal ring of mass elements embedded in a 
1D background stratification \citep{ss91} in the 3D Lagrangian
frame. The mass elements are subject to body forces including the drag 
force, Lorentz force, and buoyancy, as well as the pseudo-forces induced 
by the Coriolis and centrifugal effects. The evolution of the tube is 
considered as an isentropic process. The magnetic flux and the 
integrity of the tube with its closed structure are also conserved. 
We chose $10^3$ mass elements for the flux tube, 
which is initially in mechanical equilibrium and subject to 
a linear combination of azimuthally periodic spatial perturbations 
with wave numbers in the range $1\leqslant m \leqslant 5$. Their 
magnitudes are 
$10^{-4}$ in units of the local pressure scale height, in each of the three 
dimensions. Unstable tubes experience magnetic buoyancy 
instability and develop loops that rise up to a heliocentric radial distance 
of about $0.98R_\sun$. At this point, the simulation halts owing to the ambient 
pressure scale height becoming comparable with the tube diameter, 
violating the thin-tube criterion. 
We roughly define this stage as the `emergence' of the loop, 
though the loop is still under the surface. 
In general, the fastest-growing azimuthal wave 
mode in the non-linear simulations is consistent with the prediction 
of linear stability analysis (Fig.~\ref{fig:stab}). When  $m=2$, 
two buoyant loops form with a 180-degree phase difference, 
and one of them emerges before the other. 
The simulations are stopped at this point due to the 
thin-flux-tube criterion, so it is not possible to track the other emergence 
at the opposite longitudinal hemisphere. 
Therefore, our simulations may be somewhat underestimating the amount 
of magnetic flux that emerges at the stellar surface when $m=2$ is the 
dominant mode of instability.

\begin{figure*}
    \centering
    \includegraphics[width=.65\columnwidth]{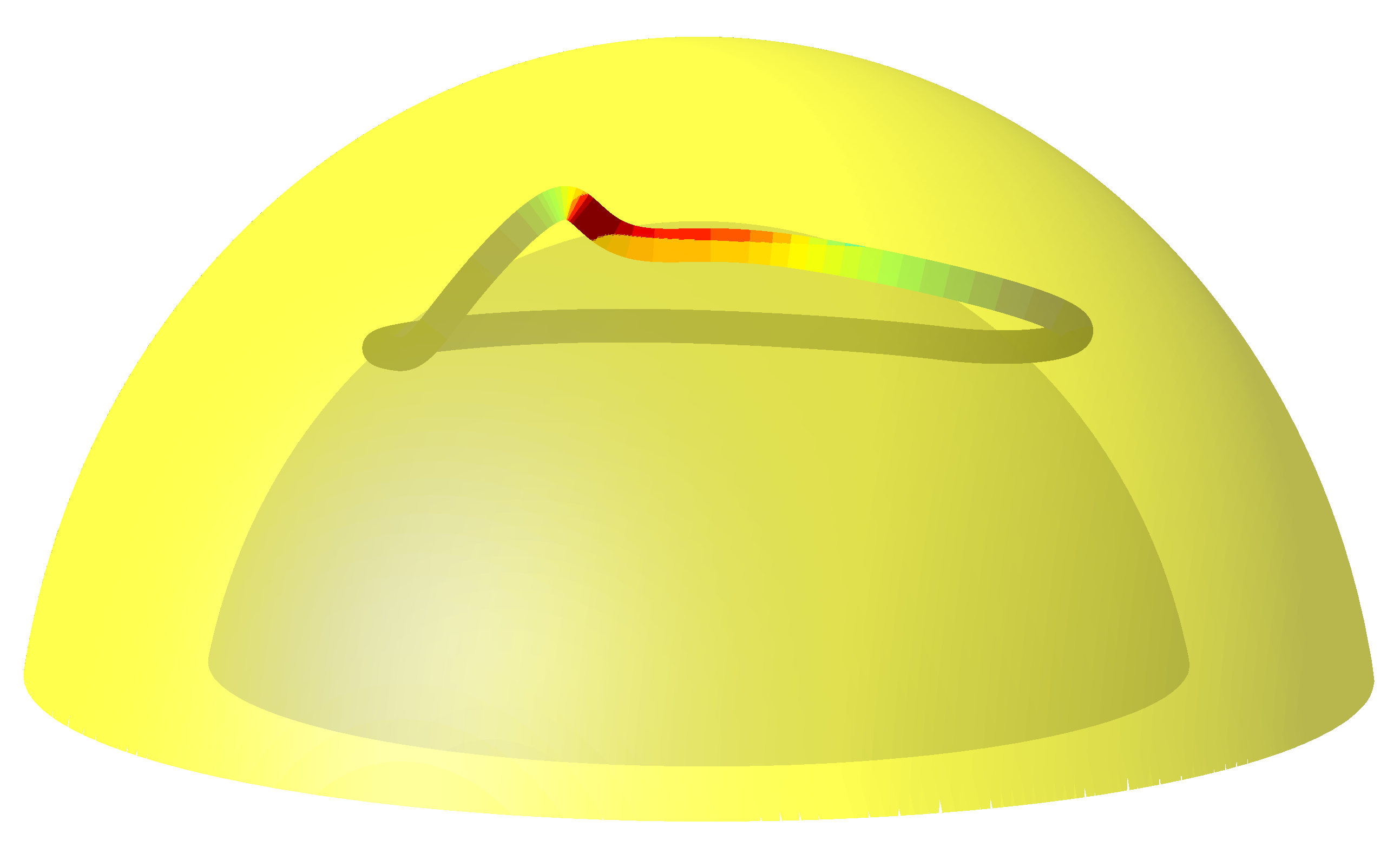}
    \includegraphics[width=.65\columnwidth]{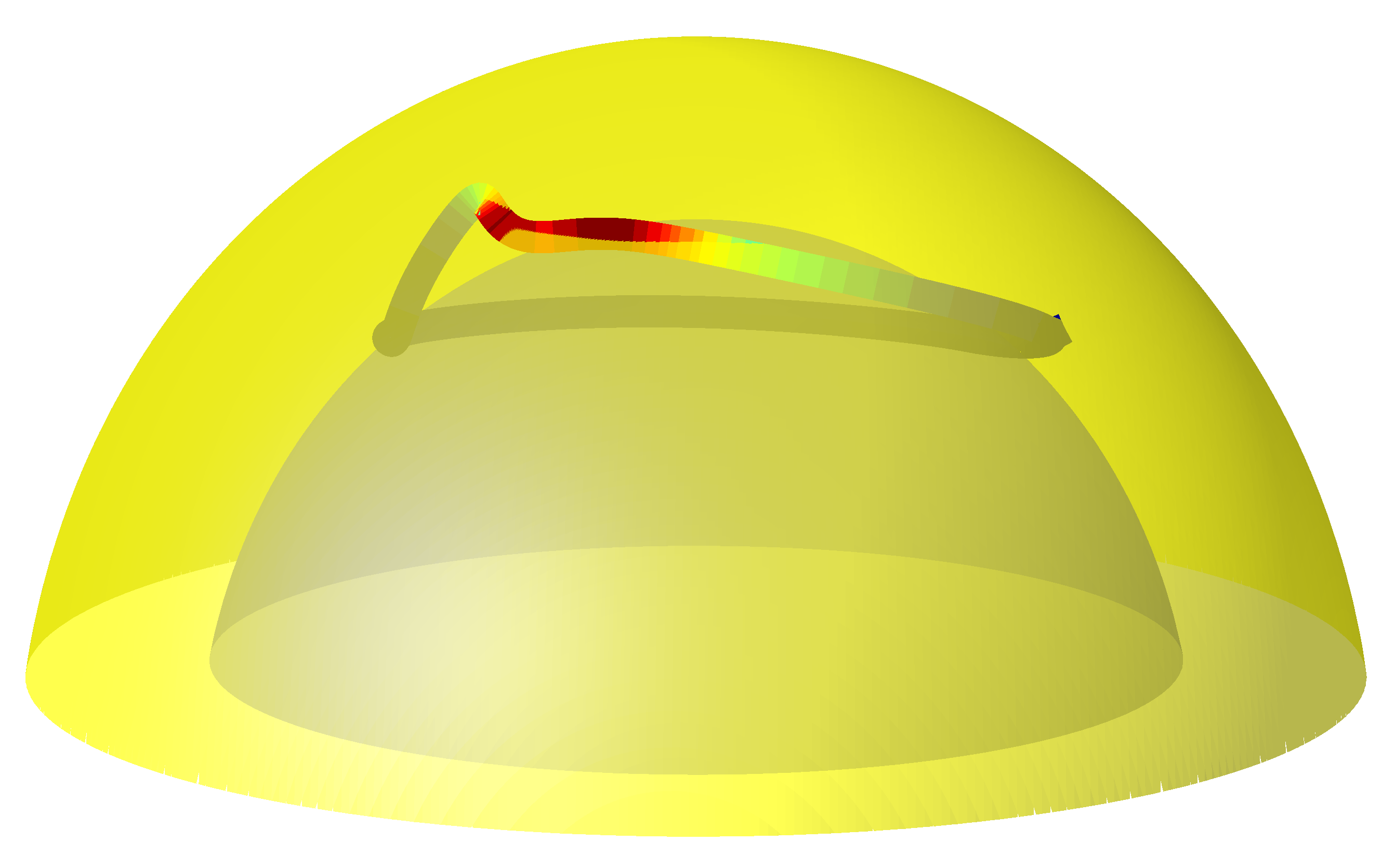}
    \includegraphics[width=.65\columnwidth]{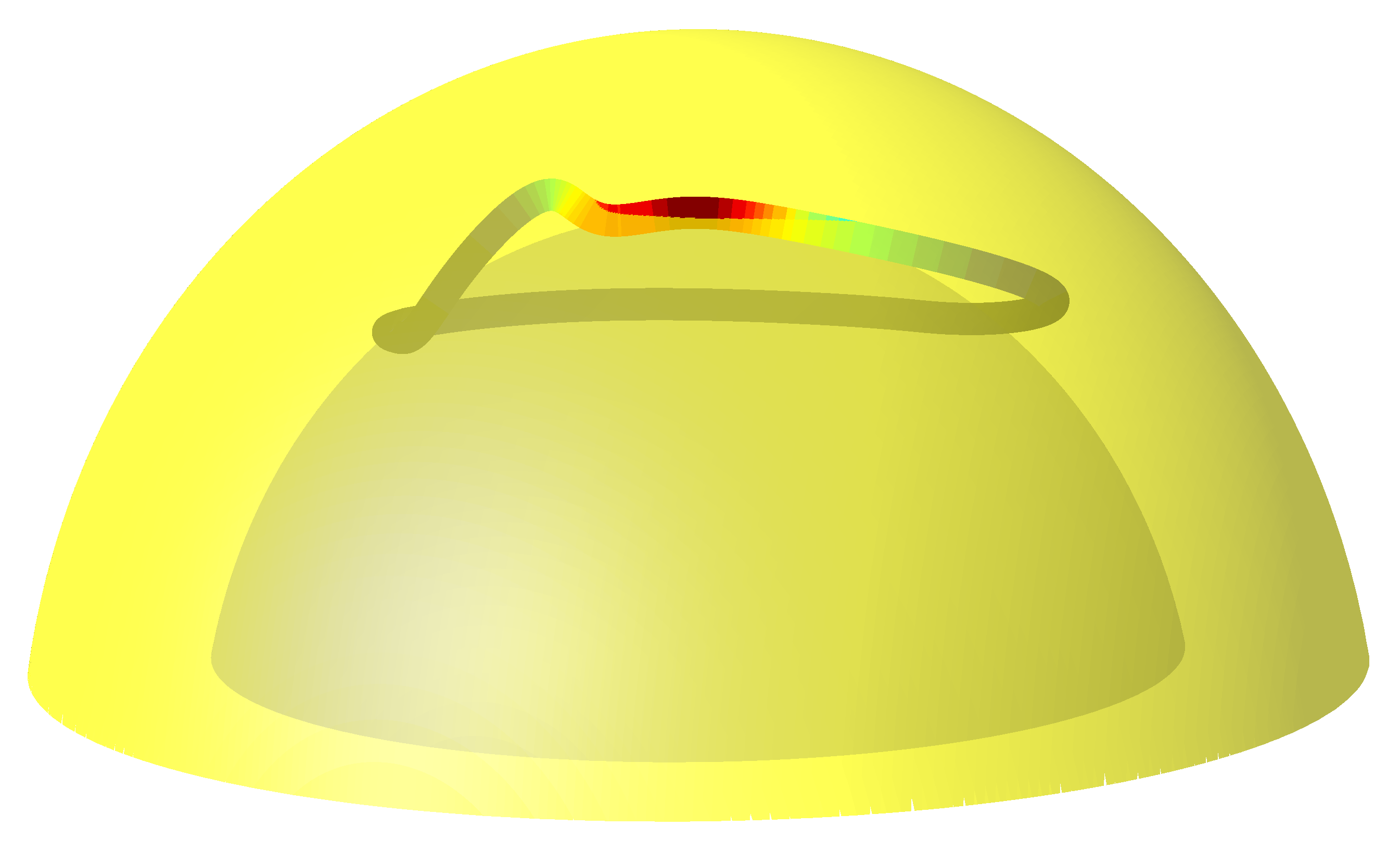} \\
    \includegraphics[width=.65\columnwidth]{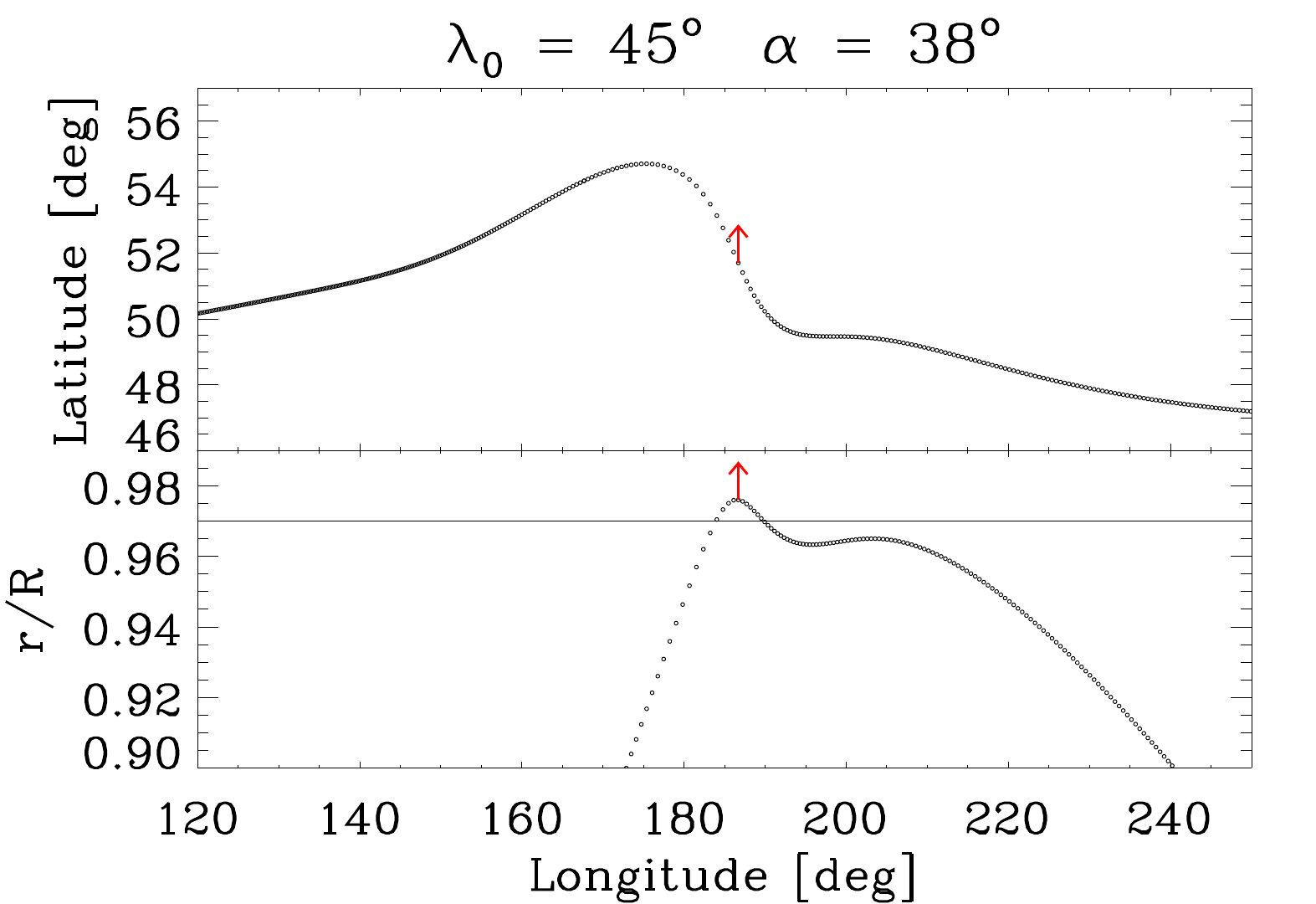}
    \includegraphics[width=.65\columnwidth]{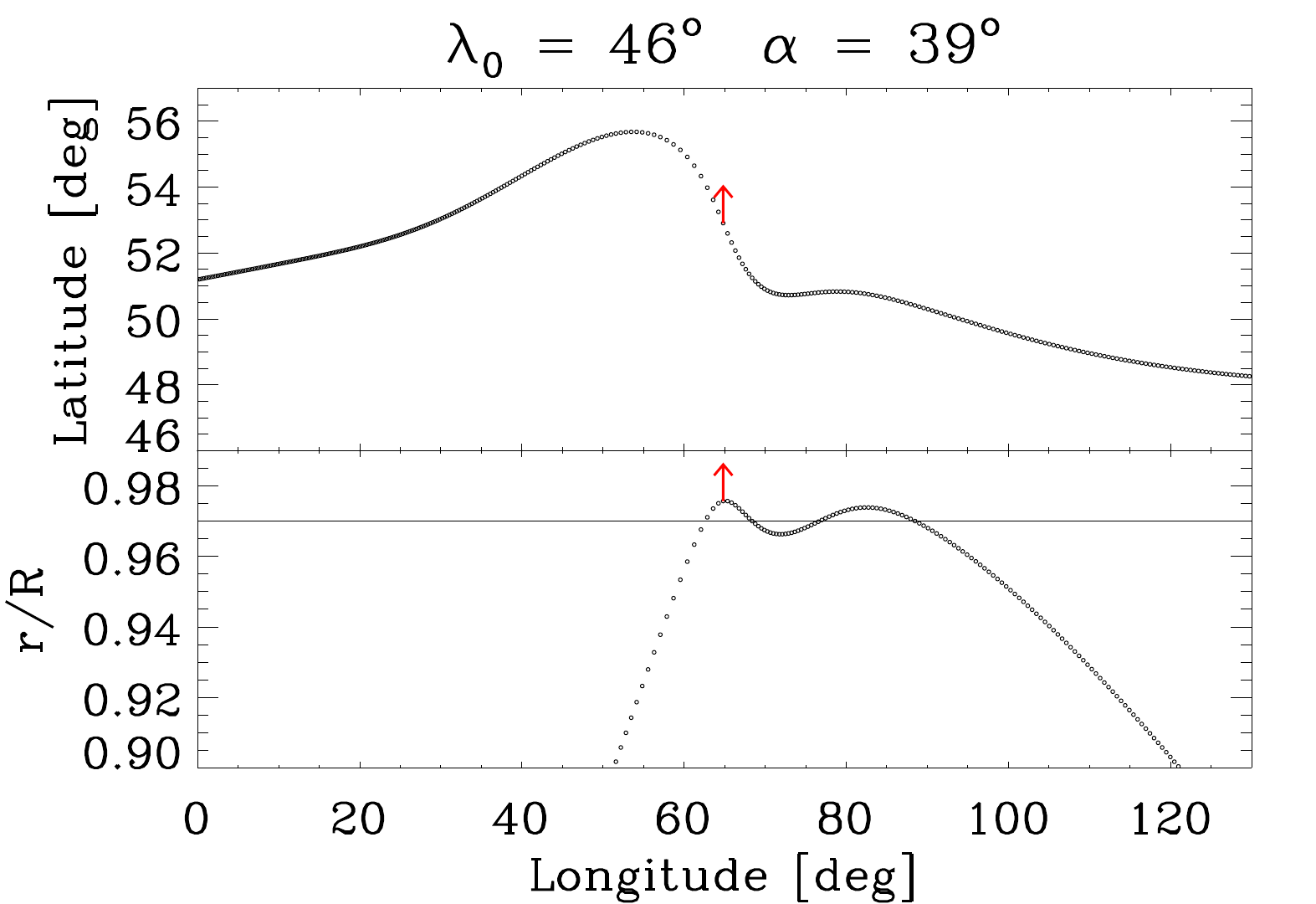}
    \includegraphics[width=.65\columnwidth]{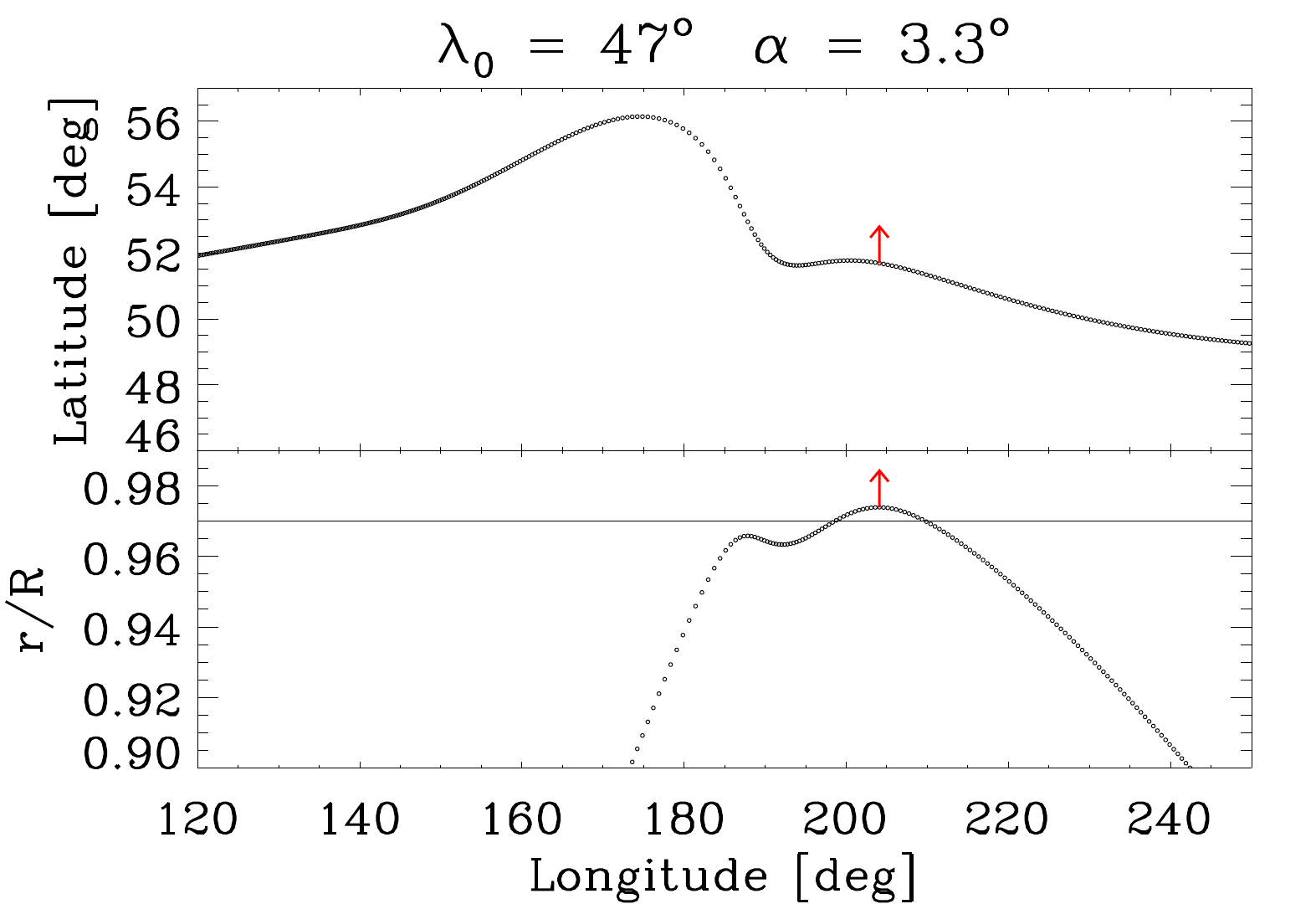}
    \caption{Geometry of three emerging flux loops with initial 
    latitudes $45^\circ$ (\emph{left}), $46^\circ$ (\emph{middle}), 
    and $47^\circ$ (\emph{right}) at the base of the convection zone. In the \emph{upper panels}, 
    the inner sphere is drawn at $0.72R_\sun$, 
    and the outer sphere at $0.97R_\sun$. The parts of the tube 
    that are beneath the outer layer are shaded in grey, whereas 
    the emerged 
    parts are brighter. For each mass element of the tube, 
    the colours denote the cross-sectional tube diameter (the redder the 
    larger). 
    \emph{Lower panels} show the latitudinal and radial projections 
    of the tubes, where each mass element is represented by a dot. The 
    horizontal line on the radial profile corresponds to the 
    location of the outer 
    sphere ($0.97R_\sun$), where the tilt angle 
    ($\alpha$) is measured from the footpoint locations. 
    The red arrows denote the apex of each tube.     }
    \label{fig:why}
\end{figure*}

The initial $(\lambda_0,B_0)$ determined by the growth time of 50 days 
and the initial radial location are set such that 
the loops yield a realistic set of emergence latitudes 
and tilt angles for the Sun. 
The time between the initial state to the emergence, that is, including the development of the instability, 
ranges from a few hundred to a thousand days and increases with 
$\Omega_\star$. We measure the emergence 
latitude from the apex of the tube at the end of the simulation. 
We determine the tilt angle by using the longitudes and latitudes of the preceding 
and follower legs of the flux loop at $0.97R_\sun$. 
The results are roughly consistent
with the tilt angles obtained from the slope of the tangent vector 
at the apex. 

Figure~\ref{fig:joys} shows, as a function 
of the emergence latitude $\lambda_e$, the tilt angles 
$\alpha$ (this dependence is called Joy's law in solar physics) and 
the latitudinal deflection $\lambda_e-\lambda_0$ as a function of the initial 
latitude, for different rotation 
rates. Simulations were made for the full range of latitudes in the 
case $\tom=1$, including latitudes where no emergence occurs 
in $(\tom,\tos)=(1,1)$
(smaller diamonds in Fig.~\ref{fig:joys}). This additional latitude range 
was needed because we scaled the mean latitude of activity with $\tos$ in 
step I (Eq.~\ref{eq:meanlat}), in accordance with the empirical solar model 
extrapolated to higher flux emergence rates. In general, both $\lambda_e$ 
and $\alpha$ increase with the rotation rate, owing to enhanced 
Coriolis force components in the rotating frame.

There are a few abrupt changes in $\alpha(\lambda_e)$, which are 
also visible in $\lambda_e-\lambda_0$. 
To understand the origin of such features, 
we first eliminated the possibility of a numerical resolution 
issue. For this, we set up tubes with higher numbers of mass elements 
(up to 4000), 
but these runs converged to very similar values of emergence latitudes 
and tilt angles. 
Following additional simulations with slightly different initial field 
strengths, we found that these jumps were robust features. They are shaped by 
physical effects, that is, they represent regimes where 
different forces and/or different wave modes become dominant. For 
instance, the local peak in the tilt angle for $\tom=1$ at 
$\lambda_e\simeq 15^\circ$ occurs at the transition of the 
fastest-growing mode from $m=1$ to $m=2$ (see Fig.~\ref{fig:stab}a). 

To investigate the nature of the largest jump for $\tom=8$, we show in 
Fig.~\ref{fig:why} the emergence phases of flux tubes starting 
at $\lambda_0=45^\circ$, $46^\circ$, and $47^\circ$, that is, 
roughly where the jump in the tilt angle at about $\lambda_{\rm e}=50^\circ$ 
occurs (see Fig.~\ref{fig:joys}). 
The plots clearly depict the transition from the case when a highly tilted 
part of the tube emerges earlier (Fig.~\ref{fig:why}a), to the case when 
a much less tilted part emerges earlier (Fig.~\ref{fig:why}c). 
The radial and azimuthal projections of the tubes mark this transition 
clearly. The relative 
phase speeds of the two competing loops vary with $\lambda_0$, such that 
the low-tilt loop intrudes the high-tilt loop above a certain 
initial latitude of about $46^\circ$. The transition from partial to full 
intrusion is responsible for the low-tilt 
plateau in Fig.~\ref{fig:joys}a, for $\lambda_{\rm e}$ between about 
$50^\circ$ and $65^\circ$. The two loops fully merge beyond 
$\lambda_{\rm e}=70^\circ$, where moderate tilts of about $22^\circ$ 
are reached (see Fig.~\ref{fig:joys}a).

As an independent test, thin flux tube simulations for $\tom=8$ have 
been kindly made by M. Weber (priv. comm.) using the code developed 
by \citet{fan93}. The only two differences with our setup were that she assumed 
rigid rotation for the stellar interior and started the tubes in the 
lower convection zone, where the superadiabaticity was positive. 
Nevertheless, 
the resulting latitude dependence of the tilt angle and the latitudinal 
deflection turned out to be qualitatively similar 
to our case for $\tom=8$. 
The distribution and amplitude of the abrupt variations 
in the tilt angle roughly agreed with each other in these two sets of 
independent simulations. 

It is quite possible that both loops forming out of a single flux tube eventually 
emerge, producing active regions with quite different tilts. Hence, regions with 
systematically different tilts can coexist at nearly the same latitude on rapidly 
rotating stars. 
However, since the thin tube approximation does not allow us to continue the computations 
further, we have simply used the tilts of the region emerging first.

\subsection{Surface flux transport}
\label{ssec:sft}

We used the solar and stellar SGRs described in Sect.~\ref{ssec:record}
as input to the SFT model 
\citep[see][for a review]{jiang14}, for which we employed 
the code developed by \citet{baumann04}. The code solves, 
with one-day steps, the magnetic induction equation at the solar/stellar 
surface, where the field is assumed to be purely radial. This
allows us to consider the field as scalar, with a sign 
representing the magnetic polarity. This is a reasonable 
assumption for the kilo-Gauss 
fields on the Sun found in active region plage, solar network, and 
sunspot umbrae, though the geometry of the highly inclined fields 
in penumbrae is not taken into account. {Because the purpose of this 
series of studies is to simulate brightness variations, we did not attempt to model 
the horizontal components of the magnetic field, for which 
the relationship with brightness variations in Sun-like stars is unclear. 
Exceptions are spot penumbrae, which can be treated as having a homogeneous 
brightness at the effective spatial resolution that can be reached in stellar observations. 
{In the current endeavour to model brightness variations, we implicitly 
assume that larger-scale horizontal fields are transients that occur only 
during flux emergence, 
and neglect their signature in the brightness distribution on a stellar disc.} 
}
\subsubsection{{ Properties of bipolar magnetic regions}}
\label{sssec:bmrs}
{In our SFT model,} the distribution of the field 
on the solar surface is represented in terms of 
spherical harmonic functions, with a maximum degree of 
$l=64$, corresponding to a resolution at the level of 
supergranular cells on the Sun. The freshly emerged BMRs are defined as two 
circular regions of opposite polarity, with a fixed upper limit of the
field strength, $B_{\rm max}$. The interpolarity 
distance (ranging between 3 and 10$^\circ$) controls 
the size of each BMR, as the characteristic radius of each polarity 
is fixed at $4^\circ$. We adopt $B_{\rm max}=374$~G, which was 
determined by 
\citet{cjss10}, who matched the variation of the total unsigned magnetic flux 
from an SFT simulation to magnetographic observations of the Sun from 
Mount Wilson and Wilcox Solar Observatories. 

\begin{figure*}
\centering
\begin{subfigure}[t]{0.01\textwidth}
(a)
\end{subfigure}
\begin{subfigure}[t]{0.4\textwidth}
\includegraphics[width=\linewidth,valign=t]{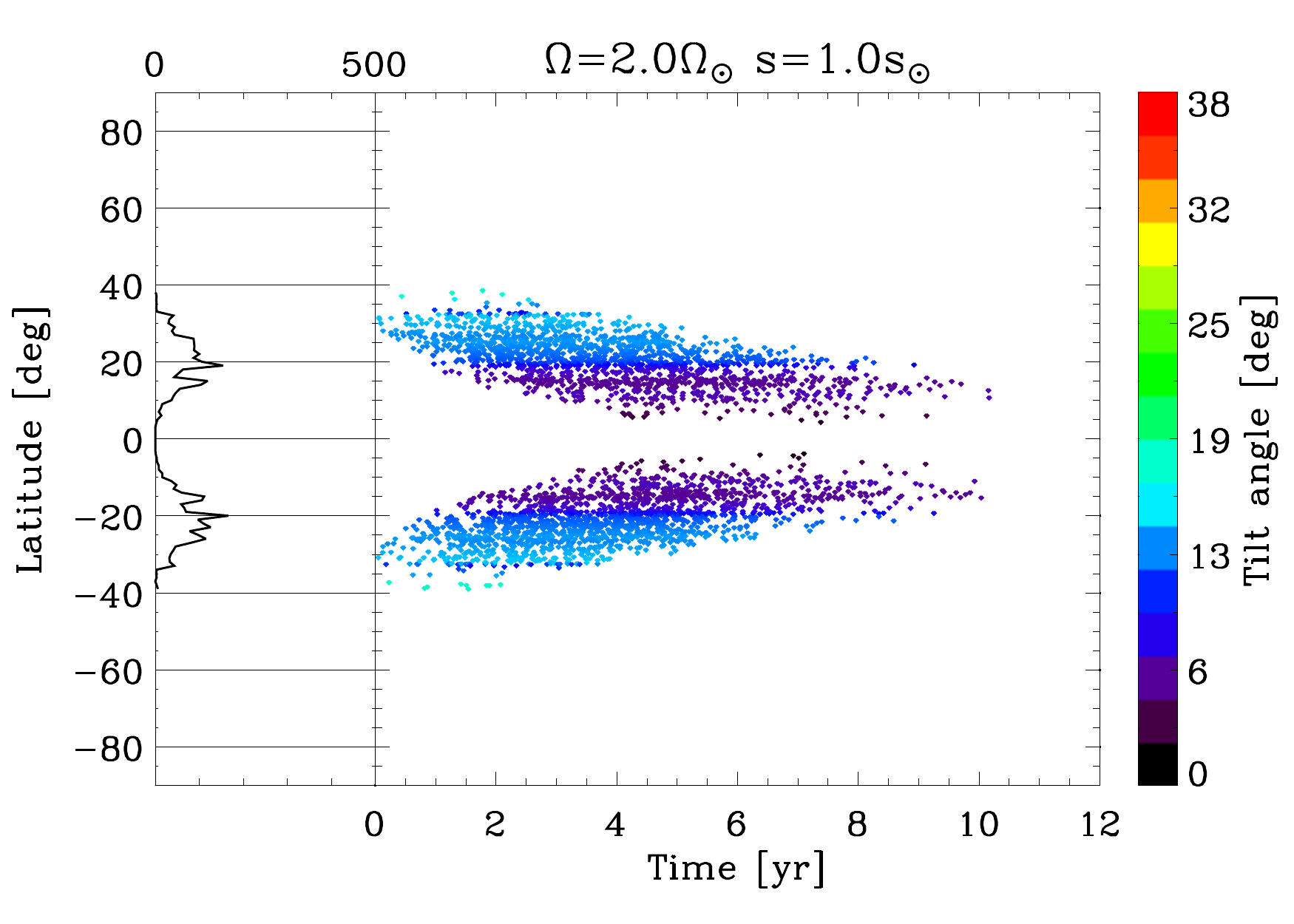}
\end{subfigure}
\begin{subfigure}[t]{0.01\textwidth}
(b)
\end{subfigure}
\begin{subfigure}[t]{0.4\textwidth}
\includegraphics[width=\linewidth,valign=t]{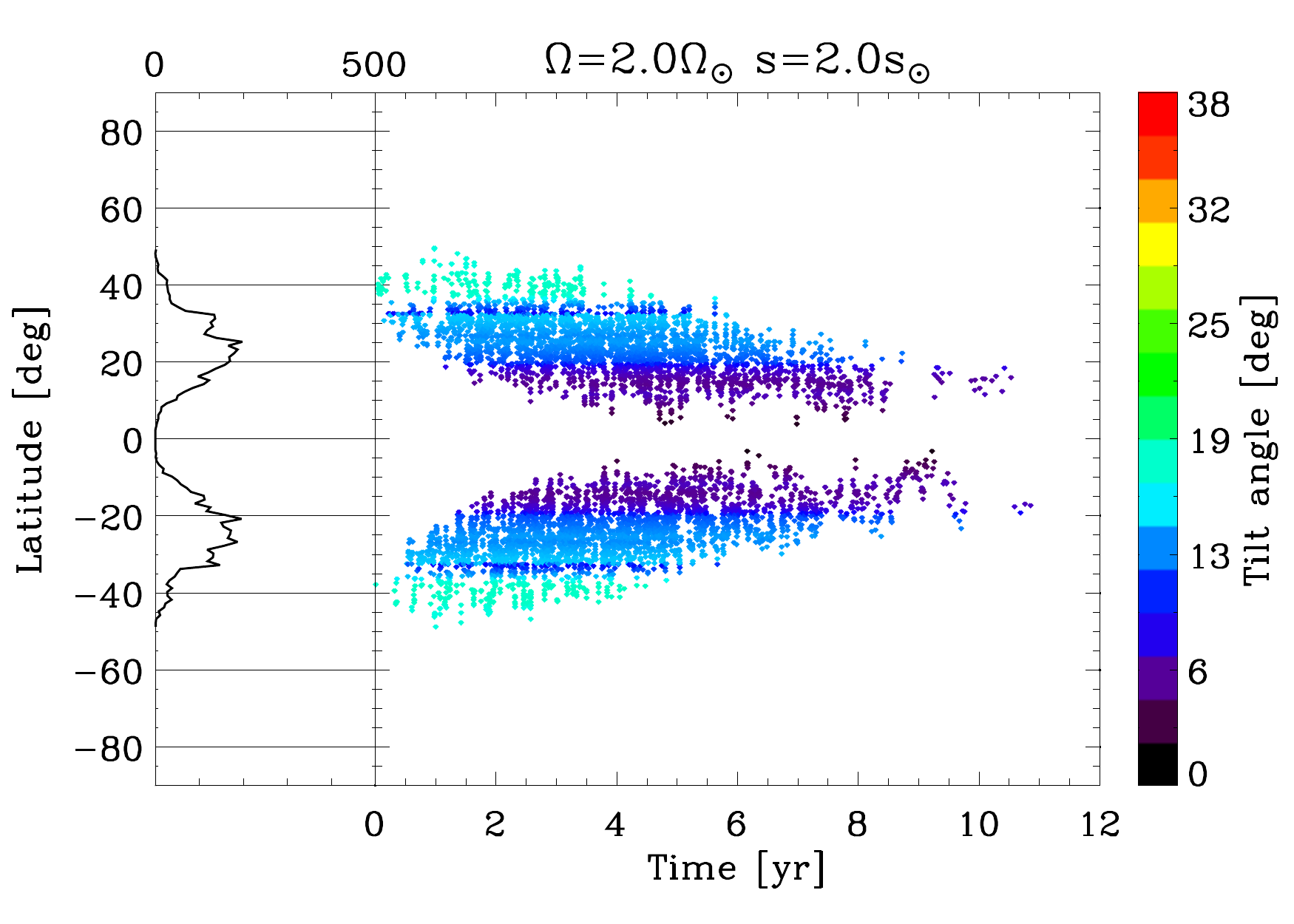}
\end{subfigure} \\
\begin{subfigure}[t]{0.01\textwidth}
(c)
\end{subfigure}
\begin{subfigure}[t]{0.4\textwidth}
\includegraphics[width=\linewidth,valign=t]{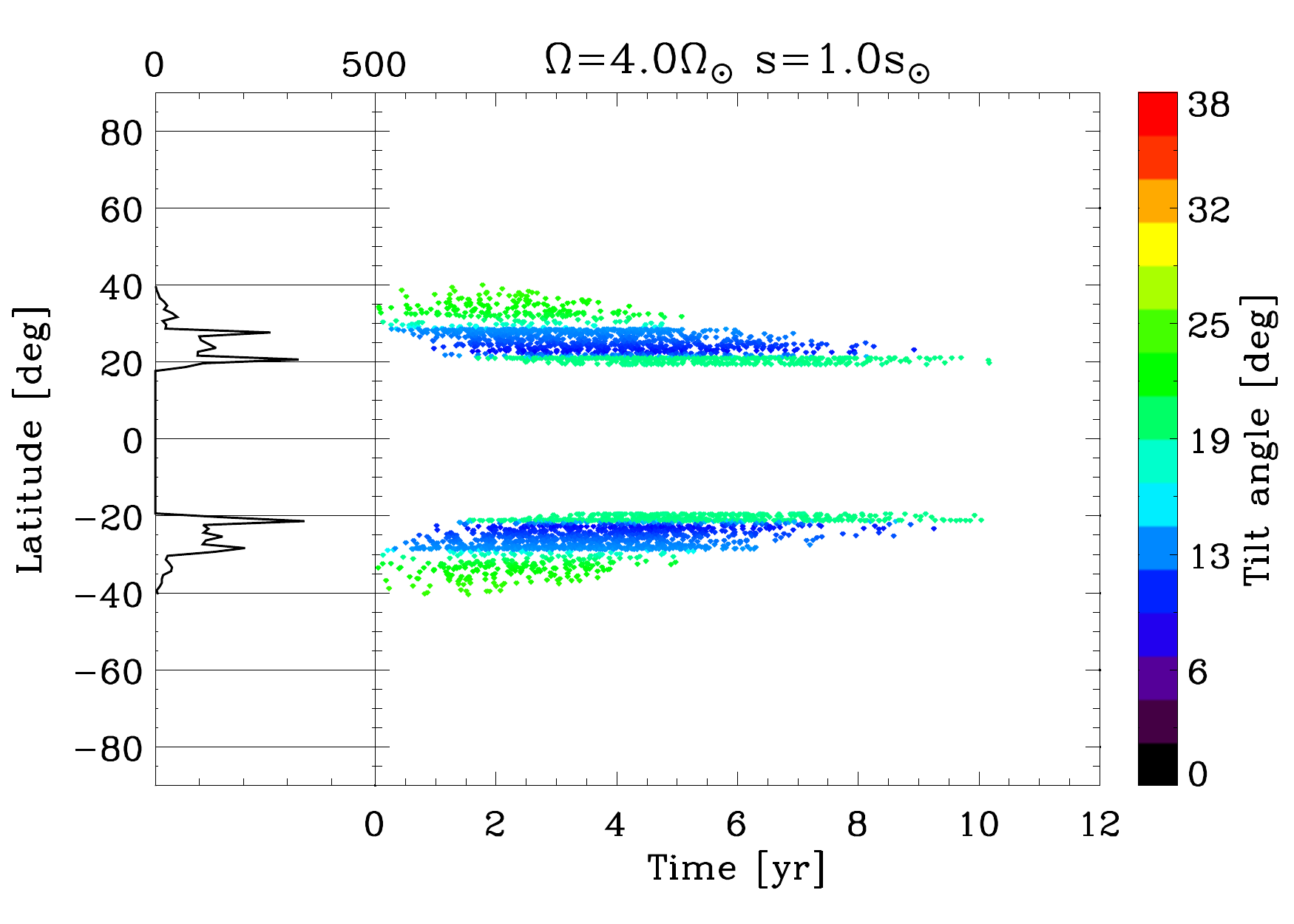}
\end{subfigure}
\begin{subfigure}[t]{0.01\textwidth}
(d)
\end{subfigure}
\begin{subfigure}[t]{0.4\textwidth}
\includegraphics[width=\linewidth,valign=t]{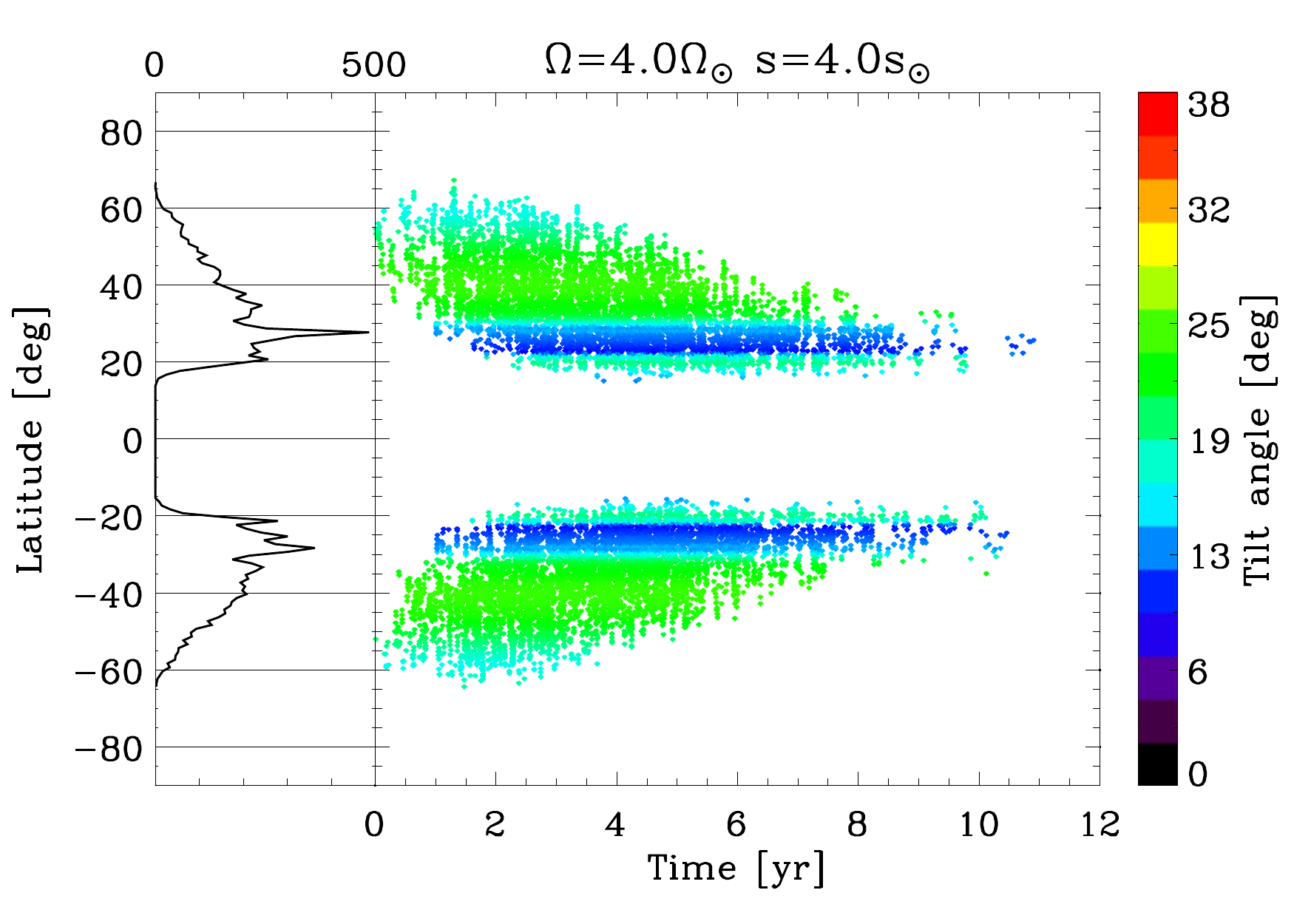}
\end{subfigure} \\
\begin{subfigure}[t]{0.01\textwidth}
(e)
\end{subfigure}
\begin{subfigure}[t]{0.4\textwidth}
\includegraphics[width=\linewidth,valign=t]{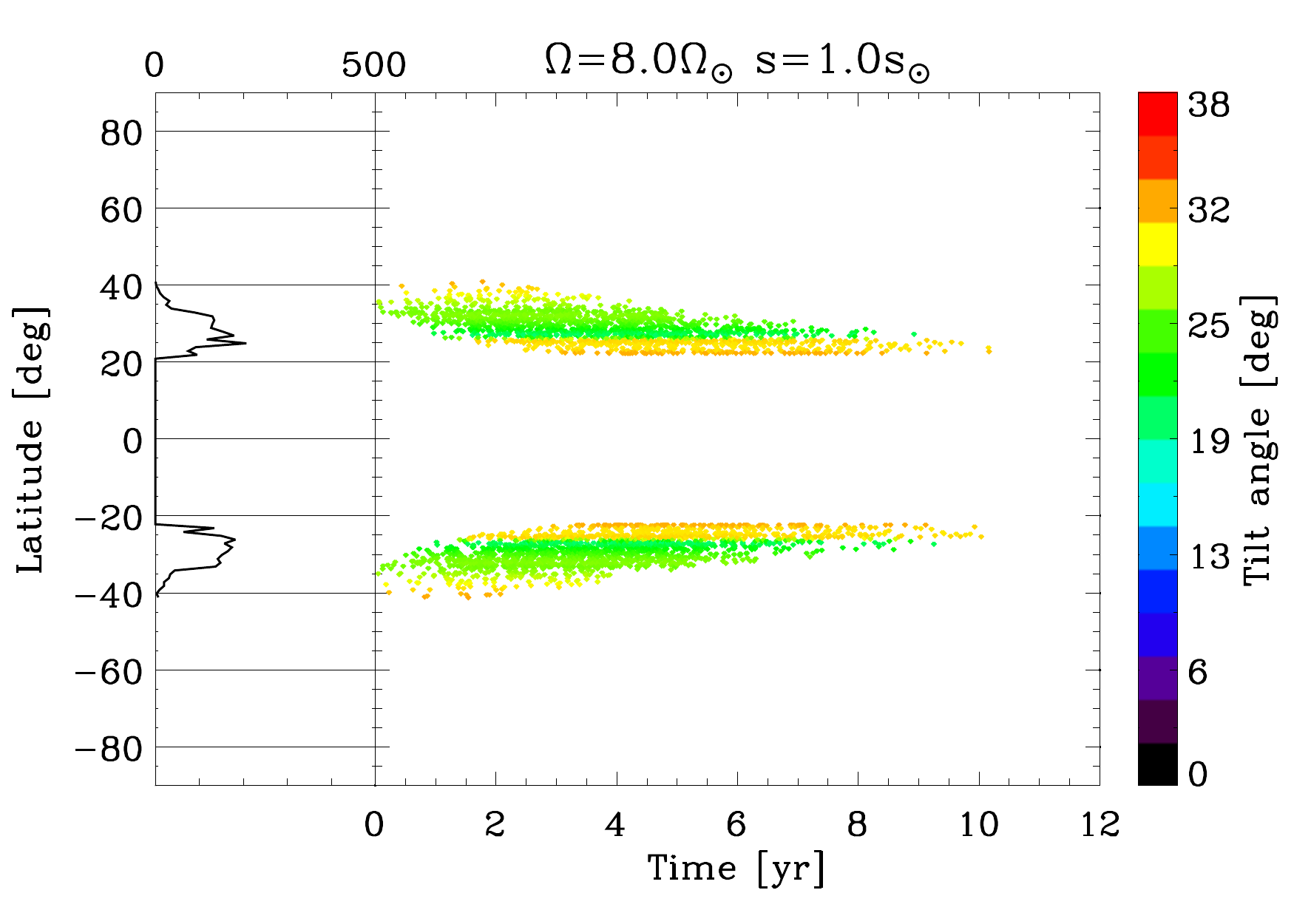}
\end{subfigure}
\begin{subfigure}[t]{0.01\textwidth}
(f)
\end{subfigure}
\begin{subfigure}[t]{0.4\textwidth}
\includegraphics[width=\linewidth,valign=t]{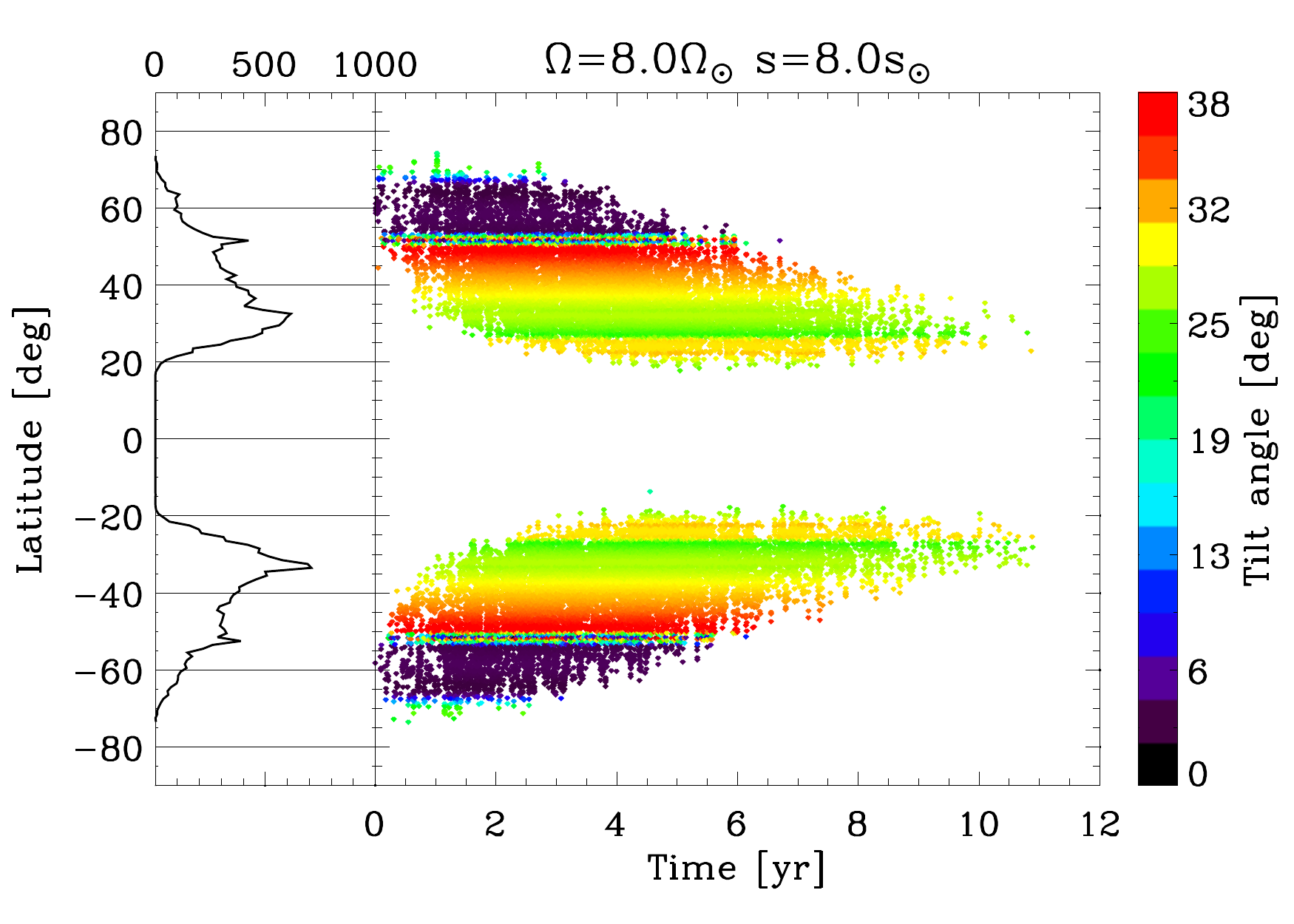}
\end{subfigure}
    \caption{As in Fig.~\ref{fig:bfly-ref}, but now for various sets of 
    $(\tom,\tos)$ as indicated at the top of each panel. The left panels 
    are for $\tos=1$ (solar emergence rate), and the right panels are 
    for $\tos=\tom$.}
    \label{fig:bfly}
\end{figure*}

\begin{figure}
    \centering
\begin{subfigure}[t]{0.01\columnwidth}
(a)
\end{subfigure}
\begin{subfigure}[t]{0.47\columnwidth}
    \includegraphics[width=\linewidth,valign=t]{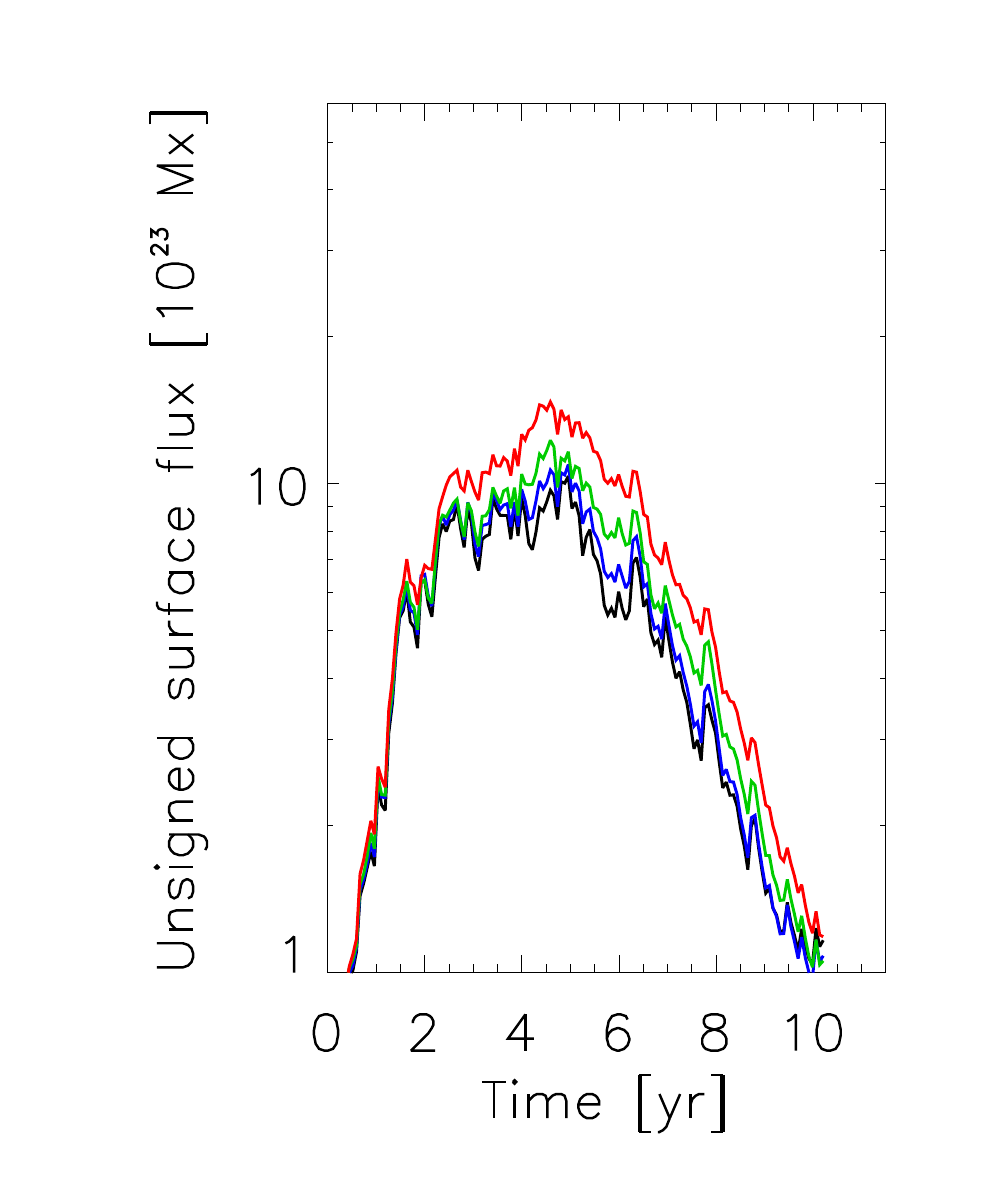}
\end{subfigure}
\begin{subfigure}[t]{0.01\columnwidth}
(b)
\end{subfigure}
\begin{subfigure}[t]{0.47\columnwidth}
    \includegraphics[width=\linewidth,valign=t]{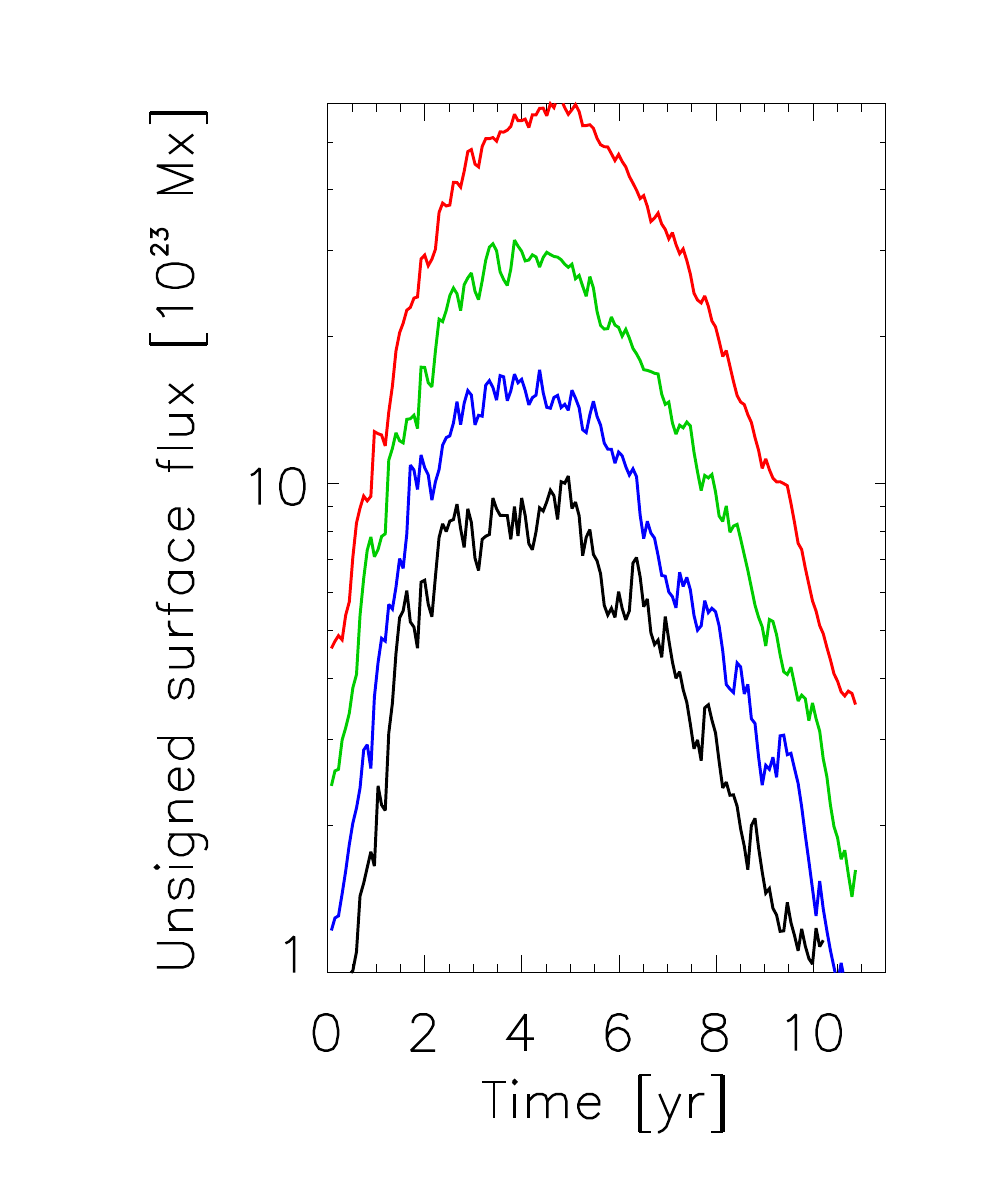}
\end{subfigure} \\
\begin{subfigure}[t]{0.01\columnwidth}
(c)
\end{subfigure}
\begin{subfigure}[t]{0.47\columnwidth}
    \includegraphics[width=\linewidth,valign=t]{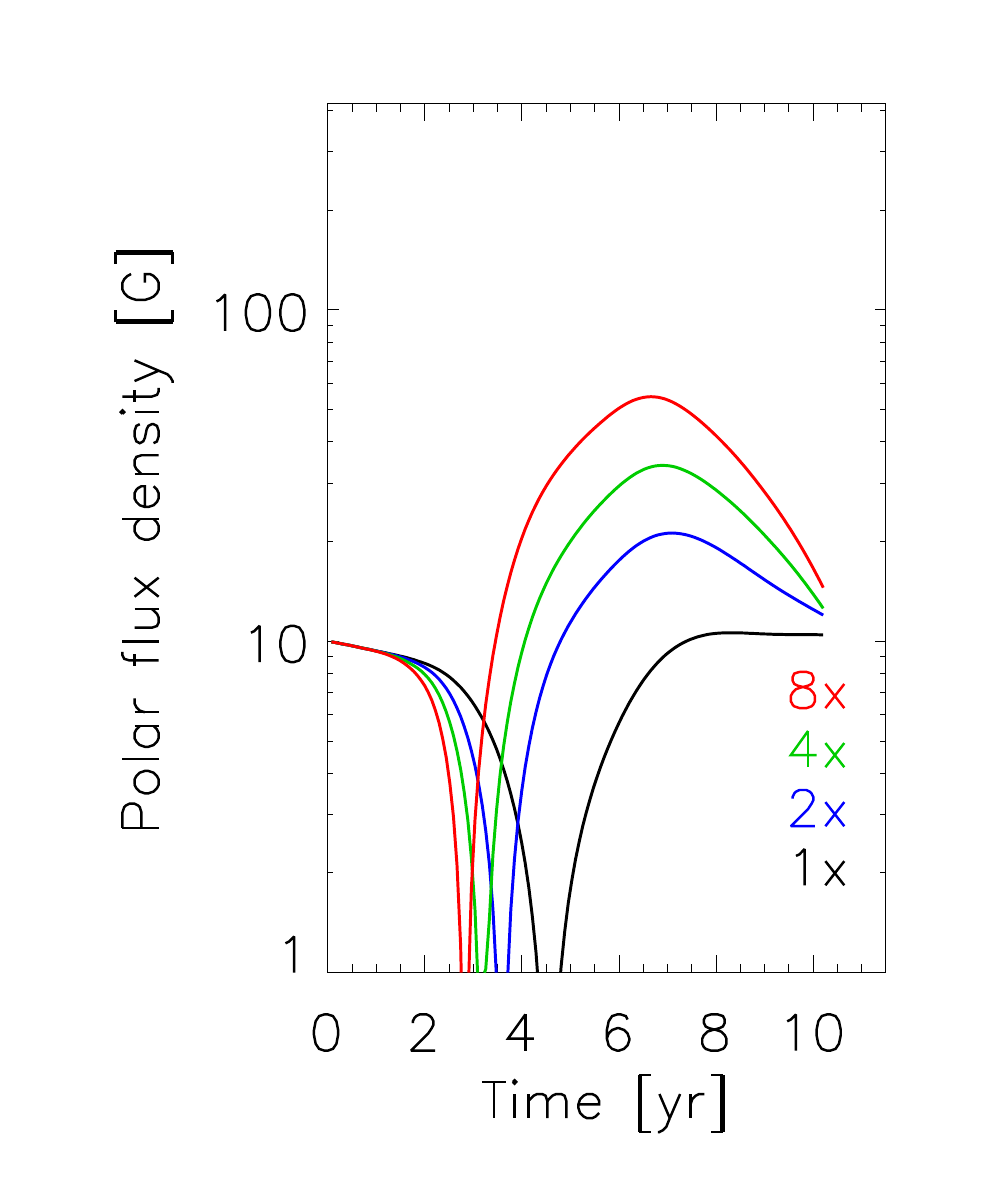} 
\end{subfigure}
\begin{subfigure}[t]{0.01\columnwidth}
(d)
\end{subfigure}
\begin{subfigure}[t]{0.47\columnwidth}
    \includegraphics[width=\linewidth,valign=t]{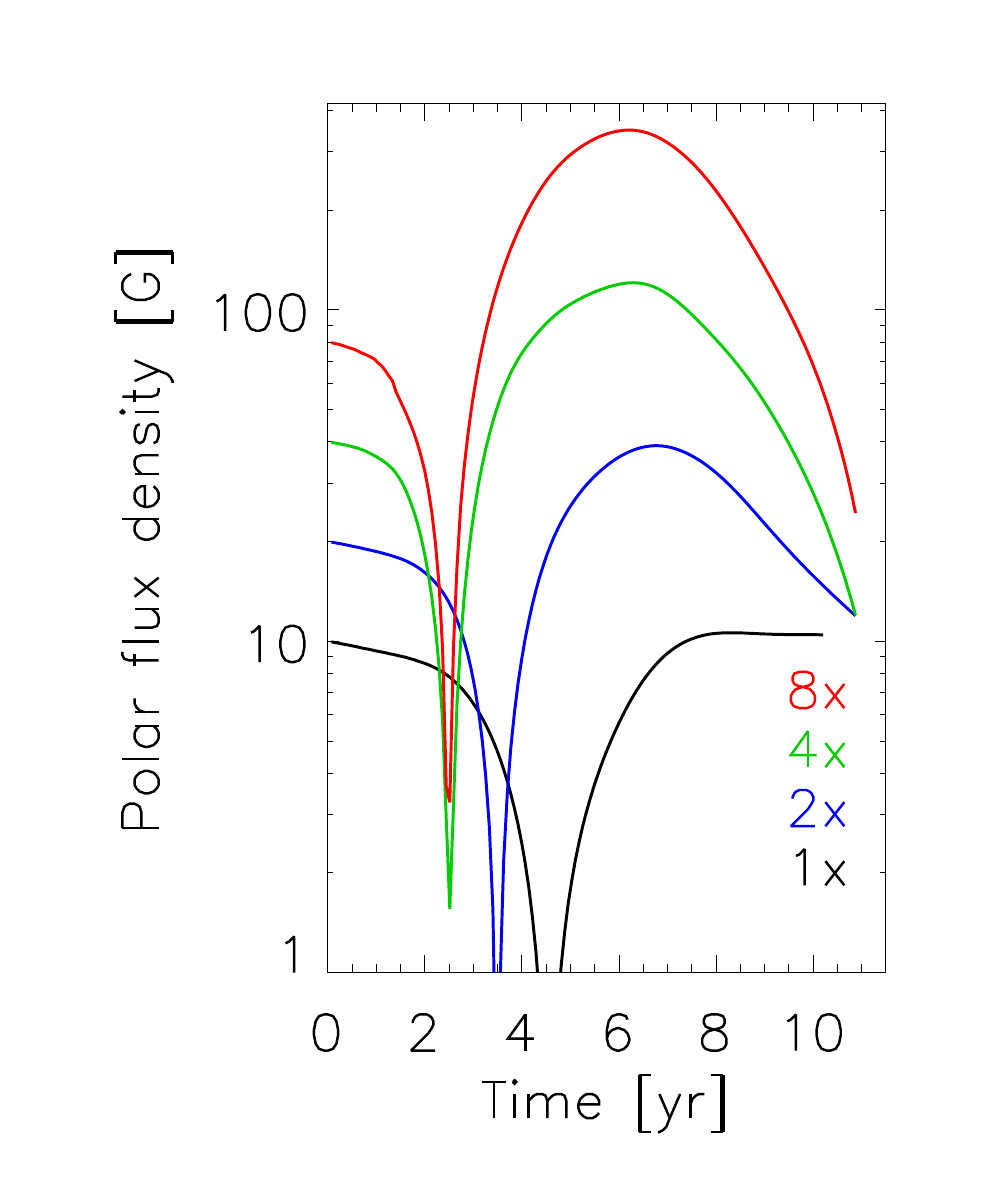}
\end{subfigure} \\
    \flushleft{(e)}
\hskip2cm\vskip-1cm
    \includegraphics[width=\linewidth]{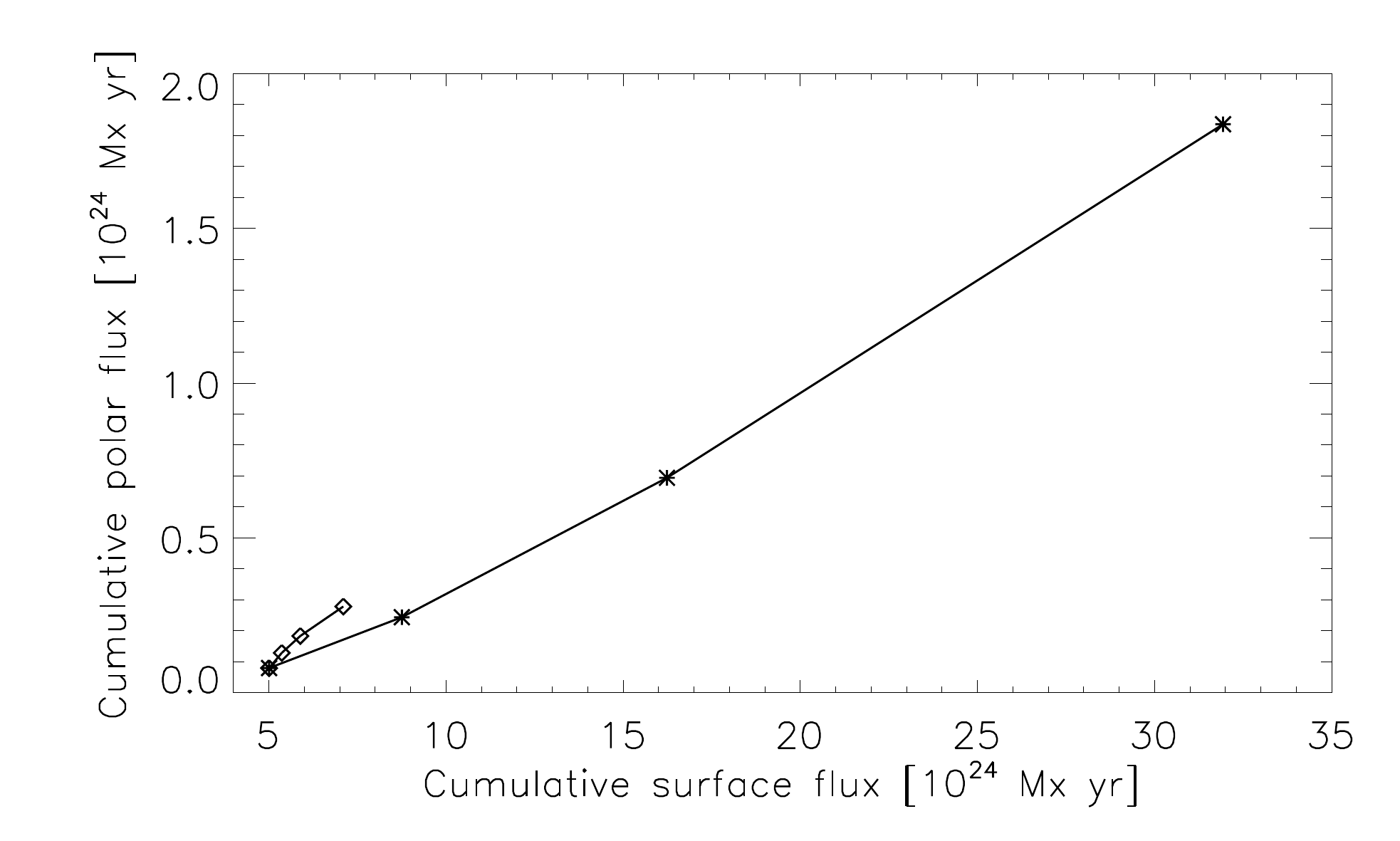}
    \caption{Total unsigned surface flux for (a) $\tos=1$ and (b) $\tos=\tom$ 
    , and the average polar flux density for latitudes poleward of $\pm 75^\circ$, 
    for (c) $\tos=1$ and (d) $\tos=\tom$. The colours denote $\tom$, as in  
    Fig.~\ref{fig:joys}. (e) Comparison of the time 
    integrals of polar flux based on the average polar field and the 
    total unsigned flux, for $\tos=1$ (diamonds) and $\tos=\tom$ 
     (asterisks). }
\label{fig:fluxvar}
\end{figure}

\subsubsection{Surface flows}
\label{sssec:flows}

The surface fields are subject to differential rotation and 
poleward meridional flow, which are assumed stationary, 
and follow the same profiles as in \citet{baumann04}. The latitudinal 
shear considered in the SFT model is very similar to Eq.~(\ref{eq:drc}) 
at $r=R_\sun$. 
The difference does not affect the resulting surface flux distributions. 
The meridional flow reaches a poleward speed of 11~m~s$^{-1}$ 
at mid-latitudes and ceases at $\lambda=\pm 75^\circ$.
For the effects of smaller-scale flows (supergranulation), 
the SFT model considers the diffusion term in the 
induction equation, with a horizontal turbulent diffusivity of 
250~km$^2$~s$^{-1}$ \citep{cjss10} and a radial diffusivity of 
25~km$^2$~s$^{-1}$ \citep{bss06}. 

\subsubsection{Initial magnetic field}
\label{sssec:initb}

As the initial condition, we assume a dipolar field 
reaching $B_{\rm max}(t=0)=\pm 10\tos$~G 
at the poles, which takes into account the possibility of stronger initial 
axial dipole moments for higher levels of flux emergence rate, $\tos$.

The resulting time series of surface maps of the 
radial magnetic field are represented in the so-called Carrington frame, 
where the latitudes of about $\pm 5^\circ$ are at rest. 
We assumed that the SFT parameters were invariant for all sets of 
$(\tom,\tos)$, {except for the initial field condition}. 

{
\subsection{Summary of assumptions}
\label{ssec:assumptions}
Here we summarise our assumptions when modelling faster-rotating, more active 
suns. 
\begin{enumerate}
\item The time-latitude distribution of flux tube eruptions at the base 
of the convection zone follows the statistical properties of the solar butterfly diagram 
(Sect.~\ref{sssec:record1}), with the cycle duration set to 11 years. 
\begin{enumerate}
\item The mean latitude of eruptions at the base follows the same  
linear scaling with the cycle strength, as observed on the solar surface 
(Sect.~\ref{sssec:record2}). 
\item The initial field strengths of erupting flux tubes correspond to a constant linear 
growth time (50 days) of the magnetic buoyancy instability (Fig.~\ref{fig:stab}) at 
the middle of the convective overshoot layer near the base (Sect.~\ref{sssec:stab}).
\end{enumerate}
\item The number of eruptions throughout the 11-year cycle scales linearly 
with the rotation rate, $\tos=\tom$, except for the comparison case $\tos=1$ 
(Sect.~\ref{sssec:record2}). 
\item The stratification and the differential rotation ($\Delta\Omega$) in the 
convection zone are kept the same as in the solar model (Sect.~\ref{ssec:rise}). 
\item The surface flux transport coefficients and the large-scale flows are the 
same as in 
the solar model (Sects.~\ref{sssec:bmrs}-\ref{sssec:flows}). 
\item The initial surface field is dipolar with a peak unsigned 
strength (at the rotational poles) equal to $10\tos$~G (Sect.~\ref{sssec:initb}). 
\end{enumerate}
}
\section{Results}
\label{sec:res}

\subsection{Butterfly diagrams}
\label{ssec:bfly}

Figure~\ref{fig:bfly} shows the 
butterfly diagrams for six sets of $(\tom,\tos)$. For 
$\tom=$2, 4, and 8, we consider either a solar ($\tos=1$, 
left panels) 
or a scaled stellar ($\tos=\tom$, right panels) flux emergence rate 
(Sect.~\ref{sssec:record2}). 

For $\tos=1$ (panels a, c, and e), 
the systematic increase of both 
the mean emergence latitude and the tilt angle  with $\tom$  is evident, 
owing to stronger Coriolis acceleration of rising 
tubes. For $\tos=\tom$ (panels b, d, and f), the imposed increase 
in $\tos$ leads to a further increase in the mean latitude 
of activity, owing not only to the Coriolis effect, but also
to the solar-like scaling of the mean latitude (Eq.~\ref{eq:meanlat}), 
as can be seen in the associated latitude histograms. 
In addition, the 
gap of inactivity around the equator with $\tom$ widens. However,  
the gaps are not very different from each 
other in cases $\tom=4$ and 8 (see also Fig.~\ref{fig:joys}b for comparison). 
The colours represent the tilt angles, as in Fig.~\ref{fig:bfly-ref}. 
The abrupt changes in the tilt angle are visible at 
some latitudes (cf. Fig.~\ref{fig:joys}a), especially for $(\tom,\tos)=(8,8)$.  

\subsection{Variation of surface magnetic activity}
\label{ssec:surface}

The synthetic emergence records presented in Sect.~\ref{ssec:bfly}
are now provided as input to the SFT model. 
In the first set of simulations, we keep the solar 
flux emergence rate ($\tos=1$) and increase the rotation rate (Sect.~\ref{sssec:tos1}). 
In the second set, we increase both quantities with $\tos=\tom$ (Sect.~\ref{sssec:tostom}).

\subsubsection{Solar flux emergence rate ($\tos=1$)}
\label{sssec:tos1}

Figure~\ref{fig:fluxvar} shows the 27-day averages of the 
unsigned flux over the stellar surface and the unsigned mean polar field 
strength (averaged over both polar caps, which are defined for 
$|\lambda| > 75^\circ$) using the same colours as in 
Fig.~\ref{fig:joys}. Taking $\lambda_e$ and $\alpha$ for $\tom>1$ 
with the solar flux emergence rate ($\tos=1$, Fig.~\ref{fig:fluxvar}a) leads to a 
total flux variation which increases only weakly with $\tom$. 
This enhancement of the flux is due to the systematic increase in the 
average tilt angles of emerging BMRs, which weakens flux cancellation 
over the polarity inversion line of each BMR. 
While the mean tilt angle $\langle\alpha\rangle$ 
increases from about $5^\circ$ to $30^\circ$, 
the latitudinal separation between the polarities increases with 
$\sin\langle\alpha\rangle$. 
The poleward deflection of rising loops indirectly contributes  
to the increase (with $\tom$) 
of flux for $\tos=1$: the emergence latitudes 
become more confined to a range where the latitudinal shear is stronger than at 
the solar activity belts. The strong differential rotation at the 
activity belt tends to disperse the opposite polarities at an even 
higher rate, immediately following any BMR emergence. 
This decreases the cancellation between the opposite polarities within 
the same BMR. Such self-cancellation accounts for a significant fraction 
of the initial flux decrease of a BMR.   

The isolated effect of faster rotation for $\tos=1$ 
(Fig.~\ref{fig:bfly} a, c, and e) is more conspicuous 
for the polar field amplitudes (Fig.~\ref{fig:fluxvar}c), for which the increasing 
tilt angle is the major contributor. 
This leads to a larger latitudinal separation between the polarities of a BMR, 
allowing the meridional flow to transport a larger amount of the 
following polarity to the pole, while the leading polarity has more time to cancel 
with the field from other BMRs.
The increasing latitudinal deflection produces the opposite effect 
for the polar field than for the total magnetic flux: by shifting 
both polarities of each BMR towards higher latitudes, it decreases
the efficiency of cross-equatorial flux cancellation. However, 
the effect remains sufficiently weak 
for up to $\tom=8$, for which most BMRs still emerge well below the 
latitude of the fastest meridional flow ($\sim37^\circ$). 
Another consequence of the increase in $\langle\alpha\rangle$ is that 
the polar 
field reverses its polarity and reaches its peak value earlier for increasing $\tom$. 

\subsubsection{Flux emergence rate scaled with rotation ($\tos=\tom$)}
\label{sssec:tostom}

For the set $\tos=\tom$ (Fig.~\ref{fig:bfly}b, d, and f), 
the amplitudes of both the total flux and the polar field increase 
more strongly with $\tom$ than in the case of $\tos=1$ (Fig.~\ref{fig:fluxvar}b,d). 
Still, the total flux and the polar field increase at a rate 
lower than the scaling $\tos=\tom$ itself. 
The polar field reversals occur earlier owing to increasing 
average tilt angles, but there is no difference in the reversal time from 
$(\tom,\tos)=(4,4)$ to (8,8). The reason is that a significant amount of 
BMRs in the case (8,8) emerge within the low-tilt plateau above $\lambda\sim 
50^\circ$ (see Figs.~\ref{fig:joys}a and \ref{fig:bfly}f). In this case, $\tos=8$ 
does not provide sufficient contribution to the global (axial) dipole moment to 
reverse it 
earlier than in the case (4,4). We note that in (8,8) the initial polar field that has 
to be reversed is stronger by a factor of two, relative to (4,4), owing to our 
assumption $B_{\rm max}(t=0)=\pm 10\tos$~G.

\begin{figure}
    \centering
    \includegraphics[width=.45\columnwidth]{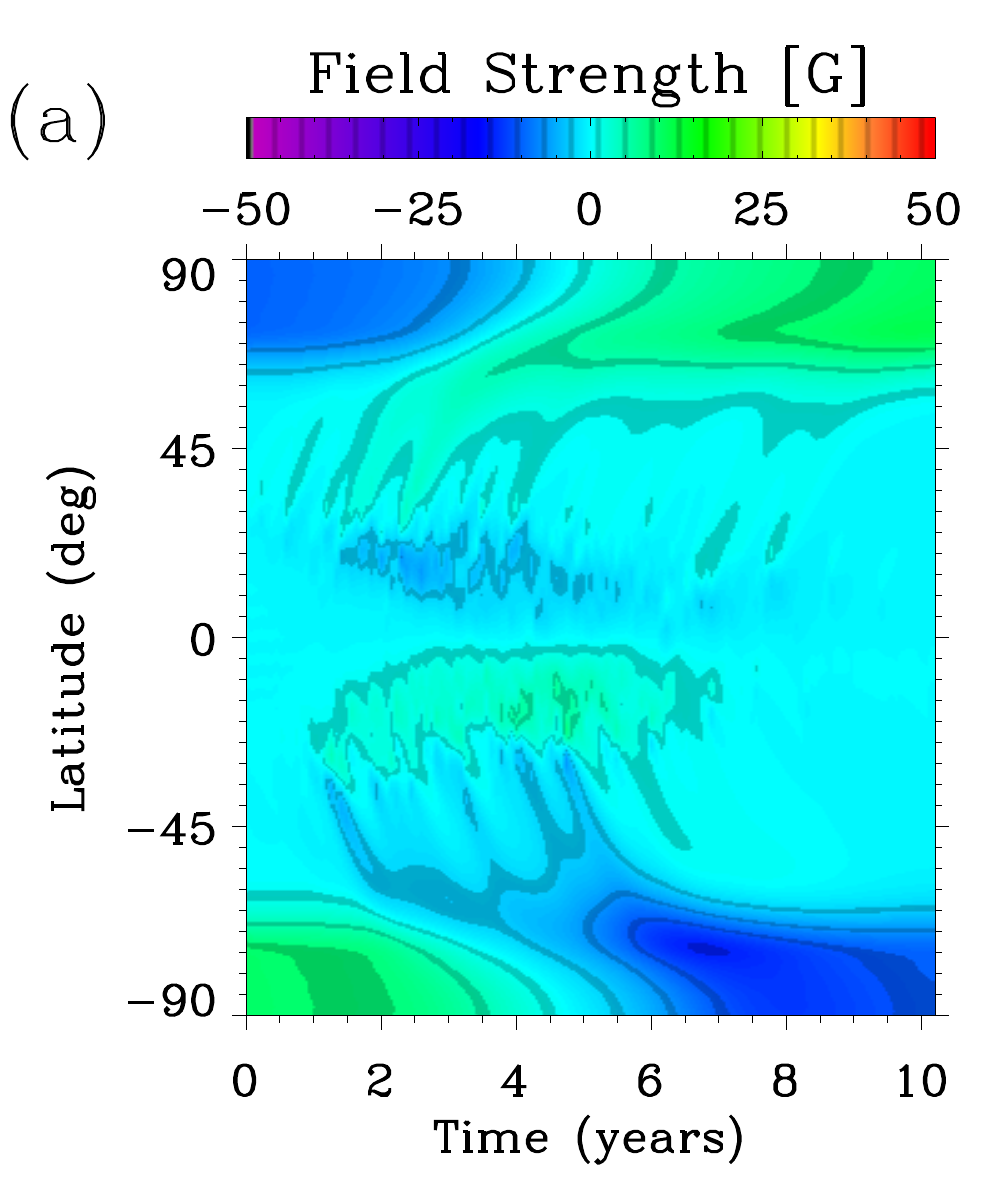}
    \includegraphics[width=.45\columnwidth]{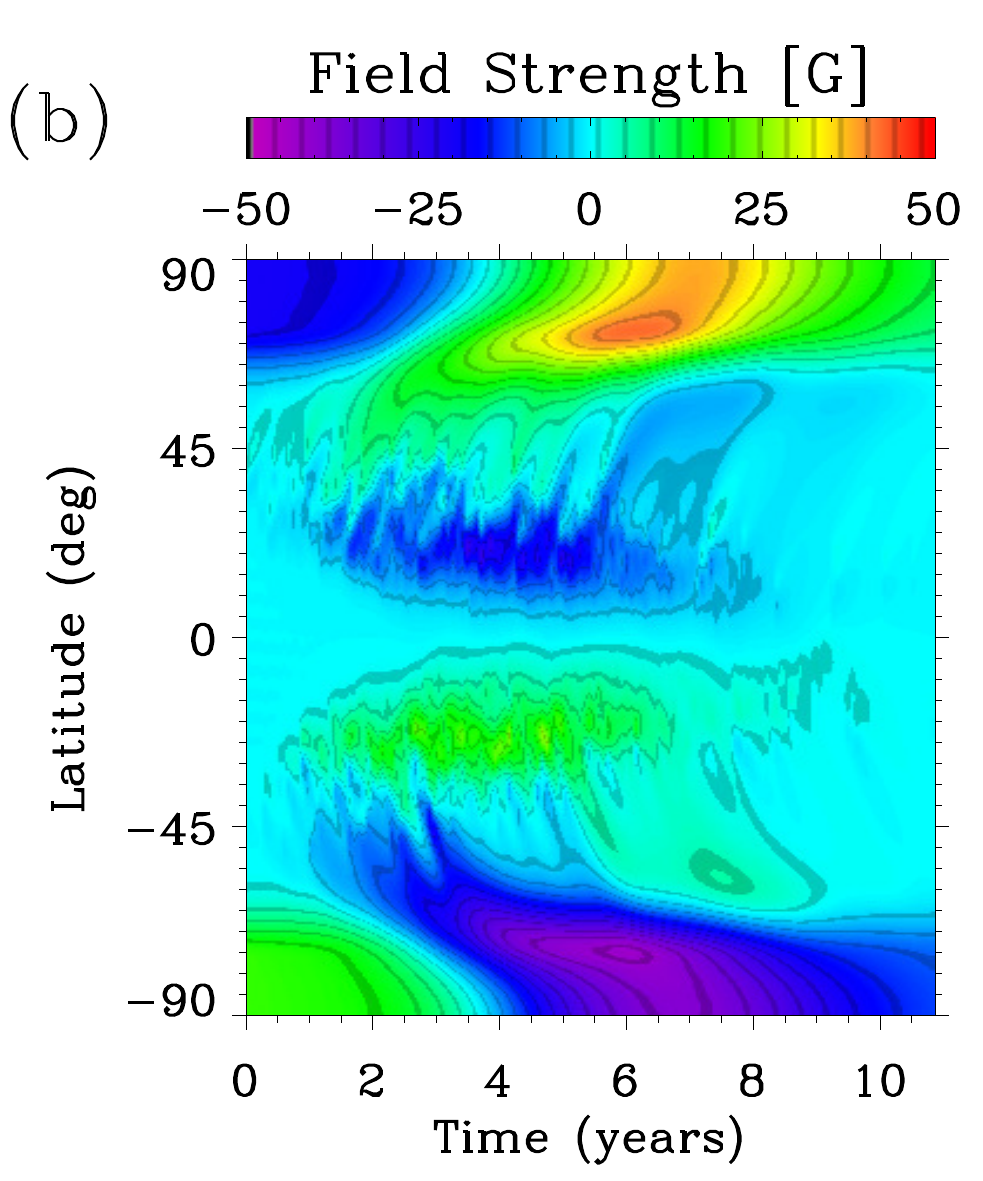} \\
    \includegraphics[width=.45\columnwidth]{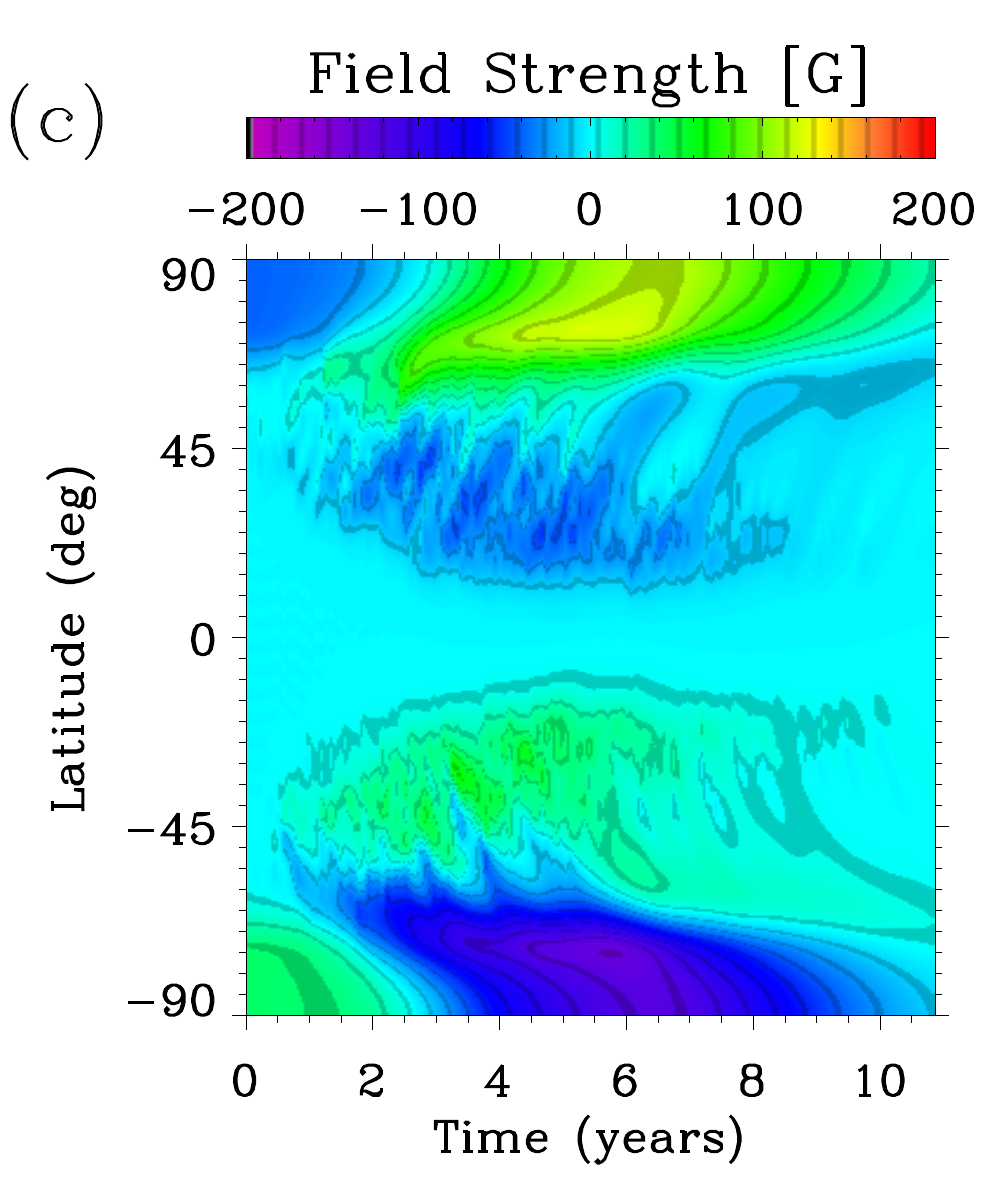}
    \includegraphics[width=.45\columnwidth]{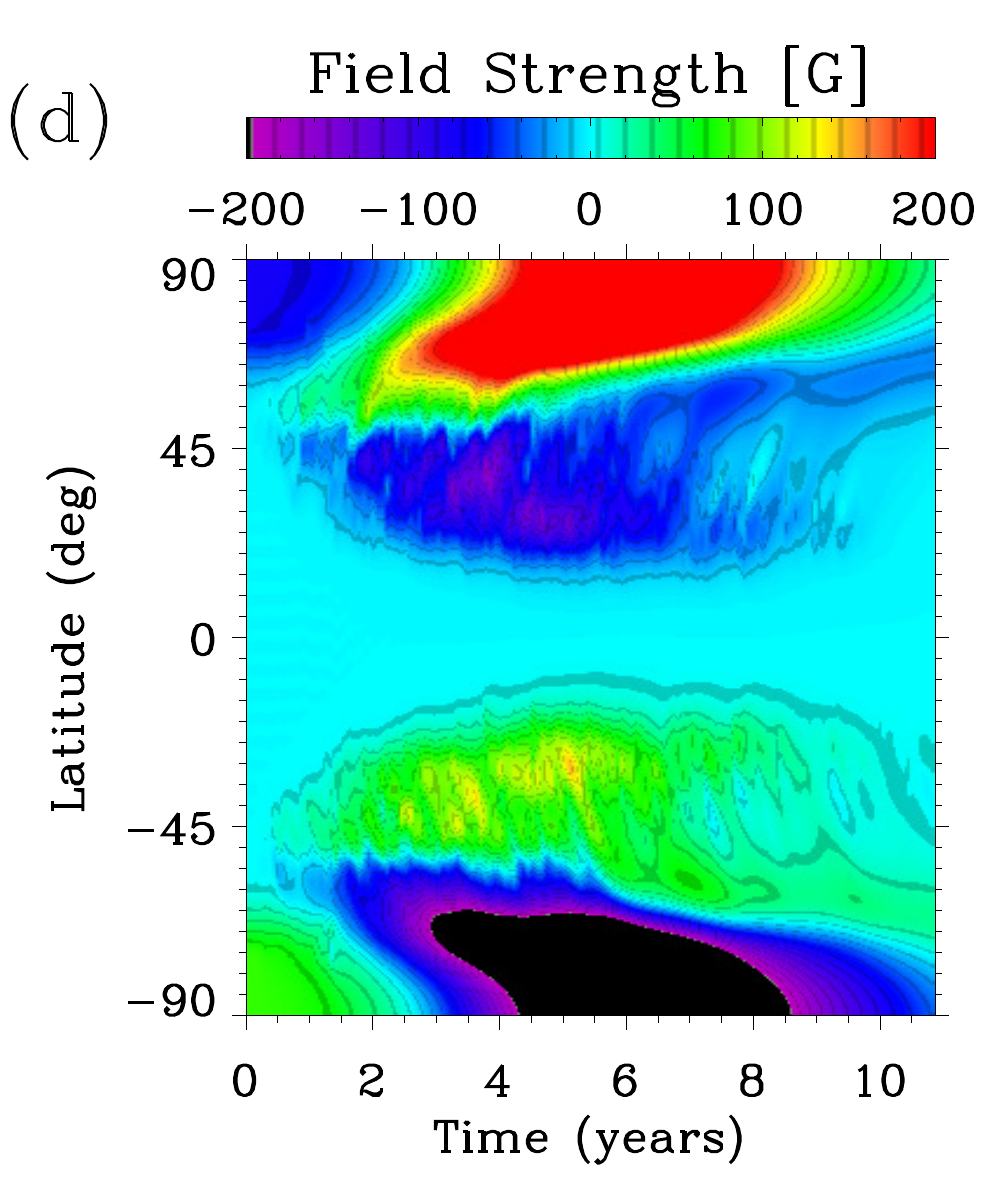}
    \caption{Time-latitude diagrams of azimuthally averaged surface magnetic field for (a) $\tom=1$, (b) $\tom=2$, (c) $\tom=4$, and (d) $\tom=8$. For all cases, $\tos=\tom$. We note the different saturation levels of the colour scale.}
    \label{fig:mbfly}
\end{figure}

\begin{figure}
    \centering
    \includegraphics[width=.45\columnwidth]{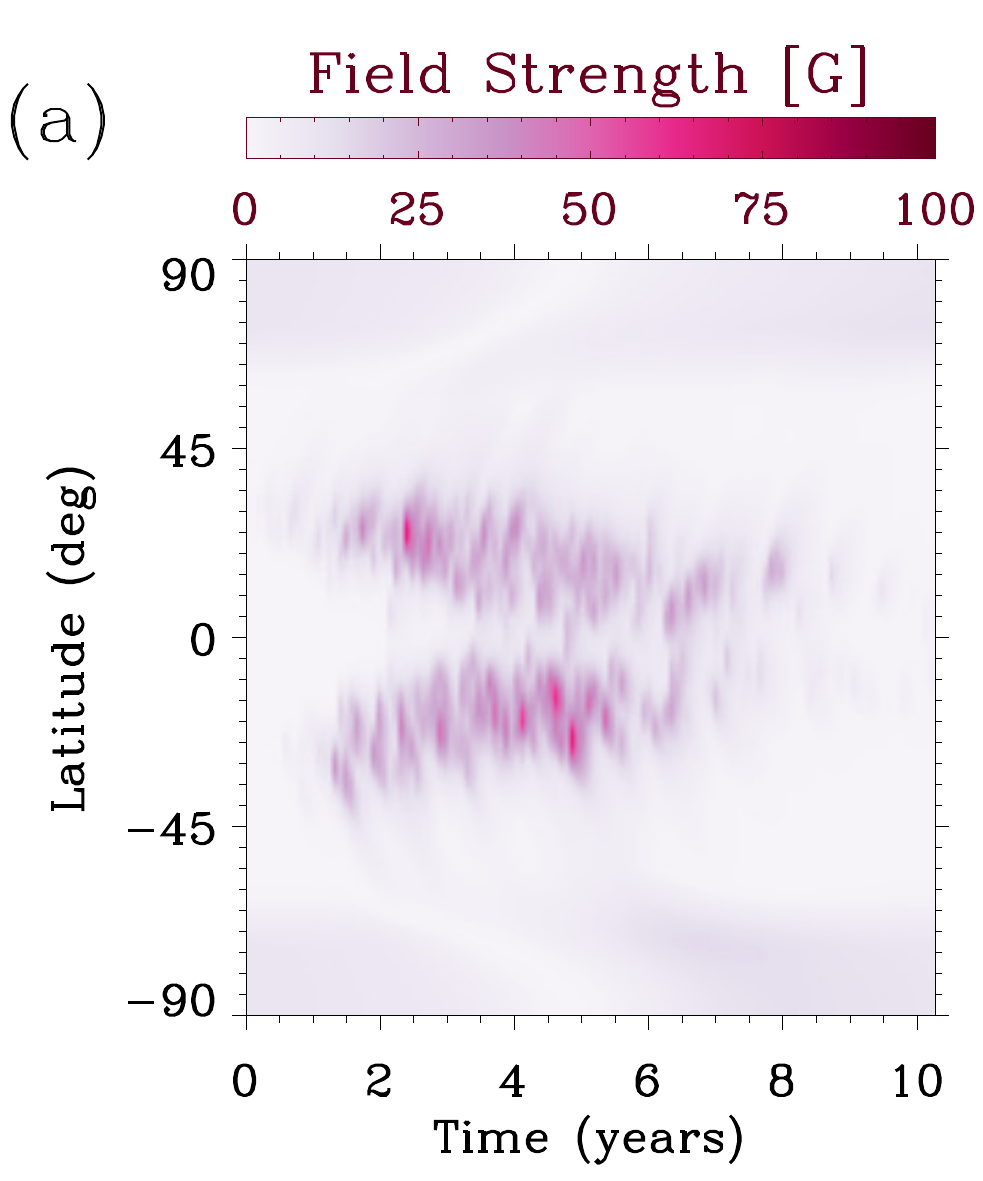}
    \includegraphics[width=.45\columnwidth]{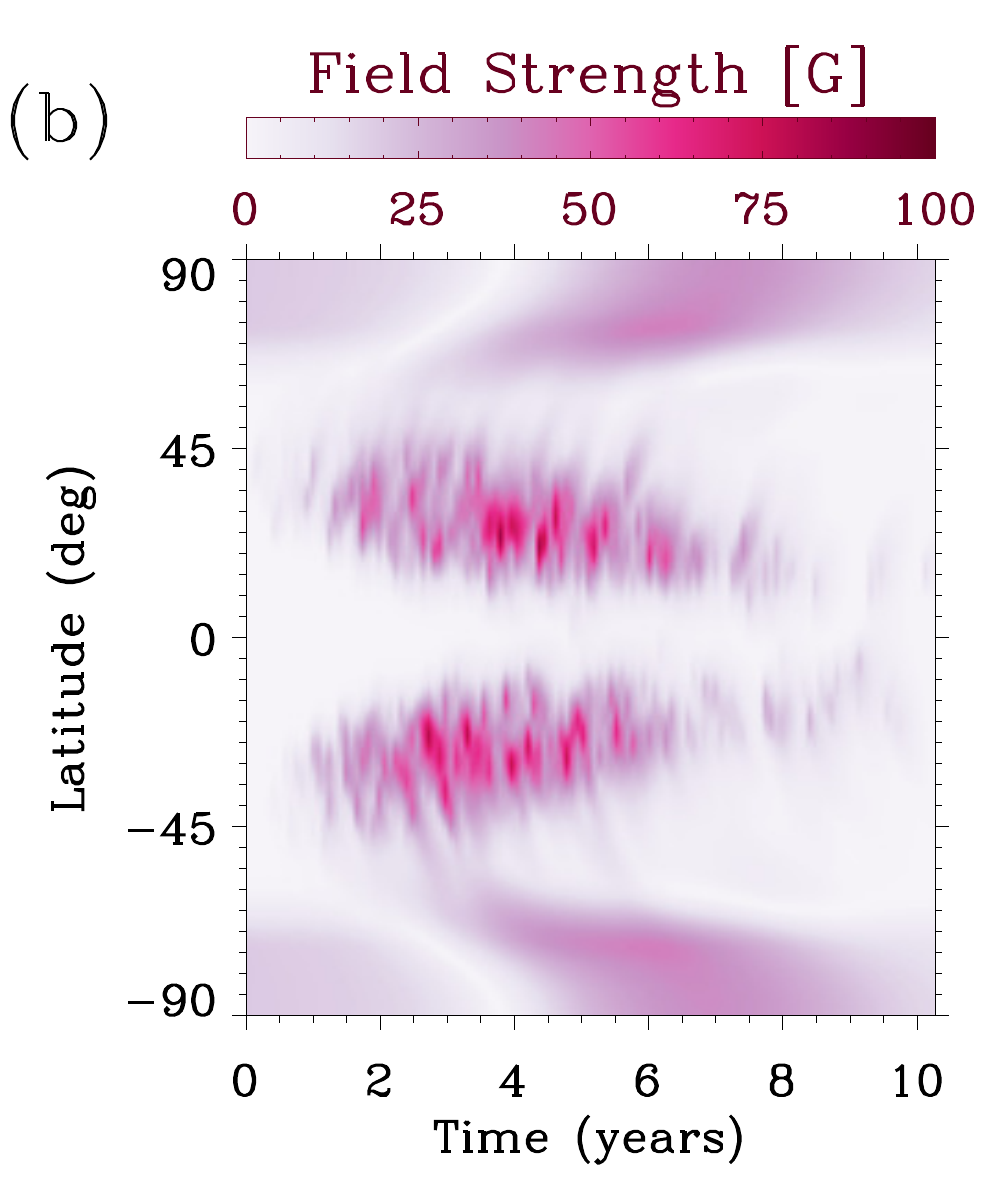} \\
    \includegraphics[width=.45\columnwidth]{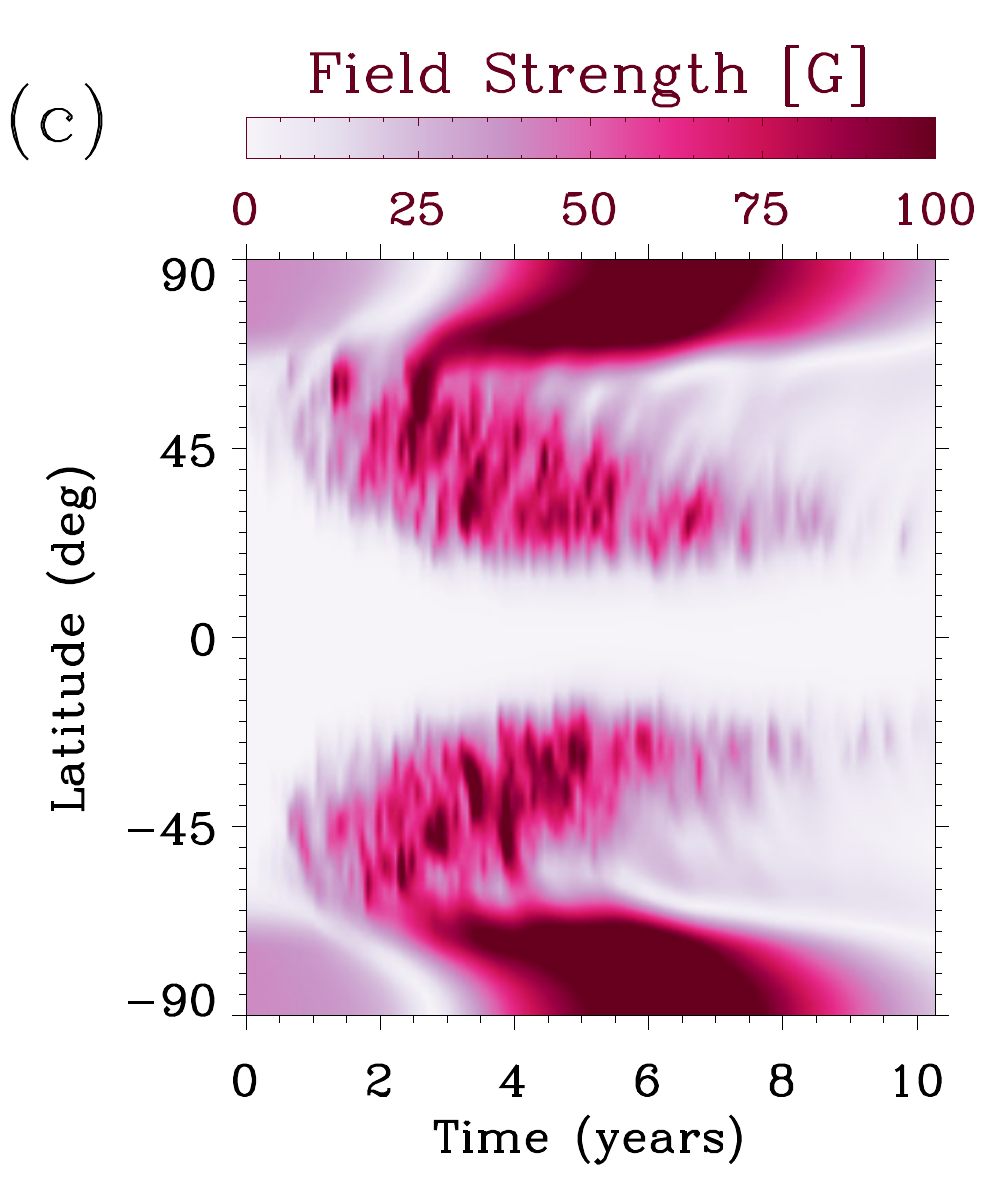}
    \includegraphics[width=.45\columnwidth]{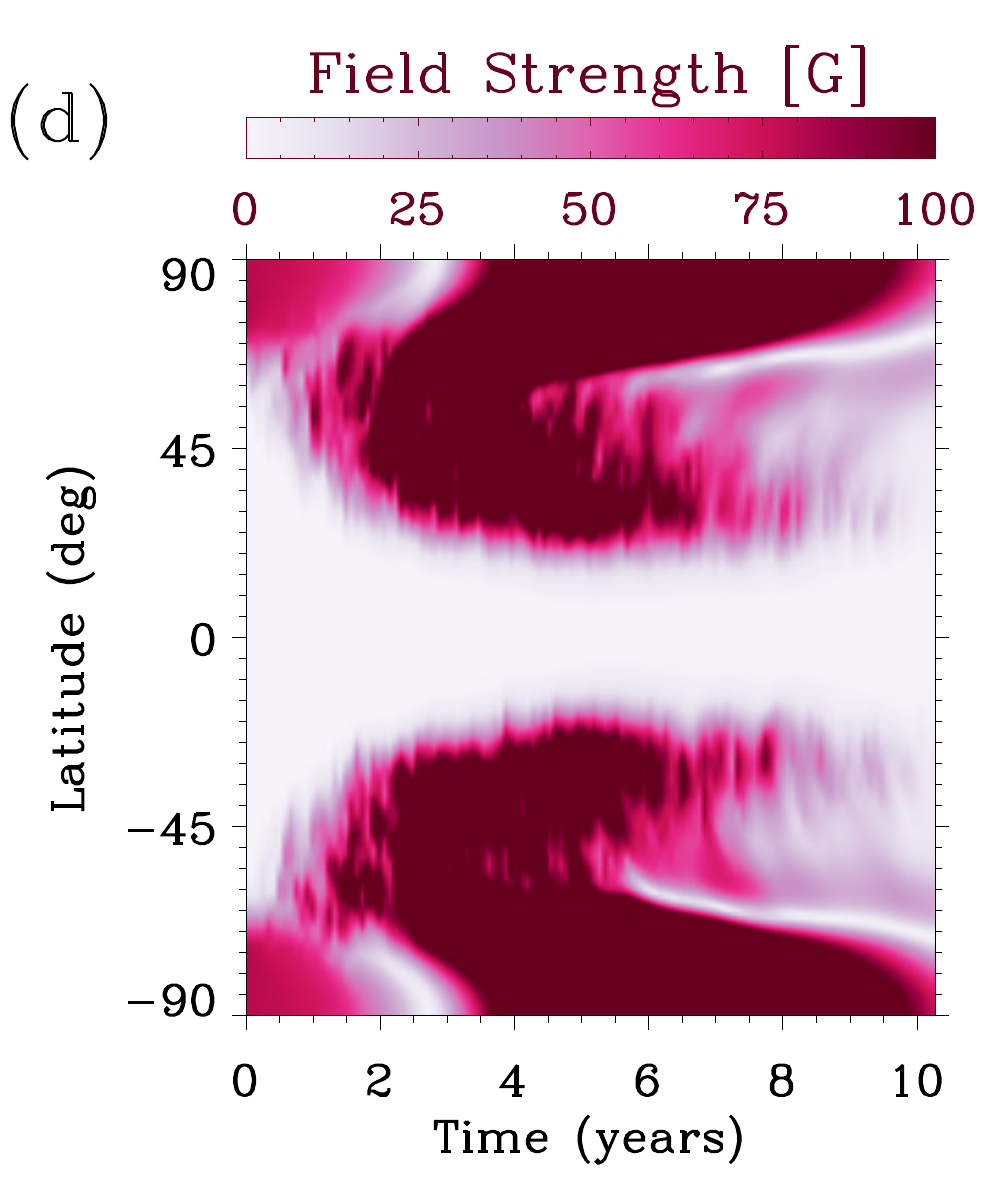}
    \caption{Time-latitude diagrams of azimuthally averaged $|B_r|$ for 
    (a) $\tom=1$, (b) $\tom=2$, (c) $\tom=4$, and (d) $\tom=8$. $\tos=\tom$ for all 
    cases. Here, the saturation levels are all 100~G.}
    \label{fig:absb}
\end{figure}

The rotational dependencies of the total surface flux and 
the polar flux are shown in Fig.~\ref{fig:fluxvar}e in terms of the time integral of 
both quantities. The dependence is almost linear 
for both $\tos=1$ and $\tos=\tom$, with similar slopes. 
In the latter case the time-integrated total flux scales as $c\tom$, 
with $c\approx 0.9$. This is expected from the 
linearity of the SFT process \citep{schrijver01}. The non-linear dependence of the 
BMR dipole moments on $\tom$ does not lead to a significant deviation here. 
Eight-times-faster rotation 
leads to a change in the cumulative polar flux of about $2\times 
10^{24}$~Mx, which is about 8 \%\ of the corresponding change in 
the cumulative surface flux. It should be noted, however, 
that the cumulative polar flux scales non-linearly with 
$\tom$, at a gradually increasing rate, owing to rotationally induced 
effects. 
The contributors to the positive correlation between the two quantities are 
now both the increased latitudinal 
separation between polarities (increasing mean tilt angle) and the imposed 
dependence $\tos=\tom$.

\subsubsection{Magnetic butterfly diagrams for $\tos=\tom$}

Figure~\ref{fig:mbfly} shows the longitudinal averages of 
the surface magnetic field for $\tos=\tom$ models, as a function of time. 
One can directly see 
the effect of increased tilt angles (\emph{i.e.} latitudinal separation 
of {the preceding} and {the follower} polarities) on the quicker formation 
of a stronger polar field. For the more rapidly rotating stars, 
the increase in both the flux 
emergence rate and the tilt angles cause the time at which the 
polar field reaches its maximum value to converge towards the time of the 
maximum total surface flux. 

Figure~\ref{fig:absb} shows the unsigned radial field at the 
surface, averaged over longitude, for the cases shown in Fig.~\ref{fig:mbfly}.
In this way, it is easier to compare the distribution of the activity belts 
with that of polar fields, because the arithmetic 
cancellation of opposite-polarity fields is avoided during longitudinal averaging. 
It is evident that mid-latitude activity strengthens considerably for $\tom=4$ 
and 8. Above $(\tom,\tos)=(4,4)$, the polar fields start to become 
comparable with the mid-latitude activity, and they also become 
dominant for an increasingly larger fraction of the cycle.

\subsection{The distribution and coverage of starspots ($\tos=\tom$)}

To define spot areas from surface magnetic maps, we followed the simplest 
approach of setting a 
threshold to the magnetic field strength. All pixels with a field strength 
above the threshold are considered to belong to spots. We determine the 
threshold value from the condition that the time average of the spot 
coverage through 
the cycle, $\langle a_s\rangle_{\rm cyc}$, roughly corresponds to the
cycle-averaged umbral spot coverage observed on the Sun, which is about 0.002. 
Using this criterion, 
we have found a threshold of {187}~G, corresponding to about 
{50} \%\ of $B_{\rm max}$ defined in Sect.~\ref{ssec:sft}. 
We have then used this threshold to determine 
the latitudinal distribution of the spot coverage for all four rotation rates 
with $\tos=\tom$. 

\begin{figure}
    \centering
    \begin{subfigure}[t]{0.01\columnwidth}
        (a)
    \end{subfigure}
    \begin{subfigure}[t]{0.47\columnwidth}
        \includegraphics[width=\linewidth,valign=t]{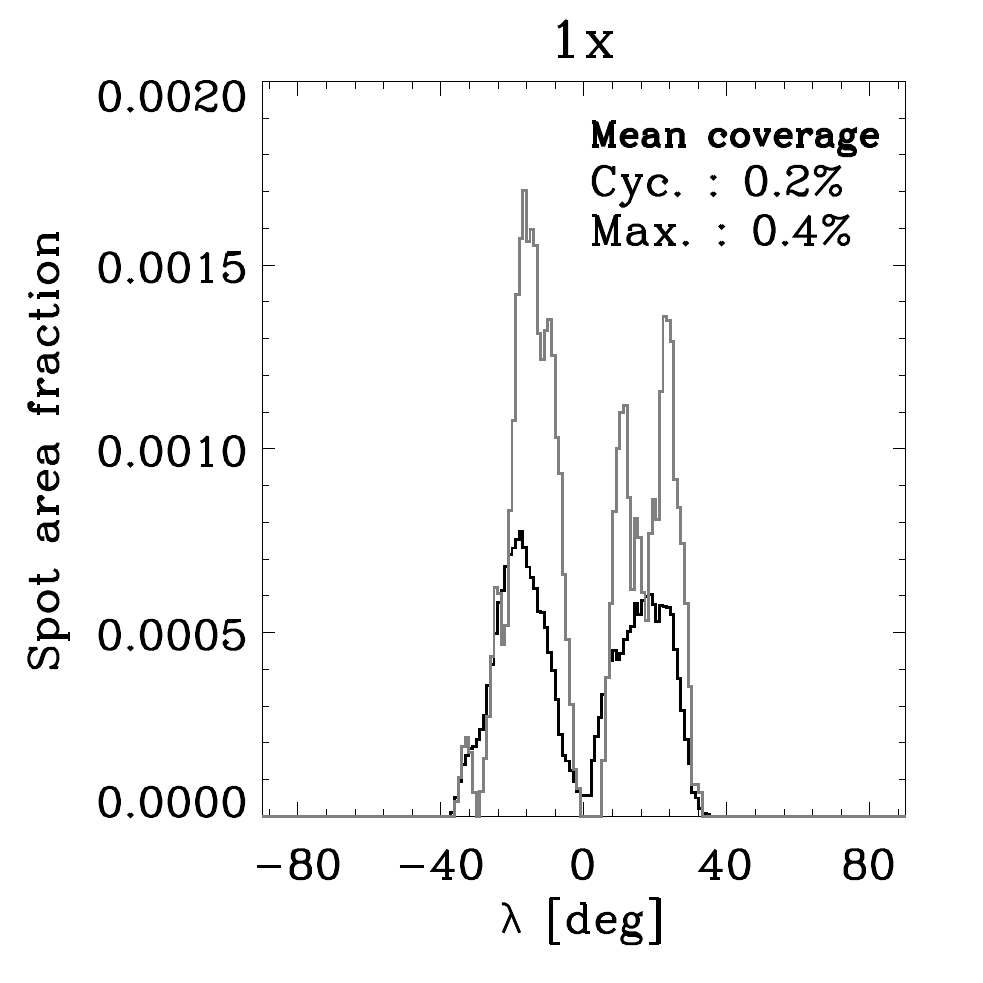}
    \end{subfigure}
    \begin{subfigure}[t]{0.01\columnwidth}
        (b)
    \end{subfigure}
    \begin{subfigure}[t]{0.47\columnwidth}
        \includegraphics[width=\linewidth,valign=t]{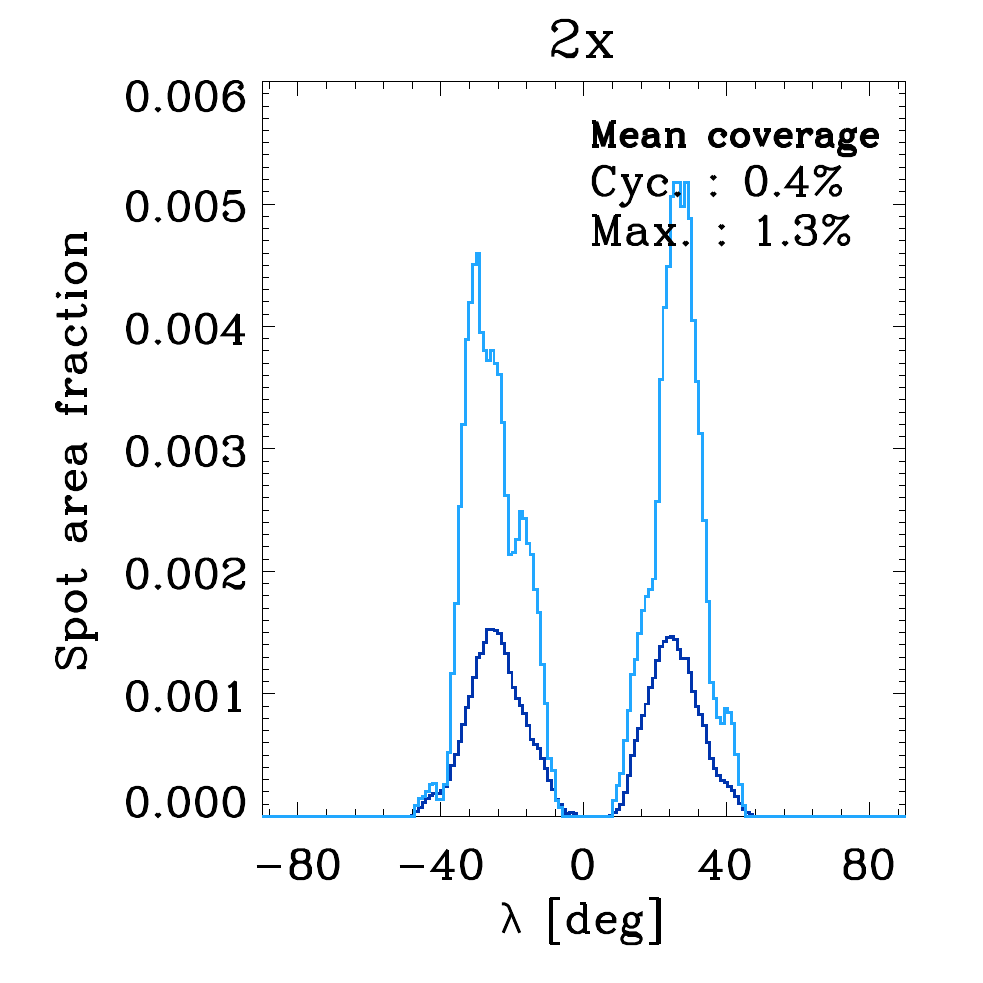} 
    \end{subfigure} \\
    \begin{subfigure}[t]{0.01\columnwidth}
        (c)
    \end{subfigure}
    \begin{subfigure}[t]{0.47\columnwidth}
        \includegraphics[width=\linewidth,valign=t]{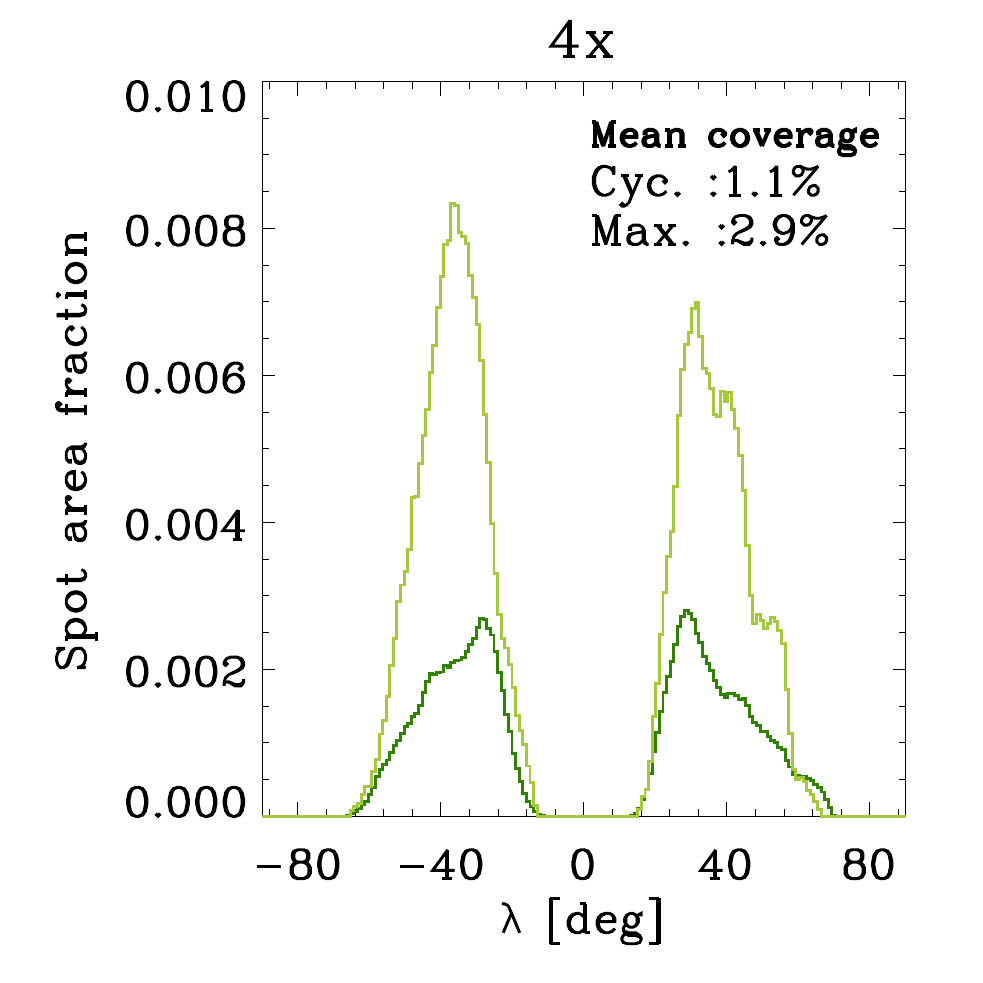}
    \end{subfigure}
    \begin{subfigure}[t]{0.01\columnwidth}
        (d)
    \end{subfigure}
    \begin{subfigure}[t]{0.47\columnwidth}
        \includegraphics[width=\linewidth,valign=t]{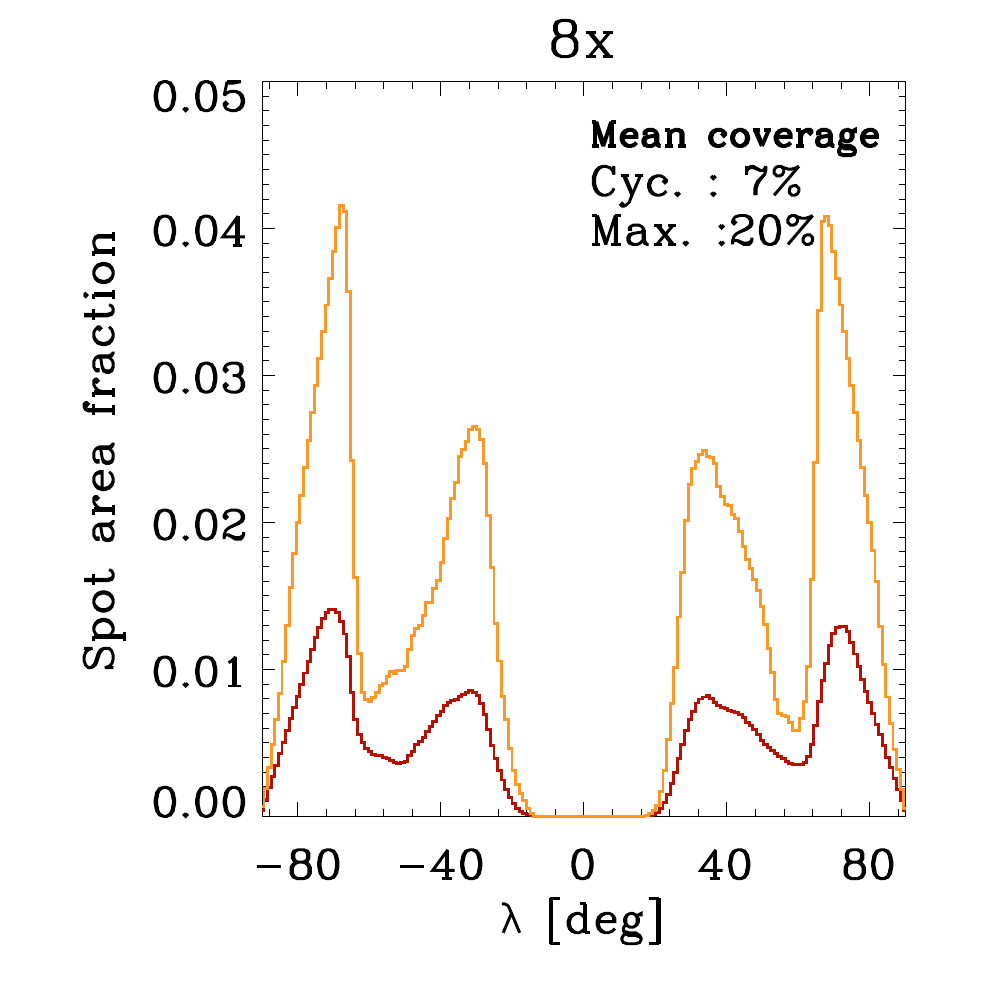}
    \end{subfigure}
    \flushleft{(e)}
\hskip2cm\vskip-1cm
    \includegraphics[width=\linewidth]{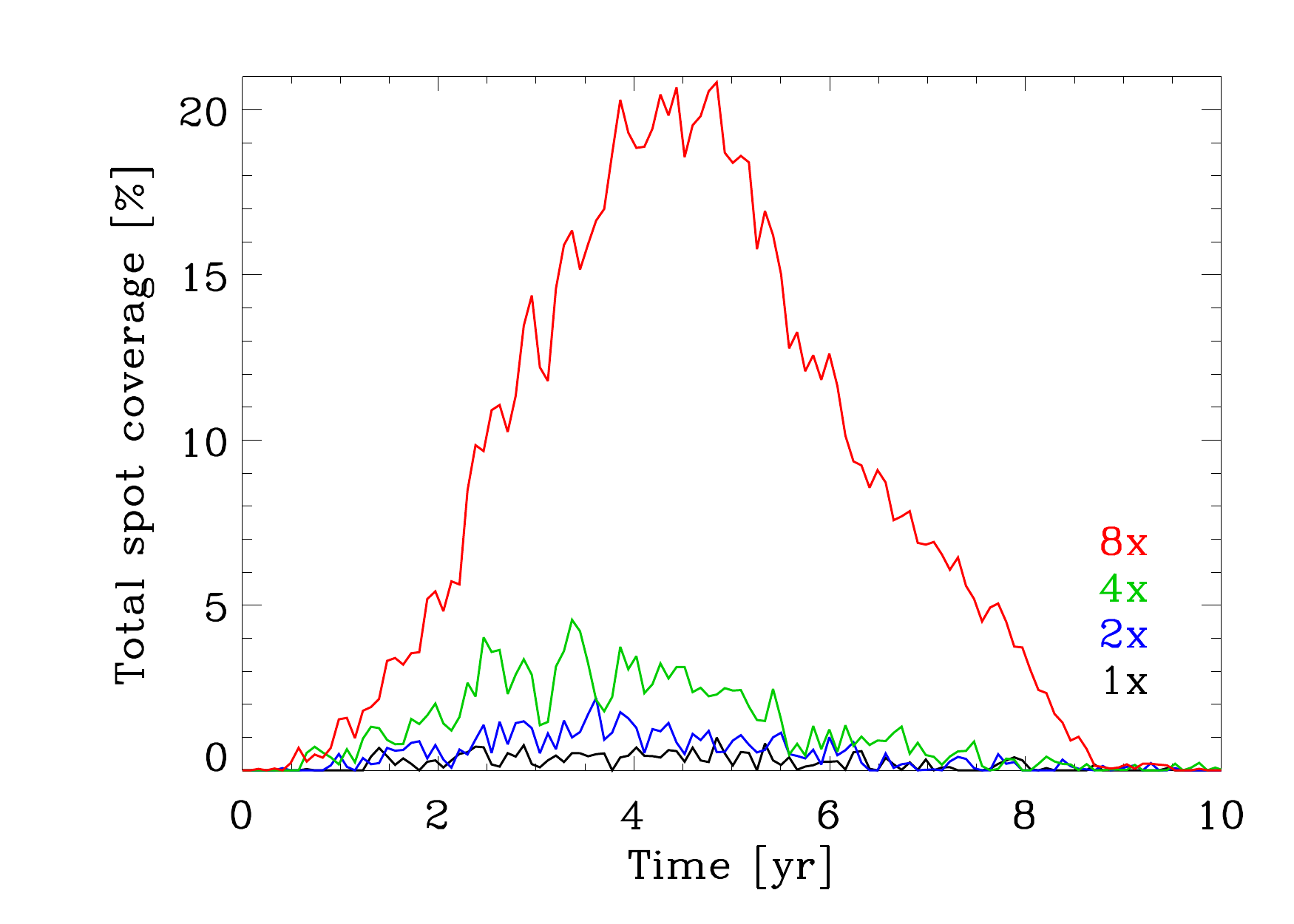}
    \caption{{Time-}averaged latitudinal distributions of the fraction of surface 
    area covered by starspots for (a) $\tom=1$, (b) $\tom=2$, (c) $\tom=4$, and (d) $\tom=8$, where $\tos=\tom$, {given at the top of each panel. The  
    colours correspond to the cases in Figs.~\ref{fig:joys} and \ref{fig:fluxvar}; lighter 
    ones are averages over a one-year window centred at the activity maximum 
    and darker 
    curves represent cycle averages. Time-averaged surface coverages for 
    maxima and whole-cycles are given inside each plot. Panel (e) shows the 
    variation of the total spot coverage for each case.} } 
    \label{fig:spotlats}
\end{figure}

Figure~\ref{fig:spotlats}{a-d} shows {time-averaged} latitudinal profiles 
of spot occupancy as a fraction of the stellar surface area.  
The area fractions were calculated by counting the pixels 
above the threshold, taking into account their areas on a 
spherical grid. In most observational studies, 
a factor of $\cos\lambda$ is used when estimating the fractional spot 
area per latitude bin. This means that our profiles are comparable with 
such latitudinal profiles presented in the literature 
\citep[e.g.][]{jarvinen07,waite17}. 
When the spot occupancy would be given as a fraction of the latitudinal 
band area, however, the polar spot of the case $(\tom,\tos)=(8,8)$ would 
lead to a much larger coverage near the pole than at mid-latitudes.
{The} spot coverage over the whole stellar 
surface, {averaged over the whole} activity cycle {(darker curves)}, 
$\langle a_s\rangle_{\rm cyc}$, 
increases from the solar value of 0.2\% to 0.4\% for $\tom=2$, 
1.2\% for $\tom=4$, and 10\% for $\tom=8$. 
{For comparison, one-year averages centred at the activity maximum are 
also given (lighter colours).}
There is a marked tendency for the mean 
latitude and the latitudinal spread of starspots to increase with 
$\tom$ and $\tos$. The mean latitude at each 
hemisphere increases, owing to the Coriolis deflection of 
rising flux and the enhanced poleward transport of highly 
tilted source regions. The increase in the latitudinal 
spread is led by ($i$) higher flux emergence rate, 
($ii$) the scaling of mean latitude with $\tos$ (Eq.~\ref{eq:latscale}), and 
($iii$) weaker flux cancellation between opposite polarities of BMRs, 
owing to a higher average tilt angle. 
It is evident from Fig.~\ref{fig:spotlats}d ($\tom=8$) that, 
through the flux transport at sufficiently high $\tom$ and 
$\tos$, spotted regions are formed 
at even higher latitudes than their emergence latitudes. 
Such starspots are formed by signed magnetic flux being 
transported towards higher latitudes and concentrated there.
{Figure~\ref{fig:spotlats}e shows the variation of the surface coverage 
of starspots for all the four cases above. 
The cycle means are about one-third of the annual means around maxima, 
except for the the case (1,1), for which the ratio is about 0.5. 
}

\begin{figure}
    \centering
    \begin{subfigure}[t]{0.02\columnwidth}
        a
    \end{subfigure}
    \begin{subfigure}[t]{0.44\columnwidth}
        \includegraphics[width=\linewidth,valign=t]{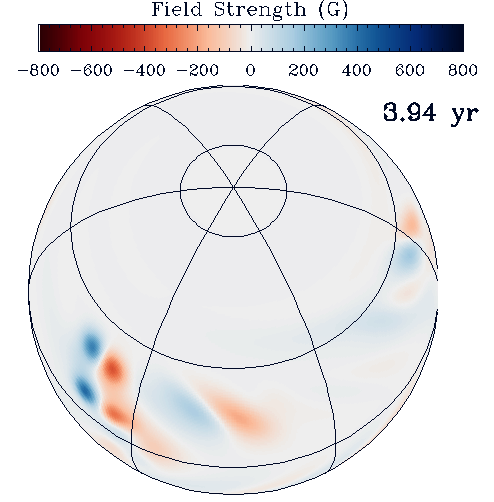}
    \end{subfigure}
    \begin{subfigure}[t]{0.02\columnwidth}
        b
    \end{subfigure}
    \begin{subfigure}[t]{0.44\columnwidth}
        \includegraphics[width=\linewidth,valign=t]{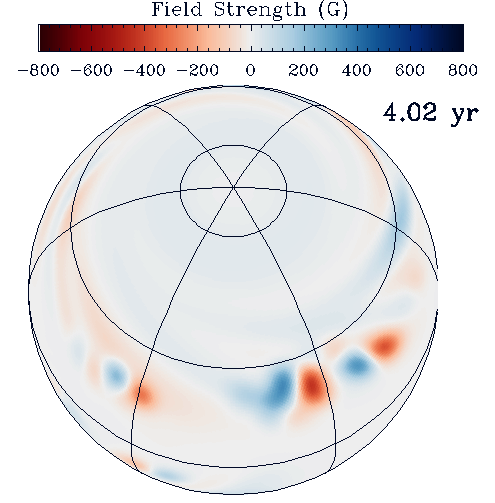}
    \end{subfigure} \\
    \begin{subfigure}[t]{0.02\columnwidth}
        c
    \end{subfigure}
    \begin{subfigure}[t]{0.44\columnwidth}
        \includegraphics[width=\linewidth,valign=t]{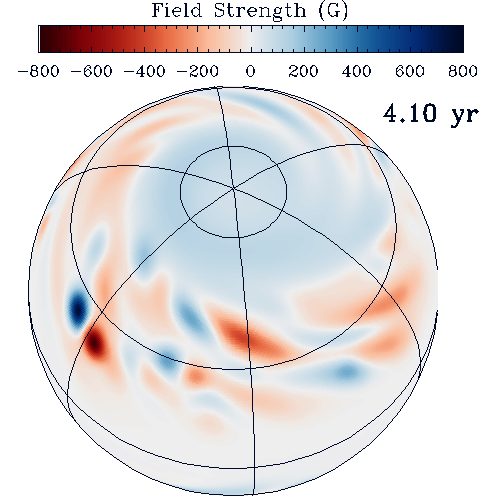}
    \end{subfigure}    
    \begin{subfigure}[t]{0.02\columnwidth}
        d
    \end{subfigure}
    \begin{subfigure}[t]{0.44\columnwidth}
        \includegraphics[width=\linewidth,valign=t]{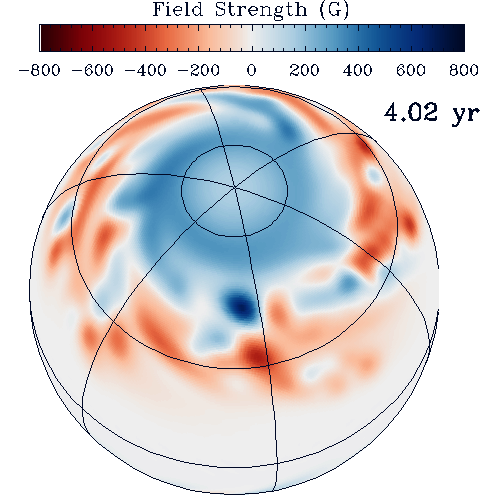}
    \end{subfigure}
    \caption{Snapshots of magnetic field strength from the runs for 
    $\tom=1,2,4,8$ (panels a to d), corresponding to the cycle 
    maximum in each case. 
    The inclination angle of the rotation axis with respect to the 
    line of sight is $30^\circ$. The latitudinal circles are 
    drawn at $\lambda=37.5^\circ$, where the poleward flow speed 
    has a maximum, and at $\lambda=75^\circ$, above which it is assumed
    to be zero.}
    \label{fig:snap}
\end{figure}

Figure~\ref{fig:snap} shows snapshot surface distributions of signed 
field from the maximum phases of activity, 
corresponding to each of the considered rotation rates, 
for the case $\tos=\tom$. 
The polar fields become stronger for higher 
$(\tom,\tos)$, while the activity belts move {towards higher latitudes}. 
A strong unipolar polar cap forms for $\tom=8$, 
but it is surrounded by patches of opposite-polarity field, owing to 
the leading polarities of freshly emerged, tilted BMRs. A conspicuous feature 
{is visible} for $\tom=8$: 
$dB/d\lambda$ changes its sign at $\lambda\simeq 75^\circ$, where the 
poleward meridional flow piles up the field diffusing from lower latitudes. 
In the later stages of the cycle, this ring-like structure diffuses 
to form a circular region peaking at the rotational pole. 

\begin{figure}
    \centering
    \begin{subfigure}[t]{0.02\columnwidth}
        a
    \end{subfigure}
    \begin{subfigure}[t]{0.44\columnwidth}
        \includegraphics[width=\linewidth,valign=t]{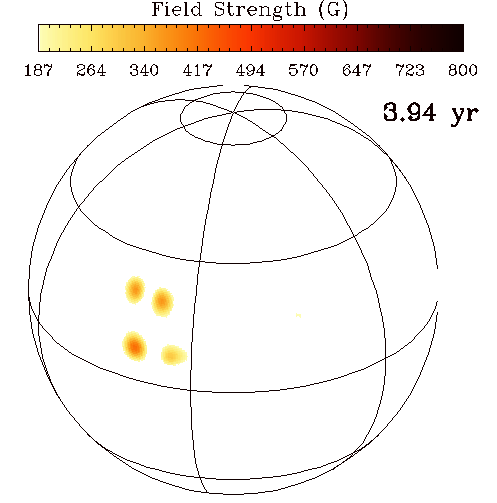}
    \end{subfigure}
    \begin{subfigure}[t]{0.02\columnwidth}
        b
    \end{subfigure}
    \begin{subfigure}[t]{0.44\columnwidth}
        \includegraphics[width=\linewidth,valign=t]{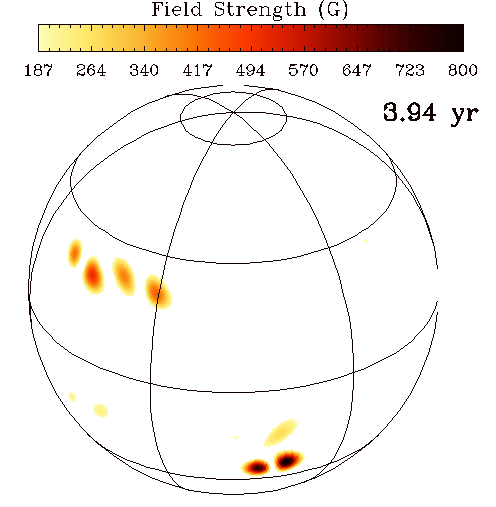}
    \end{subfigure} \\
    \begin{subfigure}[t]{0.02\columnwidth}
        c
    \end{subfigure}
    \begin{subfigure}[t]{0.44\columnwidth}
        \includegraphics[width=\linewidth,valign=t]{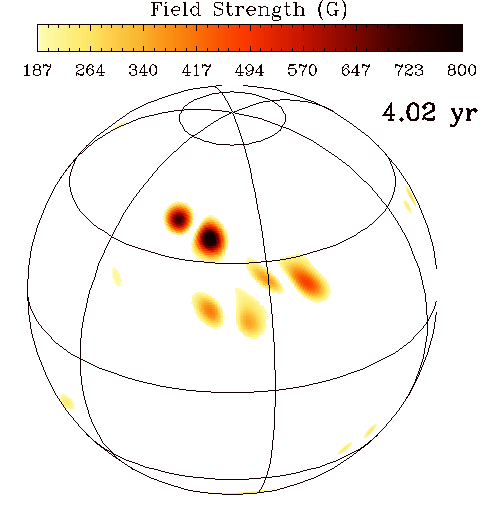}
    \end{subfigure}    
    \begin{subfigure}[t]{0.02\columnwidth}
        d
    \end{subfigure}
    \begin{subfigure}[t]{0.44\columnwidth}
        \includegraphics[width=\linewidth,valign=t]{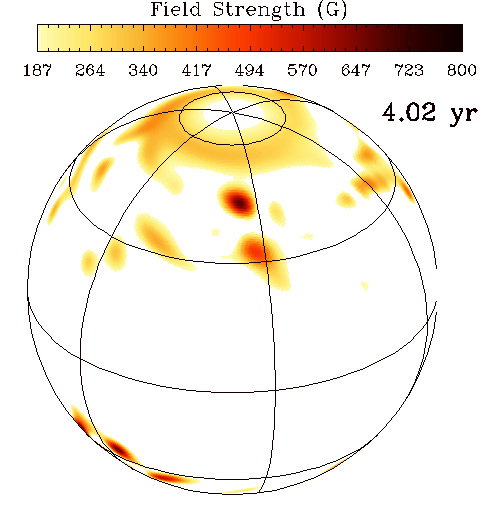}
    \end{subfigure}
    \caption{As in Fig.~\ref{fig:snap}, but for the \emph{unsigned} 
    magnetic field strength with an inclination of $60^\circ$. 
    The maps show the distribution of starspots, 
    which are defined as all pixels above the threshold of {187~G}.}
    \label{fig:snapabs}
\end{figure}

In Fig.~\ref{fig:snapabs} we show the unsigned magnetic 
field strength, {filtered with the threshold of 187~G}. These snapshots can be 
seen as approximations of intensity images {by omitting} 
the limb-darkening {and} facular brightening effects. 
The strengthening of the polar field and of mid-latitude activity 
with increasing $\tom$ {is} 
visible. Based on the spot threshold field strength, 
the polar spot forms only for $\tom=8$, and it shows the same ring-shaped 
pattern as in Fig.~\ref{fig:snap} with occasional plumes in the direction of the equator. 
Here, the solar case 
exhibits too large `sunspots', though the total area 
fraction is solar-like. This is because the resolution of the SFT 
model is not high enough to resolve typical sunspots. In addition, 
the saturation level 
for the unsigned field is only 400~G, about an order of magnitude 
below the average field strength of sunspot umbrae. Consequently, the 
simulated stellar images presented here are meant to represent a 
medium-resolution picture of spots on Sun-like stars. 
The resolution 
is therefore between those of solar full-disc white-light images and those 
produced by Doppler imaging \citep[e.g.][]{solunruh04}. A more rigourous 
treatment of spot areas is beyond the scope of this study, and will 
be employed in the next paper for proper modelling of brightness 
variations. 

\begin{figure}
\centering
\includegraphics[width=\columnwidth]{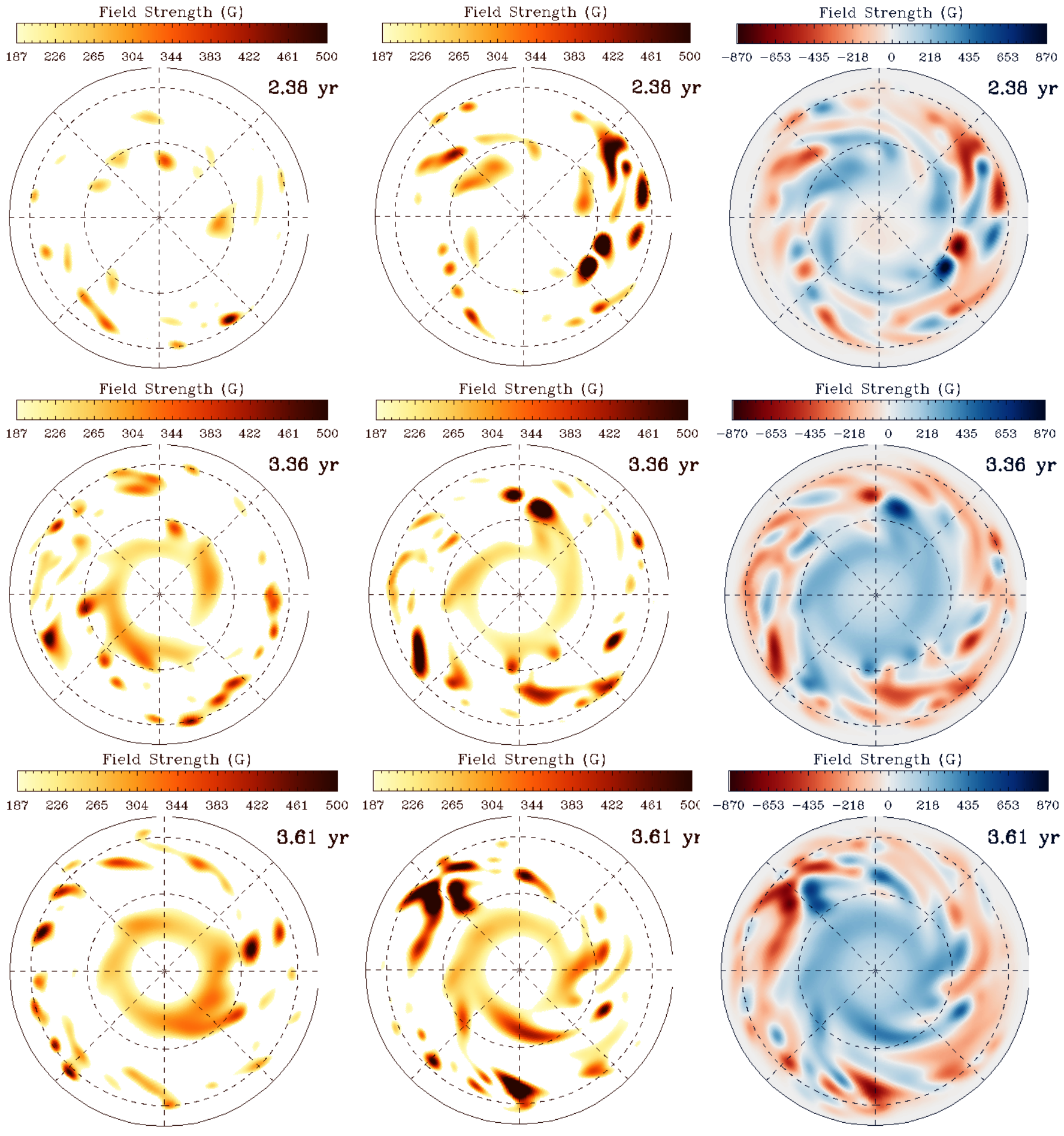}
\caption{{Pole-on views of {the} radial field, representing spot 
distributions for unnested 
(left {column}) and nested ({middle column}) cases for $(\tom,\tos)=(8,8)$. 
{The right column shows the corresponding signed magnetic field strength for 
the nested case, {with a colour saturation at $\pm 870~G.$}}
Dotted {circles represent} 
the {latitudes} at $30^\circ$ and $60^\circ$. We note that the spots are defined 
above 187~G. }}
\label{fig:snapshots}
\end{figure}

{
It is known that 
{faster-rotating} G stars generally show stronger {brightness variability} 
\citep{mcquillan14}. 
{In this context, nesting of active regions can have an influence on the 
variability amplitudes on timescales comparable with the rotation period. 
This is because the surface distribution of spots would become less 
homogeneous in longitude (rotational phase), when spots tend to emerge within nests. }
As a visual demonstration of the effect of nesting, we show in 
Fig.~\ref{fig:snapshots} pole-on 
views ($i=0$) of our case $(\tom,\tos)=(8,8)$ at {three different} phases of the 
activity cycle, for random {(unnested)} and strongly nested 
cases, where the nesting probability was chosen to be 
$p=0.7$ (same as in all the 
previous figures). {We also display the signed field distributions corresponding 
to the nested case, for the overall field geometry and strength to be evaluated.} 
Without forced nesting, the spot distribution appears more 
axisymmetric. With nesting, the {highly non-axisymmetric} 
spot {distribution} is likely to {induce 
larger-amplitude brightness variations on rotational 
timescales}. 
}

\section{Discussion}
\label{sec:dis}

We have developed a two-part model, to provide time-resolved maps of the 
radial magnetic field on Sun-like stars with rotation rates in the range
$\Omega_\sun\leqslant\Omega\leqslant8\Omega_\sun$, which corresponds to a 
(sidereal) equatorial rotation period range of {25} to 3 days. 
The 
platform developed here will 
be used in the forward modelling of brightness variations on 
timescales covering active region evolution, stellar rotation, 
and the activity cycle (the second paper in this series). 
It {also} has the potential to be used in synthesising spectra covering 
photospheric lines used in Doppler imaging. 

The thin flux-tube {simulations} successfully model the basic dynamical 
aspects of the emergence of large-scale flux loops in the case 
of the Sun, despite several idealisations involved \citep{cale95}. 
{Distributions of tilt angles in faster-rotating suns have not been 
well investigated so far, despite their potential effects on the distribution and  
evolution of surface magnetic flux. 
We have demonstrated in this study that these effects can be significant. 
Though we focused on the photospheric 
distribution of large-scale radial fields, the dynamical effects of 
increasing rotation rate on the emerging flux would certainly have implications for} 
coronal magnetic {structure {\citep{gibb16}}}. 



The model provides clues about how patterns of stellar activity are likely 
to change with increasing rotation {and flux-emergence rates}. 
{We assumed that the time-latitude pattern of 
flux eruptions at the base follows the solar butterfly-diagram trends. 
Because currently there is no {empirical} evidence {favouring} 
any specific 
pattern for the internal toroidal field for $\tom>1$, we preferred this simple and 
conservative approach to also test the {applicability of the} solar paradigm. 
Therefore, {the only change in 
the dynamo with increasing rotation rate is that it produces more 
toroidal flux, which reaches higher latitudes {at the base of the convection zone}. }}

{We extrapolated the empirical relation for the mean latitude of 
the solar butterfly pattern to higher levels of activity, as if {a solar cycle
had an amplitude} $\tos$ times its Cycle-22 level. This 
was done to obtain the base distributions of such very strong cycles before they rise 
to the surface, for $1\Omega_\sun$ (step II in Appendix~\ref{sec:app0}). 
This step is necessary because observations show that sunspots appear at higher average latitudes during stronger cycles \citep{solanki08, jcss11a}.}
However, it is also possible that in real stars the deviation from 
{a} solar-like butterfly diagram {at the base of the convection zone} 
can be significant, especially for more rapid rotators. 
{In future work,} extrapolation of composite or Babcock-Leighton type 
dynamo models to 
faster rotating Sun-like stars can be employed \citep{isik11,karak14} to 
{obtain an estimate of} the butterfly diagram of the internal toroidal field more consistently {\citep[see also][]{warnecke18}}. 

Our simulations show that the polar field exhibits the following trends with 
increasing $\tom$: 
$(i)$ it strengthens relative to lower-latitude activity,  
and $(ii)$ it reverses its polarity increasingly earlier with respect to the 
activity maximum (Fig.~\ref{fig:fluxvar}d), in spite of the fact that the initial 
polar field was scaled with the rotation rate. The first trend results from 
two effects, namely higher tilt angles and stronger {activity}.
The second {tendency} can lead to an earlier amplification of the toroidal 
field by the action of differential rotation on the poloidal field.
In addition to meridional circulation and dynamo effects 
\citep{jouve10,isik11}, it can thus contribute to 
shortening the cycle period for more active stars. 

There are two competing effects {in our models} which determine 
the polar field amplitude: the increasing tilt 
angles (with rotation rate) tend to amplify the developing 
polar fields, whereas the gap of inactivity 
opening around the equator (with faster rotation) 
leads to weaker cross-equatorial preceding-polarity flux 
cancellation, which limits the growth of a strong polar cap. 
We speculate that there is a critical rotation rate at which the 
two following timescales become comparable: 
the timescale for the magnetic flux of each polarity 
within a BMR to diffuse and cancel each other, and the timescale 
for the preceding-polarity flux from a typical BMR to diffuse 
and cancel with the preceding-polarity flux from a corresponding 
BMR from the opposite hemisphere.
Beyond such a critical rotation rate, 
polar fluxes would be saturated, provided that we are using 
the same solar transport parameters. Such high rotation rates and 
activity levels are beyond the scope of this paper. 

Another {effect that} can hinder 
the formation of {circumpolar} spots is related to the {dynamo} process. 
If the cycle period decreases and cycle overlap increases significantly 
with the rotation rate, then the polar fields {resulting from surface flux transport}
may not {reach} sufficient 
{strengths} to form spots before the subsequent cycle peaks 
\citep[{see the Sun-like model with $P_{\rm rot}=2$~days in}][]{isik11}. 
However, {circum}polar spots {are indeed} observed on some 
{Doppler images of} rapidly rotating Sun-like stars. 
Future observations and modelling of cycles on Sun-like stars should  address this issue. 

Our $(\tom,\tos)=(8,8)$ case can be compared to the Sun-like (G1.5V) star 
\object{EK~Draconis}, which has $\tom\simeq 8.8$. This star also 
exhibits a polar spot and mid-latitude activity in several Doppler and 
Zeeman-Doppler images \citep{jarvinen07,waite17}. 
This qualitative agreement is gratifying. Furthermore, its 
mean umbral spot coverage is estimated to be in the range 0.25-0.40 
based on TiO-band observations by \citet{oneal04}, whereas our model for (8,8) 
gives {{a cycle mean of 0.07 and a maximum of 0.20}}. 
We consider these values to be in reasonable agreement given 
the rather simple thresholding used to determine spot areas, and also that 
we scale the amount of magnetic flux linearly with the rotation rate. 

{We applied nesting of BMR emergence as observed on the Sun to more 
active stars. The resulting SFT simulations led to substantial rotational 
asymmetry in the starspot distributions on active stars. If the observed starspots 
are in fact low-resolution manifestations of such active nests \citep{ozavci18} 
then the observed 
sizes and lifetimes of these structures may not be indicative of the intrinsic sizes 
and lifetimes of their constituent spots \citep[see][for solutions of monolithic vs. 
clustered unipolar spot diffusion]{isik07}.}

{Zeeman-Doppler imaging studies show that the azimuthal 
component of the magnetic field strengthens and becomes comparable to or 
even exceeds the radial field for very active Sun-like stars. Our SFT model 
considers only the radial component of the field. Inclusion of the horizontal 
components \citep[e.g.][]{gibb16,lehmann17} represents a 
considerable extension of our current understanding, and they are beyond the 
present scope of modelling brightness variations of moderately active Sun-like 
stars. These components will therefore be introduced in a future study. }

Our assumption that the meridional flow and differential rotation profiles 
are identical to those on the Sun has an effect on the resulting 
latitudinal distribution of the magnetic field. For instance, we expect that a meridional 
flow reaching up to latitudes higher than our assumed $75^\circ$ 
would lead to a field peaking at the poles earlier in a given cycle. 
However, such details should not dramatically change the production of a polar 
spot. 
Deviations of these profiles from the solar patterns would affect the evolution 
of flux in various latitudinal zones. 
Alternative profiles from theoretical models of flow fields as a function 
of the rotation rate can be used in our model \citep{kueker11,ko12}. 

{Finally, we note that it might be possible to reproduce the 
observed 
surface patterns of activity starting from different combinations of dynamo-generated 
toroidal field patterns, initial flux-tube conditions, and large-scale flows in the 
convective envelope. However, our aim has not been to seek the parameters and 
functions which give the best fits to the observed patterns, but to stick to the solar 
paradigm and test its validity for more active and more rapidly rotating configurations. 
In qualitative terms, our models are more or less consistent with the overall 
observed patterns of surface inhomogeneities in longitude and latitude, as well as 
surface fractions. 
We reserve quantitative comparisons with the observed stellar brightness 
variations for the follow-up studies, in which we shall estimate fractions 
of spot and facular areas from the modelled surface magnetic fields, and 
generate light curves. }

We plan to include further features into the model in forthcoming studies, 
such as inflows towards the activity belts, active longitudes, 
and random scatter in the tilt angle, 
which will be relevant for multiple-cycle models. Such parametrisations can also
be implemented as part of an extrapolation of a solar flux-transport dynamo model. 

\section{Conclusion}
{Under the assumption that the toroidal field 
at the base of the convection zone follows the extrapolated solar patterns,}
we have developed a numerical platform combining two main 
physical effects 
which are likely responsible for the observed variety of activity patterns on 
Sun-like stars: $(i)$ the rise of flux tubes under 3D effects of the relevant 
large-scale forces, and $(ii)$ the surface transport of emerging large-scale magnetic 
fields, under the effects of large-scale surface flows. We find that
for the final modelled distribution of magnetic fields on stellar photospheres, 
the following proposed processes play important roles: $(i)$ the deviation 
from radial flux-tube rise \citep{ss92}, $(ii)$ the evolution of the field on the stellar 
surface \citep{st01}, and $(iii)$ increasing BMR tilt angles \citep{isik07,isik11}. We also find that 
the onset of polar spot formation on Sun-like stars can occur 
by accretion of trailing polarity flux from 
BMRs, between four and eight times the solar rotation rate and the flux emergence rate. Our results also show that 
 {nesting of emerging bipoles can have substantial effects on 
the surface distribution of starspots.}

The models developed here can be used in forward modelling of {brightness} 
variations in magnetically active Sun-like stars. 
{At a later stage, we plan to extend them to be helpful in understanding other observations of active Sun-like stars. }

\begin{acknowledgements}
We thank Jie Jiang and Robert Cameron for sharing their code {which} 
generates semi-synthetic 
solar emergence records, and Maria Weber for {confirming} our flux-tube 
emergence results using a different code. 
{We also thank the referee for the comments that helped to 
improve this manuscript.}
EI acknowledges support by the Young Scientist Award Programme BAGEP-2016 
of the Science Academy, Turkey. This work has been partially supported by the BK21 
plus program through the National Research Foundation (NRF) funded by the Ministry 
of Education of Korea. AS acknowledges funding from the European Research Council under the European Union Horizon 2020 research and innovation 
programme (grant agreement No. 715947). 
\end{acknowledgements}

\bibliographystyle{aa} 
\bibliography{mainrev2} 

\begin{appendix}

\section{Synthetic input solar cycle}
\label{sec:app}

\subsection{Emergence frequency}
\label{ssec:appa}
To model the temporal profile of the monthly group sunspot number, $R_G$, 
we follow CJS16 and use the function devised by \citet{hathaway94} with additional 
Gaussian random noise, $\Delta R_G$: 
\begin{equation}
    R_G(t)= \frac{at^3}{\exp(t^2/b^2)-c} + \Delta R_G(t),
    \label{eq:rg}
\end{equation}
where $t$ is the time in months and $a$ is the amplitude. We fit the first 
function on the RHS of Eq.~(\ref{eq:rg}) to sunspot group data for cycle 
22 from RGO, which yields $a=0.00336$. 
The quantities $b(a)$ and $c$ control the length and 
the asymmetry of the cycle, respectively. For $b$, we adopt 
Eq.~(4) of \citet{hathaway94} and $c=0.71$. 
The standard deviation 
of the probability distribution of $\Delta R_G(t)$ around $R_G(t)$ is approximated 
by $0.5R_G(t)$. This level was estimated by measuring the deviations of the 
observed $R_G$ of Cycles 21-23 from the corresponding fits of the first term on 
the RHS of Eq.~(\ref{eq:rg}). 
Finally, the number of BMRs emerging in a given month 
is obtained as $R_G/2.75$, based on a calibration carried out by CJS16, 
through fits to the monthly number of groups observed for Cycles 21-24. 

\subsection{Sunspot group latitudes}
Next, we set up a synthetic butterfly diagram in the input model, 
following the procedure of JCSS11 (also followed by CJS16). We take the
average latitude of spot groups at a given temporal phase bin 
$i$ of the cycle to be described by the quadratic function 
\begin{equation}
    \langle\lambda\rangle^i=\left[26.4-34.2(i/30)+16.1(i/30)^2\right]
    \langle\lambda\rangle/\langle\lambda\rangle_{12-20},
    \label{eq:lat}
\end{equation}
where $1\leqslant i\leqslant 30$ 
(the cycle is split into 30 temporal bins), $\langle\lambda\rangle_{12-20}=14.6^\circ$ 
is the average latitude over solar Cycles 12-20, and $\langle\lambda\rangle$ 
is the average latitude of sunspot groups over the cycle. This was obtained 
by JCSS11 for a given cycle using a 
linear fit to data from all the available cycles, as 
\begin{equation}
    \langle\lambda\rangle = 12.2+0.022S_\sun,
    \label{eq:latscale}
\end{equation}
where $S_\sun$ is the cycle amplitude. To represent the Sun, 
we set $S_\sun=156$ from the maximum 
of the twelve-month running mean of the observed $R_G$ of Cycle 22 
(see Table~1 of JCSS11). 
The second term on the RHS of Eq.~(\ref{eq:latscale}) models the 
observation that stronger 
solar cycles have a higher mean latitude of emergence \citep{solanki08}. 
For the latitudinal spread around $\langle\lambda\rangle$, 
the model assumes a Gaussian distribution with a standard deviation 
$\sigma^i$ that varies with the cycle phase, following the 
quadratic function 
\begin{equation}
    \sigma^i = \left[0.14+1.05(i/30)-0.78(i/30)^2\right]\lambda^i,
    \label{eq:sigma}
\end{equation}
where $1\leqslant i\leqslant 30$ as in Eq.~(\ref{eq:lat}).

\subsection{Sunspot group areas}
Sunspot group areas ($A$) are randomly picked from either a
power law or a log-normal distribution, depending on the size (in 
millionths of a solar hemisphere), $\mu$H, as given by
\begin{equation}
n(A)=
\begin{cases} 
    0.3A^{-1.1} & ; A < 300~{\rm\mu H} \\
    0.003\exp\left[-\frac{\left(\ln A-\ln 45\right)^2}{2\ln 3}\right] &
    ; A \geqslant 300~{\rm\mu H}.
\end{cases}
\label{eq:areadist}
\end{equation}

To include the dependence of the mean group area on the cycle phase, we adopt the 
relation given by JCSS11, but with $1\leqslant i \leqslant 20$, 
\begin{equation}
    A_i = 115+396(i/20)-426(i/20)^2.
\end{equation}

\section{The link between the base and the surface}
\label{sec:app0}
\begin{figure}
    \centering
    \includegraphics[width=.7\columnwidth]{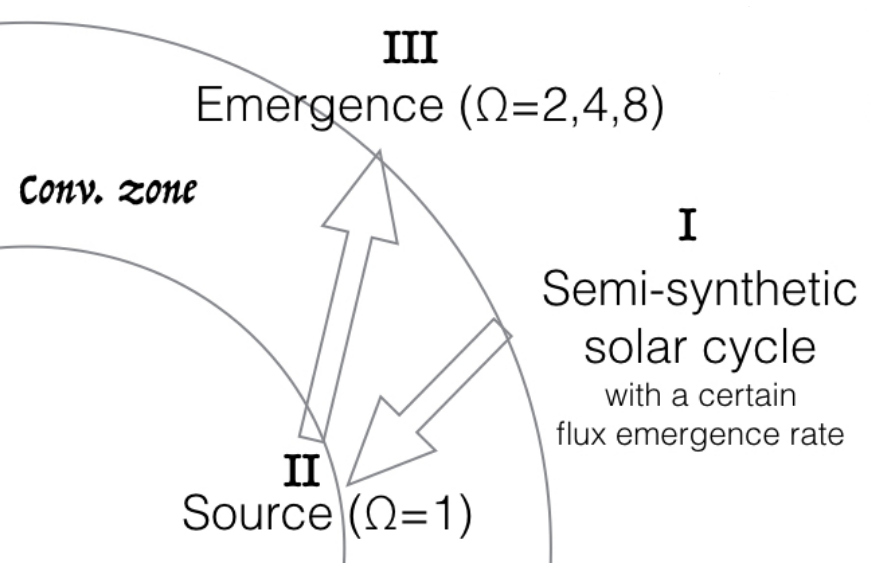}
    \caption{A schematic representation of the mappings 
    between the surface and the base of the convection zone in the model. 
    $\Omega$ denotes the equatorial rotation rate in solar units. The 
    roman numerals refer to the steps of the procedure.}
    \label{fig:scheme}
\end{figure}

To synthesise starspot emergence records, we follow the algorithm 
outlined below. The steps {I to III} correspond to those in Fig.~\ref{fig:scheme}.
\begin{enumerate}
    \item[I.] Generate a random realisation of {the spot group record} (SGR) with a 
    certain flux emergence rate $\tos$ (see Sects.~\ref{sssec:record1}--\ref{sssec:record2}). 
        {The mean latitude of this SGR depends on $\tos$ (Eq.~\ref{eq:meanlat}).} 
    \item[II.]  Map the latitudes obtained in 
    step I {down} to the base of the convection zone, 
    by interpolating the flux tube rise table for the solar rotation rate, $\tom=1$ 
    (Sect.~\ref{sssec:lookup}). 
    \item[III.] {Take} the time series {resulting from} 
    step II as the base {latitudes} of the flux tubes leading to spot groups, {for a given 
    rotation rate, $\tom$}. 
    Generate the emergence latitudes and tilt angles for $\tom$, by 
    interpolating within the flux-tube rise table (Sect.~\ref{sssec:lookup}). 
    \item[IV.] {To simulate the effect of active region nesting,  
    modify the longitudes and latitudes obtained in the previous steps, 
    using a probabilistic approach (Appendix~\ref{sec:nests}). }
\end{enumerate}

\section{Nests of activity}
\label{sec:nests}
{To simulate the observed tendency of flux emergence in the vicinity of recent 
emergence, we modified the longitudes and 
latitudes of BMRs in our standard starspot record, using a probabilistic approach. 
We first set a generic probability $0<p<1$ for each BMR with coordinates 
$(\lambda,\phi)$ to be part of a nest. 
{In this study, we chose 
a rather high probability of $p=0.7$ to clearly demonstrate the effects of nesting.}
This value has made the cycle variation of the equatorial 
dipole moment much more similar to the observed variations \citep[][see Fig.~3]
{wang14} in comparison to the unnested case. 

If a uniformly chosen random number $E_i\sim U[0,1]$ is less than 
$p$, then the coordinates of the $i$th BMR of the unnested record is considered 
as a potential nest centre, $(\lambda_c,\phi_c)_i$. 
If the next random number $E_{i+1}<p$, then the $(i+1)$th BMR belongs to the 
nest of the $i$th BMR. 
The coordinates of such a BMR is then given a 
modified pair of coordinates $(\lambda_m,\phi_m)_{i+1}$, which are drawn from 
a normal distribution 
centred around $(\lambda_c,\phi_c)_i$ with {widths} 
$2^\circ$ in latitude and $3^\circ$ 
in longitude, close to the empirical values obtained by \citet{pc02} for Cycle 23. 
The procedure is continued iteratively until for the $(i+k)$th BMR 
$E_{i+k}\geqslant p$ holds. Such a BMR is considered as 
an isolated spot group, while its original coordinates are kept unchanged. 
}

\end{appendix}

\end{document}